\pdfoutput=1
\documentclass[usenatbib]{mn2e}
\usepackage{apjfonts}
\usepackage{amssymb}
\usepackage{amsmath}
\usepackage{ctable}
\usepackage{multicol}
\usepackage{verbatim}
\usepackage{fixltx2e} %% forces figure order to remain fixed (otherwise figure*'s can move out-of-order)
\newcommand{\be}{\begin{equation}}
\newcommand{\ee}{\end{equation}}

\newcommand{\msun}{M_{\sun}}

\newcommand{\paperone}{SH}

\newcommand{\partialAB}[2]{\frac{\partial{#1}}{\partial{#2}}}
\newcommand{\omprime}{\varpi}
\newcommand{\omdimless}{\tilde{\omega}}
\newcommand{\kdimless}{\tilde{k}}
\newcommand{\BV}{Brunt-V\"ais\"al\"a}

\newcommand{\Intermediatemodename}{Quasi-sound} %{Intermediate}
\newcommand{\intermediatemodename}{quasi-sound} %{intermediate}
\newcommand{\intermediatemodesubscript}{\rm QS} %{\rm med}
\newcommand{\Slowmodename}{Quasi-drift} %{Slow}
\newcommand{\slowmodename}{quasi-drift} %{slow}
\newcommand{\slowmodesubscript}{\rm QD} %{\rm slow}

\newcommand{\driftvel}{{\bf w}_{s}}%{{\boldsymbol{\alpha}}}
\newcommand{\driftvelmag}{\tilde{w}_{s}}%{{\alpha}}
\newcommand{\driftvelhat}{\hat{{\bf w}}_{s}}%{\hat{\boldsymbol{\alpha}}}
\newcommand{\driftveli}[1]{{\bf w}_{s,\,#1}}%{{\boldsymbol{\alpha}}}
\newcommand{\driftvelmagi}[1]{\tilde{w}_{s,\,#1}}%{{\boldsymbol{\alpha}}}

\newcommand{\coeffTSrho}{\zeta_{s}}
\newcommand{\coeffTSrhoonly}{\zeta_{\rho}}
\newcommand{\coeffTSPonly}{\zeta_{P}}
\newcommand{\coeffTSv}{\zeta_{w}}
\newcommand{\tildeCoeffTSv}{\tilde{\zeta}_{w}}

\newcommand{\gradscale}{\Lambda}
\newcommand{\gradscalerho}{\Lambda_{\rho}}
\newcommand{\gradscaleP}{\Lambda_{P}}
\newcommand{\gradscalew}{\Lambda_{w}}
\newcommand{\gradscalemu}{\Lambda_{\mu}}

\newcommand{\coulombcoeff}{a_{C}} % Numerical coefficient for Coulomb drag 
\newcommand{\epsteincoeff}{a_{\gamma}} % Numerical coefficient for Epstein drag (used this in Letter)
\newcommand{\iimag}{i} % \imath has no dot... This looks weird to me... 
\newcommand{\omegaZ}{\varomega} % \omega_{0} was used in various expressions, but I had used this as my general ``resonant'' e-value in the other paper

\newcommand{\apjl}{ApJL}
\newcommand{\apjs}{ApJS}

\newcommand{\aap}{A\&A}

\newcommand{\plotsidesize}[2]{\centering \leavevmode \includegraphics[width={#2\textwidth}]{#1}}
\newcommand{\acknowledgments}{\begin{small}\section*{Acknowledgments}\end{small}}
\newcommand\altaffilmark[1]{$^{#1}$}
\newcommand\altaffiltext[1]{$^{#1}$}
\title[The Acoustic RDI]{The Resonant Drag Instability (RDI): Acoustic Modes
\vspace{-0.5cm}}

\vspace{-0.2cm}
\author[Hopkins \&\ Squire]{
\parbox[t]{\textwidth}{ 
Philip F.~Hopkins\altaffilmark{1}, \&\ 
Jonathan Squire\altaffilmark{1}
} 
\vspace*{6pt} \\
\altaffiltext{1}{TAPIR, Mailcode 350-17, California Institute of Technology, Pasadena, CA 91125, USA} 
\vspace{-0.5cm}
}

\date{Submitted to MNRAS, July 2017\vspace{-0.6cm}}
\begin{document}
\maketitle
\label{firstpage}

\vspace{-0.2cm}
\begin{abstract}
\vspace{-0.2cm}

Recently, Squire \&\ Hopkins (2017) showed any coupled dust-gas mixture is subject to a class of linear ``resonant drag instabilities'' (RDI). These can drive large dust-to-gas ratio fluctuations even at arbitrarily small dust-to-gas mass ratios $\mu$. Here, we identify and study both resonant and new non-resonant instabilities, in the simple case where the gas satisfies neutral hydrodynamics and supports  acoustic waves ($\omega^{2}=c_{s}^{2}\,k^{2}$). The gas and dust are coupled via an arbitrary drag law and subject to external accelerations (e.g.\ gravity, radiation pressure). If there is any dust drift velocity, the system is unstable. The instabilities exist for {\em all} dust-to-gas ratios $\mu$ and their growth rates depend only weakly on $\mu$ around resonance, as $\sim\mu^{1/3}$ or $\sim \mu^{1/2}$ (depending on wavenumber). The behavior changes depending on whether the drift velocity is larger or smaller than the sound speed $c_{s}$. In the supersonic regime a ``resonant'' instability appears with growth rate increasing {\em without limit} with wavenumber, even for vanishingly small $\mu$ and values of the coupling strength (``stopping time''). In the subsonic regime non-resonant instabilities always exist, but their growth rates no longer increase indefinitely towards small wavelengths. The dimensional scalings and qualitative behavior of the instability do not depend sensitively on the drag law or equation-of-state of the gas. The instabilities directly drive exponentially growing dust-to-gas-ratio fluctuations, which can be large even when the modes are otherwise weak. We discuss physical implications for cool-star winds, AGN-driven winds and torii, and starburst winds: the instabilities alter the character of these outflows and could drive clumping and/or turbulence in the dust and gas.

\end{abstract}

\begin{keywords}
instabilities --- turbulence --- ISM: kinematics and dynamics --- star formation: general --- 
galaxies: formation --- planets and satellites: formation\vspace{-0.5cm}
\end{keywords}

\vspace{-1.1cm}
\section{Introduction}
\label{sec:intro}

Astrophysical fluids are replete with dust, and the dynamics of the dust-gas mixture in these ``dusty fluids'' are critical to astro-chemistry, star and planet formation, ``feedback'' from stars and active galactic nuclei (AGN) in galaxy formation, the origins and evolution heavy elements, cooling in the inter-stellar medium, stellar evolution in cool stars, and more. Dust is also ubiquitous as a source of extinction or contamination in almost all astrophysical contexts. As such, it is critical to understand how dust and gas interact, and whether these interactions produce phenomena that could segregate or produce novel dynamics or instabilities in the gas or dust.

Recently, \citet{squire.hopkins:RDI} (henceforth \paperone) showed that there exists a general class of previously unrecognized instabilities of dust-gas mixtures. The \paperone\ ``resonant drag instability'' (RDI) generically appears whenever a gas system that supports some wave or linear perturbation mode (in the absence of dust) also contains dust moving with a finite drift velocity $\driftvel$ relative to the gas. This is unstable at a wide range of wavenumbers, but the fastest-growing instabilities occur at a ``resonance'' between the phase velocity ($v_{p} = \omega_{0}/|{\bf k}|$) of the ``natural'' wave that would be present in the gas (absent dust), and the dust drift velocity projected along the wavevector direction ($\driftvel\cdot \hat{\bf k} \approx v_{p}$).\footnote{Equivalently, we can write the resonance condition as $\driftvel\cdot{\bf k} \approx \omega_{0}$, where $\omega_{0} = v_{p}\,|{\bf k}|$ is the natural frequency a wave would have in the gas, absent dust drag. Note this is a resonance condition for a given (single) Fourier mode -- it does not require two different modes actually be present.} Some previously well-studied instabilities -- most notably the ``streaming instability'' of grains in protostellar disks \citep{youdin.goodman:2005.streaming.instability.derivation}, which is related to a resonance with the disk's epicyclic oscillations (i.e.\ has maximal growth rates when $\driftvel \cdot {\bf k} \approx \Omega$) --  belong to the  general RDI category. These instabilities {\em directly} generate fluctuations in the dust-to-gas ratio and the relative dynamics of the dust and gas, making them potentially critical for the host of phenomena above (see, e.g., \citealt{chiang:2010.planetesimal.formation.review} for applications of the disk streaming instability).

The relative dust-gas drift velocity $\driftvel$ and the ensuing instabilities can arise for a myriad of reasons. For example, in the photospheres of cool stars, in the interstellar medium of star-forming molecular clouds or galaxies, and in the obscuring ``torus'' or narrow-line region around an AGN, dust is accelerated by absorbed radiation from the stars/AGN, generating movement relative to the gas. Similarly, in a proto-stellar disk, gas is supported via pressure, while grains (without such pressure) gradually sediment. In both cases, a  drag force, which couples the dust to the gas, then causes the dust to accelerate the gas, or vice versa. While there has been an extensive literature on such mechanisms -- e.g.,\ radiation-pressure driven winds -- there has been surprisingly little focus on the question of whether the dust can stably transfer momentum to gas under these conditions. We will argue that these process are all inherently unstable. 

Perhaps the simplest example of the RDI occurs when one considers ideal, inviscid hydrodynamics, where the only wave (absent dust) is a sound wave. This ``acoustic RDI'' has not yet been studied, despite having potentially important implications for a wide variety of astrophysical systems. In this paper, we therefore explore this manifestation of the RDI in detail. We show that homogenous gas, coupled to dust via some drag law, is generically unstable to a spectrum of exponentially-growing linear instabilities, regardless of the form of the dust drag law, the magnitude of the drift velocity, the dust-to-gas ratio, the drag coefficient or ``stopping time,'' and the source of the drift velocity. 
This includes both the ``resonant'' instabilities above as well as several non-resonant instabilities which have not previously been identified. If the drift velocity exceeds the sound speed, the ``resonance'' condition is always met and the growth rate increases without limit at short wavelengths.

We present the basic derivation and linearized equations-of-motion in \S~\ref{sec:deriv}, including various extensions and caveats (more detail in Appendices). In \S~\ref{sec:general.modes}, we then derive the stability conditions, growth rates, and structure of the unstable modes for arbitrary drag laws, showing in \S~\ref{sec:draglaws}  how this specifies to various physical cases (Epstein drag, Stokes drag, and Coulomb drag).
The 
discussion of \S~\ref{sec:general.modes}--\S~\ref{sec:draglaws}  is necessarily rather involved, covering a variety of different unstable
modes in different physical regimes, and the reader more interested in applications
may wish to read just the general overview in  \S~\ref{sec:general overview of modes}, the
discussion of mode structure in \S~\ref{sec:mode.structure}, and skim through  relevant drag
laws of \S~\ref{sec:draglaws}.
We briefly discuss the non-linear regime (\S~\ref{sec:nonlinear}), scales where our analysis breaks down (\S~\ref{sec:breakdown}), and the relation of these instabilities to those discussed in previous literature (\S~\ref{sec:previous.work}), before considering applications to different astrophysical systems including cool-star winds, starbursts, AGN obscuring torii and narrow-line regions, and protoplanetary disks (\S~\ref{sec:applications}). We conclude in \S~\ref{sec:summary}.

\vspace{-0.5cm}
\section{Basic Equations \&\ Linear Perturbations}
\label{sec:deriv}

\begin{figure*}
\plotsidesize{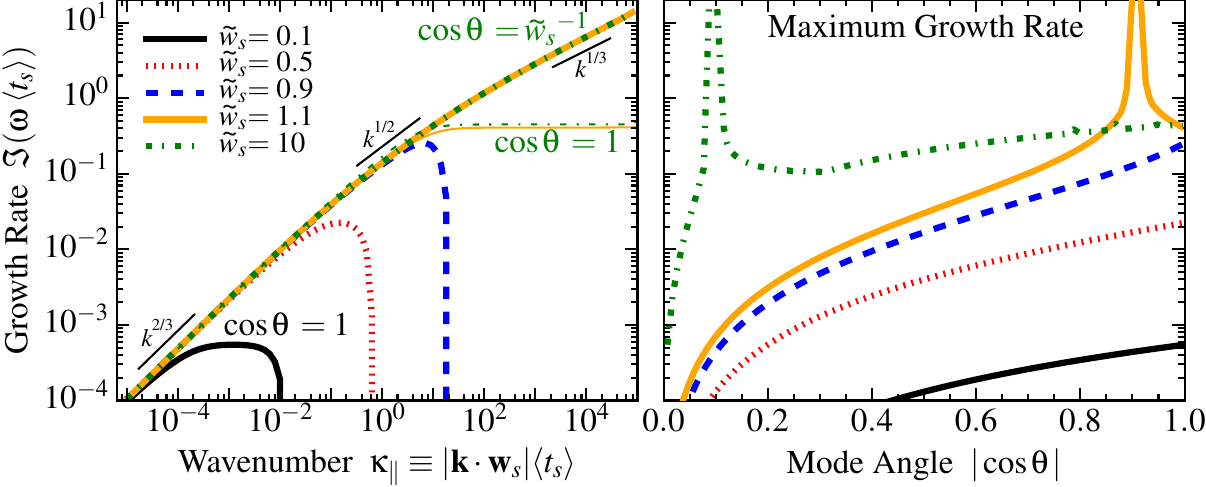}{0.99}
    \vspace{-0.25cm}
    \caption{Linear growth rates of the acoustic RDI. We show the growth rate $\Im{(\omega)}$ of the fastest-growing unstable mode (in units of the equilibrium dust drag timescale or ``stopping time'' $\langle t_{s} \rangle$; Eq.~\eqref{eqn:general}), for dust moving through gas with drift/streaming velocity $\driftvel$ (Eq.~\eqref{eqn:mean.v.offset}). For convenience we define the dimensionless $\driftvelmag \equiv |\driftvel|/c_{s}$ as the ratio of $\driftvel$ to the sound speed (\S~\ref{sec:deriv}).
   Here we assume a mean dust-to-gas mass ratio $\mu=0.1$ (Eq.~\eqref{eqn:mean.v.offset}), constant drag coefficient ($\coeffTSrho=\coeffTSv=0$; Eq.~\eqref{eqn:ts.general}), and a homogeneous background (\S~\ref{sec:pressure.gradients}). 
    {\em Left:} Growth rate vs.\ wavenumber ${\bf k}$ (\S~\ref{sec:general.modes}), in terms of the dimensionless $\kappa_{\|} \equiv {\bf k}\cdot \driftvel\,\langle t_{s} \rangle = |{\bf k}|\,|\driftvel|\,\langle t_{s} \rangle \cos{\theta}$ (Eq.~\eqref{eqn:kappa.definition}), and angle $\cos{\theta}\equiv \hat{\bf k} \cdot \driftvelhat$ between the wavevector ${\bf k}$ and $\driftvel$. For ``subsonic'' cases with $\driftvelmag < 1$, modes are unstable at long wavelengths (see \S~\ref{sec:long.wavelength}) with growth rates $\propto \kappa_{\|}^{2/3}$ (Eq.~\eqref{eqn:longwave.mode}) then saturate at a maximum growth rate, and are stabilized at high-$k$ (\S~\ref{sec:subsonic.long.wavelenghts}). We show the fastest-growing angle $\cos{\theta}=1$ for $\driftvelmag<1$. Note that up to their saturation value, the different-$\driftvelmag$ cases behave identically. 
    For ``supersonic cases'' with $\driftvelmag \ge 1$, all $k$ are unstable; at most angles the growth rate saturates at a constant value (the ``\intermediatemodename'' mode in \S~\ref{sec:intermediate}), but for $\cos{\theta}=\pm1/\driftvelmag$ the ``resonant'' RDI appears (\S~\ref{sec:resonance}), where the drift velocity in the direction $\hat{\bf k}$ is resonant with the natural response frequency of the system (a sound wave), and the growth rates increase without limit as $\propto \kappa_{\|}^{1/2}$ (Eq.~\eqref{eqn:longwave.mode.midk}) and $\propto \kappa_{\|}^{1/3}$ (Eq.~\eqref{eqn:omega.resonant}) at intermediate and high $\kappa_{\|}$, respectively. 
    {\em Right:} Maximum growth rate (over all $k$) as a function of angle. For $\driftvelmag<1$ this is maximized at finite growth rate, at $\cos{\theta}=\pm1$; for $\driftvelmag \ge1$, the maximum growth rates diverge around the ``resonant angle.'' 
    \vspace{-0.25cm}
        \label{fig:growth.rate.demo}}
\end{figure*}

\vspace{-0.05cm}
\subsection{General Case with Constant Streaming}
\label{sec:free.streaming}

Consider a mixture of gas and a second component which can be approximated as a pressure-free fluid  (at least for {\em linear} perturbations; see \citealt{youdin.goodman:2005.streaming.instability.derivation} and App.~A of \citealt{Jacquet:2011cy}), interacting via some generalized drag law. We will refer to this second component as ``dust'' henceforth. For now we  consider an ideal, inviscid gas, so the system is described by mass and momentum conservation for both fluids: 
\begin{align}
\nonumber \partialAB{\rho}{t} + \nabla\cdot ({\bf u}\,\rho) &= 0,\\
\nonumber \left(\partialAB{}{t} + {\bf u}\cdot\nabla \right){\bf u} &=  -\frac{\nabla P}{\rho} +  {\bf g} + \frac{\rho_{d}}{\rho}\,\frac{({\bf v}-{\bf u})}{t_{s}},\\
\nonumber \partialAB{\rho_{d}}{t} + \nabla\cdot ({\bf v}\,\rho_{d}) &= 0,\\
\label{eqn:general}  \left(\partialAB{}{t} + {\bf v}\cdot\nabla \right){\bf v} &= -\frac{({\bf v}-{\bf u})}{t_{s}} +  {\bf g} + {\bf a},
\end{align}
where ($\rho,\,{\bf u}$) and ($\rho_{d},\,{\bf v}$) are the density and velocity of the gas and dust, respectively; ${\bf g}$ is the external acceleration of the gas while ${\bf g}+{\bf a}$ is the external acceleration of dust (i.e., ${\bf a}$ is the difference in the dust and gas acceleration), and  $P$ is the gas pressure. We assume a barotropic equation 
of state with sound speed $c_{s}^{2}=\partial P/\partial \rho$ and polytropic index $\gamma$ (see \S~\ref{sec:epstein} for further details). The dust experiences a drag acceleration ${\bf a}_{\rm drag} = -({\bf v}-{\bf u})/t_{s}$ with an arbitrary drag coefficient $t_{s}$, known as the ``stopping time''  (which  can be a function of other properties). The term in $t_{s}$ in the gas acceleration equation is the ``back-reaction'' -- its form is dictated by conservation of momentum.

The equilibrium (steady-state), spatially-homogeneous solution to Eq.~\eqref{eqn:general} is the dust and gas accelerating together at the same rate, with a constant relative drift velocity $\driftvel$:
\begin{align}
\nonumber \rho^{h} &= \langle \rho \rangle = \rho_{0}, \\ 
\nonumber \rho_{d}^{h} &= \langle \rho_{d} \rangle = \rho_{d,\,0} \equiv \mu\,\rho_{0}, \\ 
\nonumber {\bf u}^{h} &= \langle {\bf u} \rangle = {\bf u}_{0} + \left[{\bf g} + {\bf a}\,\left(\frac{\mu}{1+\mu}\right) \right]\,t,\\
\nonumber {\bf v}^{h} &= \langle {\bf v} \rangle = \langle {\bf u} \rangle + \driftvel,\\
\label{eqn:mean.v.offset} \driftvel &\equiv \frac{{\bf a}\,\langle t_{s} \rangle}{1+\mu} = \frac{{\bf a}\, t_{s}^{h}(\rho^{h},\,\driftvel,\,...)}{1+\mu},
\end{align}
where we define the total mass-ratio between the two fluids as $\mu\equiv \langle \rho_{d} \rangle / \langle \rho \rangle$, and $\langle t_{s} \rangle = t_{s}(\langle \rho \rangle,\,\langle {\bf v} \rangle,\,...)$ is the value of $t_{s}$ for the homogeneous solution.\footnote{Eq.~\ref{eqn:general} also admits {\em non-equilibrium} but spatially homogeneous solutions with an additional initial transient/decaying drift $\Delta{\bf w}_{0} = {\bf w}_{0}\,\exp{(-t/\langle t_{s} \rangle)}$ (Eq.~\ref{eqn:mean.v.offset} with $\langle {\bf u} \rangle \rightarrow {\bf u}_{0} + [{\bf g}+{\bf a}\,\mu/(1+\mu)]\,t - (\mu/(1+\mu))\,\Delta{\bf w}_{0}$, $\langle {\bf v} \rangle \rightarrow \langle {\bf u} \rangle + \driftvel + \Delta{\bf w}_{0}$). If we consider modes with growth timescales $1/\Im{(\omega)} \gg \langle t_{s} \rangle$, then $\Delta{\bf w}_{0}\rightarrow 0$ decays rapidly and our analysis is unchanged by such initial transient drifts; alternatively if $1/\Im{(\omega)} \ll  \langle t_{s} \rangle$, then $\Delta{\bf w}_{0}\approx {\bf w}_{0}$ is approximately constant and our analysis is identical with the replacement $\driftvel \rightarrow \driftvel + {\bf w}_{0}$.}
  Note that $\langle t_{s} \rangle$ can depend on $\driftvel$, so Eq.~\eqref{eqn:mean.v.offset} is in general  a non-linear equation for $\driftvel$. 
Let us  also define the normalized drift speed $\driftvelmag \equiv |\driftvel|/c_{s}$, which  is a key parameter in determining stability properties and will be used extensively below. (Note that this definition of $\driftvelmag$ differs from that of \paperone: this dimensionless version is more convenient throughout this work because of our focus on the acoustic resonance; see \S~\ref{sec:General dispersion relation}.)% (shown explicitly below). Also note that in steady-state, both dust and gas accelerate together with the same constant rate $={\bf g} + {\bf a}\,\mu/(1+\mu)$ (required to prevent a net mass flux) -- this is independent of $t_{s}$ -- while the dust moves with a constant velocity offset $\driftvel$ (along the direction of ${\bf a}$) relative to the gas, which depends on $t_{s}$, ${\bf a}$, and $\mu$. 

We now consider small perturbations $\delta$: $\rho = \rho^{h} + \delta\rho$, ${\bf u} = {\bf u}^{h} + \delta {\bf u}$, etc., and adopt a free-falling frame moving with the homogeneous gas solution $\langle {\bf u} \rangle$ (see App.~\ref{sec:accel.frame} for details). Linearizing Eq.~\eqref{eqn:general}, we obtain,
%: in the new frame we have the homogeneous solutions $({\bf u}^{h})^{\prime} \rightarrow 0$, $({\bf v}^{h})^{\prime} \rightarrow \driftvel$, with a fictitious acceleration $= -({\bf g} + {\bf a}\,\mu/(1+\mu))$. Now inserting the homogenous solution and linearize to obtain the perturbation equations: 
\begin{align}
\nonumber \partialAB{\delta\rho}{t} =& -\rho_{0}\,\nabla\cdot \delta{\bf u},\\
\nonumber \partialAB{\delta{\bf u}}{t} =& -c_{s}^{2}\,\frac{\nabla \delta \rho}{\rho_{0}}
+ \mu\,\frac{(\delta {\bf v}-\delta{\bf u})}{\langle t_{s} \rangle} \\
\nonumber &- \mu\,\frac{\driftvel}{\langle t_{s} \rangle}\,\left( \frac{\delta t_{s}}{\langle t_{s} \rangle} + \frac{\delta \rho}{\rho_{0}} - \frac{\delta \rho_{d}}{\mu\,\rho_{0}} \right), \\ 
\nonumber \left( \partialAB{}{t} + \driftvel\cdot\nabla \right)\delta \rho_{d} =& -\mu\,\rho_{0}\,\nabla\cdot \delta{\bf v},\\
\label{eqn:linearized} \left( \partialAB{}{t} + \driftvel\cdot\nabla \right)\delta {\bf v} =& -\frac{(\delta {\bf v}-\delta{\bf u})}{\langle t_{s} \rangle} + \frac{\driftvel\,\delta t_{s}}{\langle t_{s} \rangle^{2}},
\end{align}
where all coordinates here now refer to those in the free-falling frame, and we have defined $\delta t_{s}$ as the linearized perturbation to $t_{s}$; i.e.\ $t_{s} \equiv \langle t_{s} \rangle + \delta t_{s}(\delta \rho,\,\delta {\bf v},\, ...) + \mathcal{O}(\delta^{2})$.

We now Fourier decompose each variable, $\delta \propto \exp{[\iimag\,({\bf k}\cdot {\bf {x}} - \omega\,{t})]}$, and define the parallel and perpendicular components of ${\bf k} \equiv k_{\|}\,\driftvelhat + k_{\bot}\,\hat{\bf k}_{\bot}$. Because of the symmetry of the problem, the solutions are independent of the orientation of ${\bf k}_{\bot}$ in the plane perpendicular to $\driftvelhat$. The density equations trivially evaluate to $\delta \rho = \rho_{0}\,\omega^{-1}\,{\bf k}\cdot \delta {\bf u}$ and $\delta \rho_{d} = \mu\,\rho_{0}\,(\omega - \driftvel\cdot {\bf k})^{-1}\,{\bf k}\cdot \delta {\bf v}$, and the momentum equations can be written
\begin{align}
\nonumber \omega\,\delta{\bf {u}} + \mu\,(\omega - \driftvel\cdot{\bf k})\,\delta{\bf {v}}
&= \frac{( c_{s}^{2}\,\langle t_{s} \rangle\,{\bf k} - \iimag\,\mu\,{\driftvel} )\,{\bf k}\cdot\delta{\bf {u}}}{\omega\,\langle t_{s} \rangle} \\ 
\nonumber &\ \ \ \ \ \ \ \ \  + \frac{(\iimag\,\mu\,{\driftvel})\,{\bf k}\cdot\delta{\bf {v}}}{(\omega-\driftvel\cdot{\bf k})\,\langle t_{s} \rangle}, \\ 
\label{eqn:linearized.fourier} \iimag\,{\driftvel}\,\frac{\delta{t_{s}}}{\langle t_{s} \rangle} &= \langle t_{s} \rangle\,(\omega - \driftvel\cdot{\bf k})\,\delta{\bf {v}} + \iimag\,(\delta{\bf {v}} - \delta{\bf {u}}).
\end{align}
In this form, the first equation is the total momentum equation for the sum gas+dust mixture. The next equation encodes our ignorance about $t_{s}$. 

A couple of important results are immediately clear from here and Eq.~\eqref{eqn:linearized}. After removing the homogeneous solution, ${\bf g}$ vanishes: an identical uniform acceleration on dust and gas produces no interesting behavior. 
More precisely, as derived in detail in App.~\ref{sec:accel.frame}, a transformation from the free-falling frame, which moves with velocity $\langle {\bf u} \rangle = {\bf u}_{0} + [{\bf g} + {\bf a}\,\mu/(1+\mu)]\,t$, back into the stationary frame, is exactly equivalent to making the replacement $\omega \rightarrow \omega + {\bf u}_{0}\cdot {\bf k} + (t/2)\,[ {\bf g} + {\bf a}\,\mu/(1+\mu) ]\cdot {\bf k}$. In other words, the only difference between 
working in the stationary and free-falling frames is a trivial phase-shift of the modes. This implies that
the acceleration ${\bf a}$ is  important only insofar as it produces a non-vanishing dust-gas drift velocity $\driftvel$, and any source producing the same equilibrium drift will produce the same linear instabilities.
Finally, we note that if ${\bf a}=\mathbf{0}$, then $\driftvel=\mathbf{0}$ and  the equations become those for a coupled pair of soundwaves with friction (all modes are stable or decay). This also occurs if $\delta {\bf u}$ and $\delta {\bf v}$ are strictly perpendicular to $\driftvel$.

In this manuscript, we will consider only single-wave perturbations in linear perturbation theory -- i.e.\ the dispersion relation and ensuing instabilities studied here involve a single wave at a given ${\bf k}$ and $\omega({\bf k})$, as opposed to, e.g.,\ higher-order two-wave interactions involving waves with different $\omega_{1}$, $\omega_{2}$. To be clear, although the waves we study necessarily involve both gas and dust, the drag coupling means that the two phases  cannot be considered separately.

To make further progress, we require a functional form for $t_{s}$ to determine $\delta t_{s}$. For most physically interesting drag laws, $t_{s}$ depends on some combination of the density, temperature, and velocity offset $|{\bf v}-{\bf u}|$ (more below). Therefore, for now, we consider an {\em arbitrary} $t_{s}$ of the form $t_{s} = t_{s}(\rho,\,T,\,c_{s},\,{\bf v}-{\bf u})$. We will assume there is some equation-of-state which can relate perturbations in $T$ and $c_{s}$ to $\rho$. Then the linearized form obeys,
\begin{align}
%\label{eqn:ts.general} t_{s} &= t_{s}(\rho,\,T,\,c_{s},\,{\bf v}-{\bf u}) \\ 
\label{eqn:ts.general} \frac{\delta {t_{s}}}{\langle t_{s} \rangle} &= -\coeffTSrho\,\frac{\delta{\rho}}{\rho_{0}} - \coeffTSv\,\frac{{\driftvel}\cdot\left(\delta{\bf {v}} - \delta{\bf {u}} \right) }{|\driftvel|^{2}},
\end{align}
where $\coeffTSrho$ and $\coeffTSv$ are the drag coefficients\footnote{Note that we label the $\delta \rho/\rho_{0}$ coefficient in Eq.~\eqref{eqn:ts.general} as $\coeffTSrho$ because it encodes the dependence of $t_{s}$ on density at constant entropy; see App.~\ref{sec:hydrostatic.generalized}.} that depend on the form of $t_{s}$ 
(see \S~\ref{sec:draglaws}).

\vspace{-0.5cm}
\subsection{Gas Supported By Pressure Gradients and Abitrarily-Stratified Systems}
\label{sec:pressure.gradients}

Above we considered a homogeneous, freely-falling system. Another physically relevant case is when the gas is stationary (hydrostatic), which requires a pressure gradient (with $\nabla P_{0} = \rho_{0}\,{\bf g} + \rho_{d,\,0}\,\driftvel/\langle t_{s} \rangle$). This will generally involve stratification in other properties as well (e.g.\ gas and dust density), so more broadly we can consider arbitrary stratification of the background quantities $P_{0}$, $\rho_{0}$, $\rho_{d,\,0}$, and $\driftvel$. 

As usual, if we allow such gradients, we must restrict our analysis to spatial scales shorter than the background gradient scale-length $L_{0}$ (e.g.\ $k \gg |\nabla U_{0}|/|U_{0}|\sim 1/L_{0}$, for each variable $U_{0}$), or else a global solution (with appropriate boundary conditions, etc.) is obviously needed. Moreover we must also require $|\driftvel|\,t_{s} \ll L_{0}$, or else the timescale for the dust to ``drift through'' the system scale-length is much shorter than the stopping time (and no equilibrium can develop). So our analysis should be considered local in space and time, with these criteria imposing maximum spatial and timescales over which it is applicable (with actual values that are, of course, problem-dependent). We discuss these scales with various applications in \S~\ref{sec:breakdown}. 

%If we consider length scales much shorter than the pressure-gradient scale-length $P/|\nabla P| \sim c_{s}^{2}/|{\bf g}+\mu\,({\bf g}+{\bf a})| \gg |{\bf k}|^{-1}$, then we usually expect the correction from $\nabla P$ is negligible. But because we will consider cases with, say, arbitrarily small $\mu$, we wish to confirm this intuition explicitly. 

In App.~\ref{sec:hydrostatic.generalized}, we re-derive our results, for the unstable modes considered in this paper, for hydrostatic systems with arbitrary stratification in $P_{0}$, $\rho_{0}$, $\rho_{d,\,0}$, and $\driftvel$. Provided we meet the conditions above required for our derivation to be valid (i.e.\ $k \gg 1/L_{0}$), we argue (at least to lowest order in a local approximation) that : 
\begin{itemize}
\item{\bf (1):} The existence and qualitative (e.g.\ dimensional, leading-order) scalings of all the instabilities analyzed here in the homogeneous case are not altered by stratification terms, and the leading-order corrections to both the real and imaginary parts (growth rates and phase velocities) of the relevant modes are usually expected to be fractionally small. 

\item{\bf (2):} Pressure gradients (the term required to make the system hydrostatic) enter especially weakly at high-$k$ in the behavior of the instabilities studied here. In our (simplified) analysis, the leading-order correction from stratification is from non-vanishing $\nabla \cdot \driftvel \sim \rho_{d,\,0}^{-1}\, \driftvel \cdot \nabla \rho_{d,\,0}$, i.e.\ a background dust density and drift velocity gradient along the direction of the drift. The sense of the resulting correction is simply that modes moving in the direction of the drift are stretched or compressed along with the background dust flow. This particular correction is therefore large only if the timescale for the dust to drift through the dust-density gradient-scale-length is short compared to mode growth timescales.

\item{\bf (3):} The leading-order corrections from stratification are not necessarily stabilizing or de-stabilizing (they can increase or decrease the growth rates).

\item{\bf (4):} Introducing stratification introduces new instabilities. For example, even when the {gas} is stably stratified, stratification leads to new linear modes in the gas, e.g.\ \BV\ buoyancy oscillations. As shown in \paperone, if these modes exist in the gas, there is a corresponding RDI (the \BV\ RDI studied  in \paperone), which has maximal growth rates when $\driftvel\cdot{\bf k} = \pm (k_{\bot}/k)\,N_{BV}$, i.e.\ when $\driftvel\cdot{\bf k}$ matches the \BV\ frequency $N_{BV}$. We defer detailed study of these modes to a 
companion paper, \citet{squire:rdi.ppd}, since they are not acoustic instabilities and have fundamentally different behaviors and dimensional scalings (e.g.\ resonance exists for all $\driftvelmag$, but the growth rates are always lower than those of the acoustic RDI at high-$k$ if $\driftvelmag > 1$). 
	
\end{itemize}

In what follows, we will take the homogeneous (free-falling) case to be our ``default'' reference case, for two reasons. (1) The homogeneous and stratified cases exhibit the same qualitative behaviors, instabilities, and modes in all limits we wish to study, but the mathematical expressions are considerably simpler in the homogeneous case. And (2), as discussed in  \S~\ref{sec:applications}, the situations where the acoustic RDI is of the greatest  astrophysical interest involve dust-driven winds (e.g.\ in cool stars, star-forming regions, AGN torii, etc.). Such systems are generally better approximated as being freely accelerating than in  hydrostatic equilibrium.

Of course, even in a ``free-accelerating'' system, there will still be gradients in fluid properties (e.g.\ as a wind expands and cools). So our focus on the homogeneous case is primarily for the sake of generality and mathematical simplicity, and must therefore be considered a local approximation in both space and time (see \S~\ref{sec:breakdown}).

\vspace{-0.5cm}
\subsection{Neglected physics}

\subsubsection{Magnetized Gas and Dust}
\label{sec:mhd}

In this paper, we focus for simplicity on a  pure hydrodynamic fluid. If the system is sufficiently magnetized, new wave families appear (e.g.\ shear Alfven, slow, and fast magnetosonic waves in MHD). \paperone\ show that slow and fast magnetosonic waves, just like the acoustic waves here, are subject to the RDI (even when there is no Lorentz force on the dust).  For resonant modes, when the projected  dust streaming velocity ($\driftvel\cdot \hat{\bf k}$) matches either the slow or fast wave phase velocity, the qualitative behavior is similar to the acoustic RDI studied here (\S~\ref{sec:resonance}). 
Further, like for hydrodynamic modes studied in detail below (\S~\ref{sec:general.modes}), even modes
that are not resonant can still be unstable (but, unsurprisingly, the MHD-dust system is  more complicated; see \citealt{tytarenko:two.fluid.drift.intabilities}). 

Another effect, which was not included in \paperone, is grain charge.
If the gas is magnetized and the grains are sufficiently charged, then  Lorentz forces may dominate over the aerodynamic drag laws we consider here. 
This regime is relevant to many astrophysical systems (even, e.g., cosmic ray instabilities;  \citealp{kulsrud.1969:streaming.instability,Bell.cosmic.rays}). Lorentz forces will alter the equilibrium solution, and introduce additional dependence of the mode structure on the direction of ${\bf k}$ via cross-product terms (terms perpendicular to both the mean drift and magnetic field), although they do not generally suppress (and in many cases actually enhance) the RDI.

For these reasons, we defer a more detailed study of MHD to the follow-up study, \citet{hopkins:2018.mhd.rdi}.

%However, because the ``grains'' can no longer be treated as a pressureless  fluid (particles undergo nontrivial drifts if the accelerating force is not parallel to the magnetic field), the physics of this regime is different and we leave its study  to future work.

\vspace{-0.5cm}
\subsubsection{Multi-Species Dust}
\label{sec:dust.species}

Astrophysical dust is distributed over a broad spectrum of sizes (and other internal properties), producing different $t_{s}$, ${\bf v}$, ${\bf a}$ for different species. Consider de-composing the dust into sub-species $i$. Since the dust is pressure free, the dust continuity and momentum equations in Eq.~\eqref{eqn:general} simply become a pair of equations for each sub-species $i$. Each has a continuity equation for $\rho_{d,\,i}$ (where $\rho_{d} = \sum_{i}\,\rho_{d,\,i}$) and momentum equation for ${\bf v}_{i}$, each with their own acceleration ${\bf a}_{i}$ and drag $t_{s,\,i}$, but otherwise identical form to Eq.~\eqref{eqn:general}. The gas continuity equation is identical, and the gas momentum equation is modified by the replacement of the drag term $\rho_{d}\,({\bf v}-{\bf u})/t_{s} \rightarrow \sum_{i}\,\rho_{d,\,i}\,({\bf v}_{i}-{\bf u})/t_{s,\,i}$. The homogeneous solution now features each grain species moving with $\driftveli{i}$ where $\driftveli{i} \propto {\bf a}_{i}\,t_{s,\,i}$, so the sum in the gas momentum equation becomes $\sum_{i}\,\rho_{d,\,i}\,({\bf v}_{i}-{\bf u})/t_{s,\,i} \sim \sum_{i}\,\mu_{i}\,{\bf a}_{i}$. 

The most important grain property is usually size (this, to leading order, determines other properties such as charge). For a canonical spectrum of individual dust grain sizes ($R_{d}$), the total dust mass contained in a logarithmic interval of size scales as $\mu_{i} \propto  d\mu / d\ln{R_{d}} \propto R_{d}^{0.5}$, i.e.\ most of the dust mass is concentrated in the largest grains  \citep{mathis:1977.grain.sizes,draine:2003.dust.review}. Further, for any physical dust law (see \S~\ref{sec:draglaws}), $t_{s,\,i}$ increases with $R_{d}$. In most situations, we expect $|{\bf a}_{i}|$ to depend only weakly on $R_{d}$. This occurs: (i) if the difference in dust-gas acceleration is sourced by gravity or pressure support for the gas, (ii) when the gas is directly accelerated by some additional force (e.g.\ radiative line-driving), or (iii) when the dust is radiatively accelerated by long-wavelength radiation.\footnote{If dust is radiatively accelerated by a total incident flux ${\bf F}_{\lambda}$ centered on some wavelength $\lambda$, the acceleration is ${\bf a} \approx {\bf F}_{\lambda}\,Q_{\lambda}\,\pi\,R_{d}^{2} / (c\,m_{d}) \propto Q_{\lambda}/R_{d}$, where $m_{d}\propto \bar{\rho}_{d}\,R_{d}^{3}$ is the grain mass and $Q_{\lambda}$ is the absorption efficiency which scales as $Q_{\lambda}\sim1$ for $\lambda \ll R_{d}$ and $Q_{\lambda} \sim R_{d}/\lambda$ for $\lambda \gg R_{d}$. So the acceleration scales $\propto 1/R_{d}$ for $\lambda \ll R_{d}$ and is independent of grain size for $\lambda \gg R_{d}$. For ISM dust, the typical sizes of the largest grains are $\sim 0.1\,\mu\,{\rm m} \sim 1000\,$\AA, so for many sources we expect to be in the long-wavelength limit (even in cases where sources peak at $\ll 1000\,$\AA, then gas, not dust, will typically be the dominant opacity source).} Therefore, in these cases,  all of the relevant terms in the problem are dominated by the largest grains, which  contain most of the mass. We  therefore think of the derivation here as applying to ``large grains.'' The finite width of the grain size distribution is expected to broaden the resonances discussed below (since there is not exactly one $\driftvelmagi{i}$, there will be a range of angles for resonance), but not significantly change the dynamics. Much smaller grains can effectively be considered tightly-coupled to the gas (they will simply increase the average weight of the gas). 

However, in some circumstances -- for example acceleration of grains by high-frequency radiation -- we may have $|{\bf a}_{i}| \propto R_{d}^{-1}$. In these cases, the ``back reaction'' term on the gas is dominated by small grains, however those also have the smallest $\driftvelmagi{i}$, and may therefore have slower instability growth rates. There can therefore be some competition between effects at different grain sizes, and the different sizes may influence one another via their effects on the gas. This will be explored in future numerical simulations.

\vspace{-0.5cm}
\subsubsection{Viscosity}
\label{sec:hydro.dissipation}

We neglect dissipative processes in the gas in Eqs.~\eqref{eqn:linearized}--\eqref{eqn:linearized.fourier} (e.g., bulk  viscosity).
Clearly, including this physics will create a minimum scale below which RDI modes may be damped. This is discussed more in \S~\ref{sec:breakdown}.

\vspace{-0.5cm}
\section{Unstable Modes: General Case}
\label{sec:general.modes}

In this section, we outline, in full detail, the behavior of the dispersion 
relation that results from Eq.~\eqref{eqn:linearized.fourier}.  While the completely general case must be solved numerically, 
 we can  derive analytic expressions that highlight  key scalings for all interesting physical regimes. To guide the reader, we start with a general overview of the different branches of the dispersion 
relation in \S~\ref{sec:general overview of modes}, referring to the relevant subsections
for detailed derivations. For those readers most interested in a basic 
picture of the instability, Figs.~\ref{fig:growth.rate.demo}--\ref{fig:growth.rate.mu} give
a simple overview of the dispersion relation and its fastest-growing modes. 

\vspace{-0.5cm}
\subsection{Overview of results}\label{sec:general overview of modes}

In general, the coupled gas-dust dispersion relation (Eq.~\eqref{eqn:dispersion.full} below)
admits at least two unstable modes, sometimes more. This 
leads to a plethora of different scalings, each valid in different regimes, which we study 
in detail throughout  \S~\ref{sec:General dispersion relation}--\ref{sec:mode.structure}.
The purpose of this section is then to provide a ``road map'' to help the reader to navigate the
discussion.

An important concept, discussed  above and in \paperone, is a mode ``resonance.'' This occurs here when $\driftvel \cdot \hat{\bf k} = \pm c_{s}$, and thus is always possible  (for some $\hat{\bf k}$) when $|\driftvel | \ge c_{s}$ ($\driftvelmag \ge 1$). As shown in \paperone, when $\mu\ll1$ (and $|{\bf k}|\,c_{s}\,t_{s} \gg \mu$), modes at the resonant angle are the  fastest growing, and will thus be the most important for dynamics (if they can exist).
In the context of the analysis presented below, we will see that the dispersion relation changes character at resonance, and we must therefore analyze these specific mode angles separately. The connection to the matrix-based analysis of \paperone, which treated only the modes at the resonant angle, is outlined in App.~\ref{app: matrix relationship}.
A clear illustration of the importance of the resonant angle is shown in the right-hand panel of Fig.~\ref{fig:growth.rate.demo}.

Below, we separate our discussion into the following modes (i.e., regimes/branches of the dispersion relation):
\vspace{-0.2cm}
\begin{description}
\item[\emph{\bf (i) Decoupling instability, \S~\ref{sec:decoupling}:}] If $\coeffTSv<-1$, the drag on the dust decreases with 
increasing $\driftvelmag$ sufficiently rapidly that the dust and the dust completely decouple, causing an instability which separates the two. This 
instability exists for all ${\bf k}$, but is not usually physically relevant (see \S~\ref{sec:coulomb}).

\item[\emph{\bf (ii) Long-wavelength or ``Pressure-Free'' modes, \S~\ref{sec:long.wavelength}:}] At long wavelengths, the two unstable branches of the dispersion relation merge. This instability, which has a growth rate that scales as $\Im(\omega) \propto k^{2/3}$, persists for all $\mu$, any $\driftvelmag$ (it is non-resonant), and any $\coeffTSrho$ and $\coeffTSv$ (except $\coeffTSv=0$, $\coeffTSrho=1$). This mode has a unique structure which does not resemble a modified sound wave or free dust drift, but arises because the drag forces on very large scales are larger than pressure gradient forces so the gas pressure terms become weak and the system resembles two frictionally-coupled pressure-free fluids.

\item[\emph{\bf (iii) The ``\intermediatemodename'' mode, \S~\ref{sec:intermediate}:}] At shorter wavelengths, the two branches 
of the dispersion relation split in two. We term the first of these the ``\intermediatemodename'' mode. The mode structure resembles a modified sound wave. When $\driftvelmag \gtrsim 1$, the \intermediatemodename\ mode is unstable for all $k$, with $\Im(\omega)\propto k^{0}$ (i.e., the growth rate is constant). At resonance (\S~\ref{sec:intermediate.mode.at.resonance}),
the \intermediatemodename\ mode is subdominant and its growth rate declines with increasing $k$. The \intermediatemodename\ mode is stable for subsonic streaming ($\driftvelmag<1$).

\item[\emph{\bf (iv) The ``\slowmodename'' mode, \S~\ref{sec:slow}:}] The second shorter-wavelength branch is the ``\slowmodename'' mode. The mode structure resembles modified free (undamped) grain drift. At the resonant mode angle (\S~\ref{sec:resonance}), the \slowmodename\ mode is the dominant mode in the system, with a growth rate that increases without bound as $k\rightarrow \infty$. For a mid range of wavelengths $\Im(\omega) \propto k^{1/2}$, while for sufficiently short wavelengths $\Im(\omega) \propto k^{1/3}$. At resonance, the mode structure also becomes ``sound wave-like'' in the gas, in some respects (\S~\ref{sec:mode.structure}). Away from resonance (e.g., if $\driftvelmag <1$),   the \slowmodename\ mode is either stable or its growth rate saturates at a constant value (i.e., $\Im(\omega)\propto k^{0}$), depending on $\driftvelmag$ and $\coeffTSrho/(1+\coeffTSv)$. 

\item[\emph{\bf (v) The ``uninteresting'' mode:}] For certain parameter choices a third unstable mode appears (it would be a fourth unstable mode if $\coeffTSv<-1$, when the decoupling instability also exists). We do not analyze this mode further because it always has a (significantly) lower growth rate than either the \intermediatemodename\ or \slowmodename\ modes. 
\end{description}
We also discuss the subsonic regime $\driftvelmag<1$ separately  in more detail (\S~\ref{sec:subsonic.long.wavelenghts}), so as 
to highlight key scalings for this important physical regime. 
Finally, in \S~\ref{sec:mode.structure}, we consider the structure of the eigenmodes for the fastest-growing
modes (the long-wavelength mode and the resonant version of the \slowmodename\ mode), 
emphasizing how the resonant modes directly seed large dust-to-gas-ratio fluctuations in the gas.

\vspace{-0.5cm}
\subsection{General dispersion relation}\label{sec:General dispersion relation}

Before continuing, let us define the problem. For brevity of notation, we will work in units of $\rho_{0}$, $c_{s}$, and $\langle t_{s} \rangle$ (i.e.\ length units $c_{s}\,\langle t_{s} \rangle$), viz.,  
\begin{align}
\driftvelmag &\equiv \frac{|\driftvel|}{c_{s}} \ \ , \ \  \omdimless \rightarrow \omega\,\langle t_{s} \rangle \ \ , \ \  \kdimless \rightarrow k\,c_{s}\,\langle t_{s} \rangle. \label{eqn:dimensionless vars}
\end{align}
Inserting the general form for $t_{s}$ (Eq.~\eqref{eqn:ts.general}) into Eq.~\eqref{eqn:linearized.fourier}, we obtain the dispersion relation
\begin{align}
\label{eqn:dispersion.full} 0 =&\, A_{\omega}\,B_{\omega} \\ 
\nonumber A_{\omega} \equiv&\, \mu + (\omdimless + \iimag\,\mu)\,(\omprime + \iimag) \\ 
\nonumber B_{\omega} \equiv&\, \omprime\,(\kdimless_{\|}^{2} - \kdimless^{2})\,
{\Bigl[}
\omprime^{3} + 
\omprime^{2}\{\kappa_{\|} + \iimag\,[1 + \tildeCoeffTSv(1+\mu)] \} \\
\nonumber &\, + 
\iimag\,\omprime\,\{ \kappa_{\|}\,(1+\tildeCoeffTSv) + \iimag\,\tildeCoeffTSv\,(1+\mu) \}
- \kappa_{\|}\{\mu + \tildeCoeffTSv\,(1-\mu) \}
{\Bigr]} \\ 
\nonumber &\, + 
\left[ 
\omprime^{2} + \omprime\,\{\kappa_{\|} + \iimag\,(1+\mu)\} + \iimag\,\kappa_{\|} 
\right]\,
{\Bigl[}
\omprime\,(\omprime+\iimag\,\tildeCoeffTSv)\,(\omdimless^{2}-\kdimless_{\|}^{2}) \\ 
\nonumber &\, + 
\iimag\,\mu\,
\{ 
\omprime^{3}\,\tildeCoeffTSv
+ \omprime^{2}\,\kappa_{\|}\,(1+\tildeCoeffTSv-\coeffTSrho)
- \iimag\,\kappa_{\|}^{2}\,(\tildeCoeffTSv-\coeffTSrho)
\}
{\Bigr]}
\end{align}
where
\begin{align}
\nonumber  \omprime &\equiv \omdimless - \kappa_{\|} \ \ , \ \  \tildeCoeffTSv \equiv 1 + \coeffTSv \\
\label{eqn:kappa.definition} \kappa_{\|} &\equiv (\driftvel\cdot {\bf k})\,\langle t_{s} \rangle = \driftvelmag\,\kdimless_{\|} = \driftvelmag\,\kdimless\,\cos{\theta}.
\end{align}
(Note that $\cos\theta$, the angle between  $\hat{\bf k}$ and $\driftvelhat$, was denoted $\psi_{kw}$ in \paperone\ to allow for simpler notation in the MHD case.) App.~\ref{sec:hydrostatic.generalized} gives more general expressions for stratified media.

Our task is to analyze the solutions to Eq.~\eqref{eqn:dispersion.full}.
Fig.~\ref{fig:growth.rate.demo} plots the growth rate of the fastest-growing modes at each $\kappa_{\|}$ for a range of $\driftvelmag$, determined by exact numerical solution of Eq.~\eqref{eqn:dispersion.full}. Figs.~\ref{fig:mode.structure}, \ref{fig:growth.rate.draglaw}, and \ref{fig:growth.rate.mu} show additional examples.

\vspace{-0.5cm}
\subsubsection{General considerations}

In Eq.~\eqref{eqn:dispersion.full}, $A_{\omega}$ has the uninteresting zeros $2\omdimless = \,\kappa_{\|} - \iimag\,(1+\mu) \pm [\kappa_{\|}^{2} - (1+\mu)^{2} - \iimag\,2\,\kappa_{\|}\,(1-\mu)]^{1/2}$. These are damped longitudinal sound waves which decay (${\Im}(\omega) \le 0$) on a timescale $\sim \langle t_{s} \rangle$ for all $\mu$ and $\kappa_{\|}$; they are independent of $\coeffTSrho$ and $\coeffTSv$. The interesting solutions therefore satisfy $B_{\omega}=0$, a sixth-order polynomial in $\omega$.

For fully-perpendicular modes (${\bf k}={\bf k}_{\bot}$), $B_{\omega}=0$ simplifies to $\omdimless^{2}\,(\omdimless + \iimag\,\tildeCoeffTSv\,[1+\mu])\,[\omdimless^{2}\,(\iimag\,[1+\mu]+\omdimless) - \kdimless^{2}\,(\iimag+\omdimless)]=0$; this has the solutions $\omdimless=0$, $\omdimless=-\iimag\,(1+\mu)\,\tildeCoeffTSv$, and the solutions to $\omdimless^{2}\,(\iimag\,[1+\mu]+\omdimless) - \kdimless^{2}\,(\iimag+\omdimless)=0$ which correspond to damped perpendicular sound waves and  decay (${\Im}(\omega)<0$) for all physical $\mu>0$. For the general physical situation, with $\tildeCoeffTSv > 0$, all unstable modes must thus have $k_{\parallel}\neq 0$. 

\vspace{-0.5cm}
\subsection{Decoupling Instability}\label{sec:decoupling}

Before considering the more general case with $k_{\parallel}\neq 0$, it is worth noting that 
the perpendicular ($k_{\parallel}=0$) mode above, $\omdimless=-\iimag\,(1+\mu)\,\tildeCoeffTSv$  is unstable 
if $\tildeCoeffTSv < 0$, i.e.\ $\coeffTSv < -1$. Physically, $\tildeCoeffTSv < 0$ is the statement that the dust-gas coupling becomes weaker at higher relative velocities, and instability can occur when  dust and gas de-couple from one another (the gas decelerates and returns to its equilibrium without dust coupling, while the dust moves faster and faster as it accelerates, further increasing their velocity separation). As discussed below (Sec.~\ref{sec:coulomb}) this could occur for Coulomb drag with $\driftvelmag \gg 1$; however, in this regime Coulomb drag will  never realistically dominate over Epstein or Stokes drag, so we do not expect this instability to be physically relevant.

\vspace{-0.5cm}
\subsection{Long-Wavelength (``Pressure-Free'') Instability: $\kappa_{\|}\ll \hat{\mu}$}
\label{sec:long.wavelength}

We now examine the case of long wavelengths (small $k$). If we consider terms in $\omdimless$ up to $\mathcal{O}(\kdimless)$ for $\kdimless \ll \hat{\mu}$, and expand $B_{\omega}$, we obtain $\omdimless^{3}\,\tildeCoeffTSv\,(1+\mu) = \iimag\,\mu\,(\tildeCoeffTSv-\coeffTSrho)\,\kappa_{\|}^{2}$ to leading order. For $\tildeCoeffTSv-\coeffTSrho>0$, this has two unstable roots with the same imaginary part but oppositely-signed real parts (waves propagating in opposite directions are degenerate). Solving $B_{\omega}$ up to $\mathcal{O}(\kdimless)$ gives:
\begin{align}
\nonumber \omdimless(\kappa_{\|}\ll\hat{\mu}) &\approx 
\begin{cases}
	{\displaystyle \kappa_{0} +  \frac{\pm \sqrt{3} + \iimag}{2}\,\left(1 - \frac{\coeffTSrho}{\tildeCoeffTSv}\right)^{\frac{1}{3}}\hat{\mu}^{1/3}\,\kappa_{\|}^{2/3}} \ \ & \hfill{(\coeffTSrho < \tildeCoeffTSv)} \\
	\\
	{\displaystyle \kappa_{0} +  \iimag\,\left(\frac{\coeffTSrho}{\tildeCoeffTSv}-1 \right)^{\frac{1}{3}}\hat{\mu}^{1/3}\,\kappa_{\|}^{2/3}} \ \ & \hfill{(\coeffTSrho > \tildeCoeffTSv)}  
\end{cases} \\ 
\label{eqn:longwave.mode} \kappa_{0} \equiv {\Bigl[} 1& + \mu\,\left(2 + \frac{\coeffTSrho-1}{\tildeCoeffTSv} \right) {\Bigr]} \, \frac{\kappa_{\|}}{3\,(1+\mu)}
\ \ \ , \ \ \ 
 \hat{\mu} \equiv \frac{\mu}{1+\mu}  
%\hat{\mu} &\equiv \frac{\mu}{1+\mu} 
%{k_{\|}} &= \pm \frac{|{\bf k}_{\bot}|}{\sqrt{\driftvelmag^{2}-1}}
\end{align}
Note that this mode depends only on $\kappa_{\|} = \driftvelmag\,\kdimless\,\cos{\theta}$ at this order; the dependence on $\driftvelmag$ is implicit. The growth rate rises towards shorter wavelengths, but sub-linearly. Most notably, instability exists at {\em all} dust abundances $\mu$ (and depends only weakly on that abundance, with the $1/3$ power), wavelengths $\kappa_{\|}$ (for $\kappa_{\|}\ll \hat{\mu}$), accelerations or $\driftvelmag$, and drag coefficients $\coeffTSrho$ and $\coeffTSv$.\footnote{\label{foot:when.instab.vanishes}Note that in the pathological case $\coeffTSrho = \tildeCoeffTSv = 1+\coeffTSv$, our approximation in Eq.~\eqref{eqn:longwave.mode} vanishes but an exact solution to Eq.~\eqref{eqn:dispersion.full} still exhibits low-$k$ instability, albeit with reduced growth rate. The reason is that the leading-order term on which Eq.~\eqref{eqn:longwave.mode} is based vanishes, so the growth rate scales with a higher power of $\kappa_{\|}$. Instability only vanishes completely at low-$k$ when $\coeffTSrho=1$ and $\coeffTSv=0$, exactly.}

This mode is fundamentally distinct from either a modified sound wave or a modified dust drift mode. Rather, it is essentially a one-dimensional mode of a pressure-free, two-fluid system with drift between the two phases. To see this, we note that the pressure force on the gas scales as $\nabla P \sim k\,c_{s}^{2}\,\delta\rho$, while the drift forces scale $\propto \mu$. So, at sufficiently small $\kdimless \ll \mu$, the pressure force becomes small compared to the drag force of the dust on the gas. Perturbations perpendicular to the drift are damped on the stopping time, but parallel perturbations can grow. As a result, one can recover all of the properties of this mode by simplifying to a pressure-free, one-dimensional system (${\bf k}$, $\delta {\bf u}$, $\delta {\bf v}$ parallel to $\driftvel$).

At long wavelengths in particular, one might wonder whether the presence of gradients or inhomogeneity in the equilibrium solution might modify the mode here. In App.~\ref{sec:hydrostatic.generalized}, we consider a system in hydrostatic equilibrium supported by pressure gradients, with arbitrary stratification of the background quantities $P_{0}$, $\rho_{0}$, $\rho_{d,\,0}$, $\driftvel$. We show that, within the context of a local approximation, the leading-order correction to this mode can be written as $\omega \rightarrow \omega\,(1 + \epsilon)$ with $\epsilon \sim \hat{\mu}^{1/3}\,\kappa_{\|}^{2/3}\,(\kdimless\,\mu/|\nabla\mu|)^{-1}$. But $\hat{\mu} \ll 1$, generally, and $\kappa_{\|} \ll \hat{\mu} \ll 1$ for this mode, so the correction term is small unless $\kdimless^{-1} \gg \mu/|\nabla\mu|$; i.e.\ unless we go to wavelengths much larger than the background gradient-scale length (of $\mu$). Obviously, in this case a global solution, with appropriate boundary conditions, would be needed.

\begin{figure*}
\plotsidesize{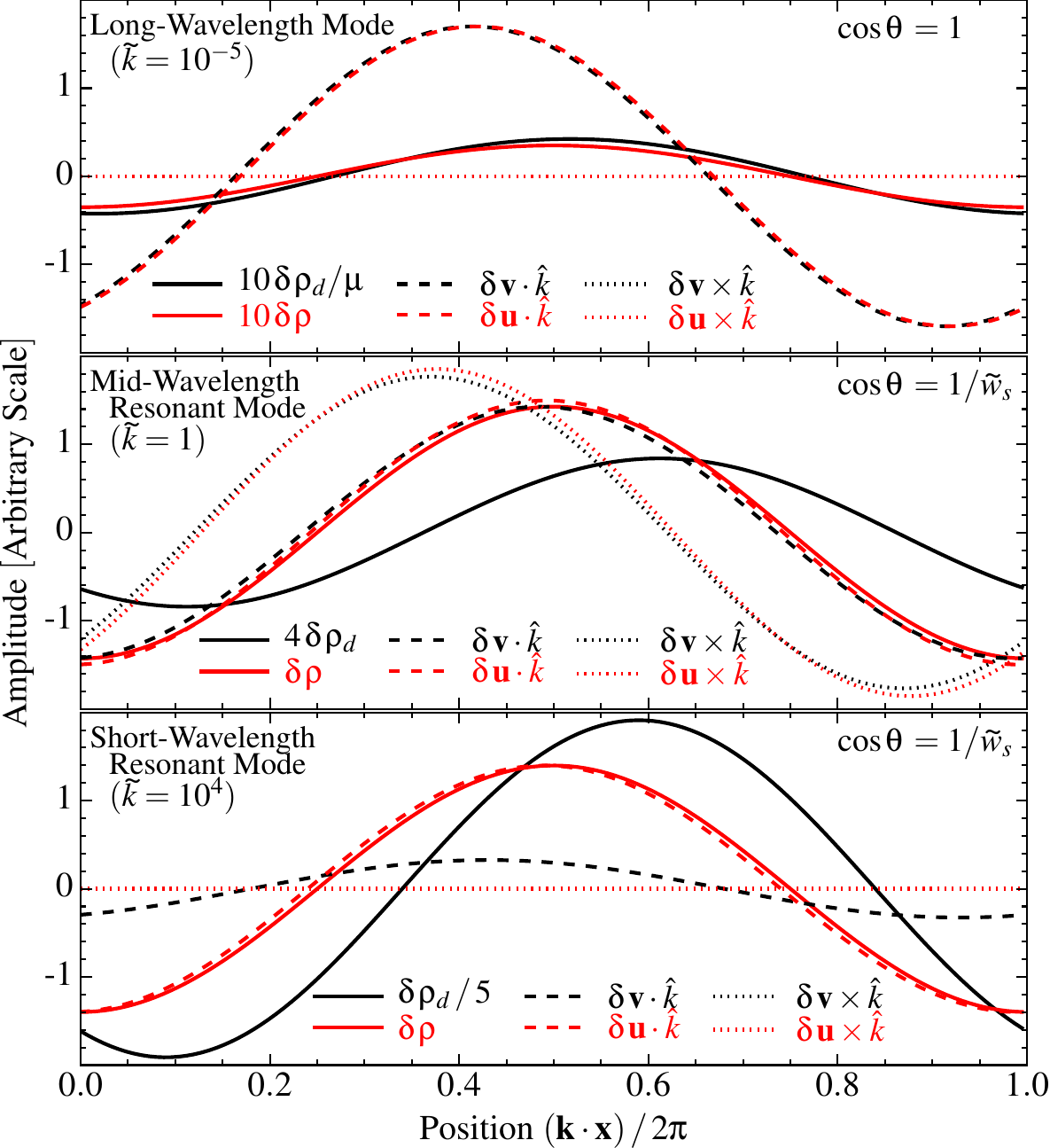}{0.75}
    \vspace{-0.25cm}
    \caption{Spatial structure of the modes in Fig.~\ref{fig:growth.rate.demo} (see \S~\ref{sec:mode.structure}). Here we take $\mu=0.01$, $\coeffTSrho=\coeffTSv=0$, $\driftvelmag=10$, and $\cos{\theta}$ shown, and plot the perturbed dust density $\delta \rho_{d}$, gas density $\delta \rho$ (in units of $\rho_{0}$, the mean density) and perturbed dust velocity $\delta {\bf v}$ and gas velocity $\delta {\bf u}$ (in units of $c_{s}$). The overall amplitude of the linear perturbation ($y$-axis normalization) is arbitrary.  For the velocities we separate them into the magnitude of the component parallel to ${\bf k}$ ($\delta {\bf v} \cdot \hat{\bf k}$), and perpendicular ($\delta {\bf v} \times \hat{\bf k}$). We show the spatial structure over one period, for a given $\kdimless \equiv k\,c_{s}\,\langle t_{s} \rangle$). In all cases, a lag between the dust and gas density perturbations arises because the dust de-celerates when moving through the denser gas, which generates a ``pileup'' and stronger dust-density peak, which in turn amplifies the gas response.
    {\em Top:} The long-wavelength mode (\S~\ref{sec:long.wavelength}) exhibits a nearly-coherent dust-gas oscillation, with $\delta \rho_{d} \approx \mu\,\delta \rho$ to leading order (the lag is higher-order). This is not a modified sound wave, however: the phase/group velocities scale $\propto k^{-1/3}$ (Eq.~\ref{eqn:longwave.mode}), the velocity and density responses are offset by a phase lag, and the gas+dust density perturbation is weak ($|\delta \rho|/\rho_{0} \ll |\delta {\bf v}|/c_{s}$; note we multiply $\delta \rho$ plotted by $10$, and $\delta\rho_{d}$ by $10/\mu$). 
    {\em Middle:} Resonant mode (\S~\ref{sec:resonance}), at intermediate wavelengths where the growth rate scales $\propto k^{1/2}$ (Eq.~\ref{eqn:longwave.mode.midk}). The wavespeed, gas density and velocity in the $\hat{\bf k}$ direction now behave like a sound wave. The dust lag is larger (phase angle $\sim \pi/6$) and because of the ``resonance,'' where the dust motion along the $\hat{\bf k}$ direction exactly matches the wavespeed, the effects above add coherently and generate a much stronger dust response with $|\delta \rho_{d}|/|\delta \rho| \sim (2\,\mu\,\kdimless)^{1/2}$, a factor $\sim (2\,\kdimless/\mu)^{1/2} \sim 20$ larger than the mean dust-to-gas ratio. Note the large perpendicular velocities also present.
    {\em Bottom:} Resonant mode, at short wavelengths (where growth rates scale $\propto k^{1/3}$; Eq.~\ref{eqn:omega.resonant}). This is similar to the intermediate-wavelength case except perpendicular velocities become negligible, the dust velocity response $\delta {\bf v}$ becomes weaker, and the dust density response becomes stronger, with $|\delta \rho_{d}|/|\delta \rho| \sim (4\,\mu\,\kdimless)^{1/3}$, a factor $\sim 1000$ larger than the mean dust-to-gas ratio $\mu$. 
        \vspace{-0.25cm}
        \label{fig:mode.structure}}
\end{figure*}

\vspace{-0.5cm}
\subsection{Short(er)-Wavelength Instabilities:  $\kappa_{\|}\gg \hat{\mu}$}

At high-$k$ there are at least two different unstable solutions. If we assume a dispersion relation of the form $\omdimless \sim \mathcal{O}(\kdimless^{1}) + \mathcal{O}(\kdimless^{\nu})$ where $\nu<1$, and expand $B_{\omega}$ to leading order in $\kdimless^{-1} \ll 1$, we obtain a dispersion relation $0 = \omdimless\,(\omdimless-\kappa_{\|})^{3}\,(\omdimless^{2} - \kdimless^{2})\,(1 + \mathcal{O}(\kdimless^{-1}))$. This is solved by $\omdimless = \pm \kdimless + \mathcal{O}(\kdimless^{\nu})$ or $\omdimless = \kappa_{\|} + \mathcal{O}(\kdimless^{\nu})$, each of which produces a high-$k$ branch of the dispersion relation. 

In the following sections, \ref{sec:intermediate}--\ref{sec:slow}, we study each of these branches 
in detail. We term the first branch, with $\omdimless = \pm \kdimless + \mathcal{O}(k^{\nu})$, the ``\intermediatemodename'' mode (\S~\ref{sec:intermediate}); to leading order this is just a soundwave (the natural mode in the gas, absent drag: $\omega = \pm c_{s}\,k$). We term the second branch, with $\omdimless = \kappa_{\|} + \mathcal{O}(\kdimless^{\nu})$, the ``\slowmodename'' mode (\S~\ref{sec:slow}); to leading order this is ``free drift'' (the natural mode in the dust, absent drag: $\omega = \driftvel\cdot{\bf k}$). 
In the analysis of each of these, we must treat modes with the  resonant angle, $\cos\theta =\pm 1/\driftvelmag$, separately, 
because the dispersion relation fundamentally changes character. %The \intermediatemodename\ mode at resonance (\S~\ref{sec:intermediate.mode.at.resonance}) is modified only in minor ways. In contrast, 
The \slowmodename\ mode at resonance  (\S~\ref{sec:resonance}) is the fastest-growing mode in the system (when $\driftvelmag>1$ and $\mu\ll1$), with growth rates that increase \emph{without bound} as $k\rightarrow \infty$. This is the resonance condition for the acoustic RDI case considered in \paperone~  (see also App.~\ref{app: matrix relationship}).

\vspace{-0.5cm}
\subsection{Short(er)-Wavelength Instability: The ``\Intermediatemodename'' Mode}
\label{sec:intermediate}

To leading-order, the \intermediatemodename\ mode satisfies $\omdimless = \pm \kdimless$ (the sound wave dispersion relation). Consider the next-leading-order term; i.e.\ assume $\omdimless = \omdimless_{\intermediatemodesubscript} = \pm \kdimless + \omegaZ + \mathcal{O}(\kdimless^{-1})$ (where $\omegaZ$ is a term that is independent of $k$) and expand the dispersion relation to leading order in $\kdimless^{-1}$ (it will transpire that the solution here is valid for all $\kdimless \gg \driftvelmag\,\mu$). This produces a simple linear leading-order dispersion relation for both the $\pm$ cases: 
\begin{align}
\label{eqn:omega.med} \omdimless_{\intermediatemodesubscript} &\approx \pm\, \kdimless - \iimag\,\frac{\mu\,(1 + \coeffTSv\,\cos^{2}{\theta}  \pm \driftvelmag\,(1-\coeffTSrho)\,\cos{\theta})}{2} 
\end{align}
Where the ``$+$'' mode applies the $+$ to all $\pm$, and vice versa.

Because both signs of $\cos{\theta}$ are allowed, it  follows that the modes are unstable (${\Im}(\omega) > 0$) if
\begin{align}
\label{eqn:eta.intermediate.mode.1} \driftvelmag\,{|}(1 - \coeffTSrho)\,\cos{\theta}{|} &> 1 + \coeffTSv\,\cos^{2}{\theta}.
\end{align}
Because $\coeffTSv$ and $\coeffTSrho$ generally are order-unity or smaller, Eq.~\eqref{eqn:eta.intermediate.mode.1} implies that $\driftvelmag \gtrsim 1$ is required for this mode to be unstable. For $\coeffTSv<1$, the more common physical case (see \S~\ref{sec:draglaws}), we also see that the condition (Eq.~\eqref{eqn:eta.intermediate.mode.1}) is first met 
for parallel modes ($\cos\theta=\pm1$) and that their growth rate (Eq.~\eqref{eqn:omega.med}) is larger than oblique modes.\footnote{For the parallel case, the general dispersion relation $B_{\omega}$ simplifies to:  $B_{\omega} \rightarrow A_{\omega}\,B_{\omega}^{\prime}$ with 
\begin{align}
\nonumber B_{\omega}^{\prime} &= \kappa_{\|}\,\driftvelmag^{2}\,\mu\,(\omdimless\,\tildeCoeffTSv -\kappa_{\|}\,\coeffTSrho) + \omprime\,{\Bigl(} 
(\omprime+\iimag\tildeCoeffTSv)\,(\omdimless^{2}\,\driftvelmag^{2}-\kappa_{\|}^{2}) \\ 
\nonumber & + \iimag\,\driftvelmag^{2}\,\mu\,(
\omdimless^{2}\,\tildeCoeffTSv + \kappa_{\|}\,\{\kappa_{\|}\,(\coeffTSrho-1) + \iimag\,\tildeCoeffTSv \}
- \omdimless\,\kappa_{\|}\,\,(\tildeCoeffTSv+\coeffTSrho - 1) {\Bigr)}
\end{align}
} Comparing the long-wavelength result in Eq.~\eqref{eqn:longwave.mode} to Eq.~\eqref{eqn:omega.med}, we see that the growth rate grows with $k$ until it saturates at the constant value given by Eq.~\eqref{eqn:omega.med} above $\kdimless \gtrsim \driftvelmag\,\mu$. For $\driftvelmag \lesssim 1$, the mode becomes stable above $\kdimless \gtrsim \driftvelmag\,\mu$. 

In App.~\ref{sec:hydrostatic.generalized} we show that up to this order in $\kdimless$, the behavior of this mode is not expected to change in hydrostatic or arbitrarily stratified media (the leading-order corrections appear at order $\sim 1/(k\,L_{0})$, where $L_{0}$ is the gradient scale-length of the system).

\vspace{-0.5cm}
\subsubsection{The \Intermediatemodename\ Mode at Resonance}\label{sec:intermediate.mode.at.resonance}

When $\driftvelmag\,\cos{\theta}=\pm1$, the behavior of the \intermediatemodename\ mode is modified (the series expansion we used is no longer valid; see \S~\ref{sec:resonance}). If we follow the same branch of the dispersion relation, then instead of the growth rate becoming constant at high-$k$, it peaks around $\kappa_{\|}\sim\hat{\mu}$ at a value ${\Im}(\omdimless) \approx \hat{\mu}/4$, and then declines with increasing $\kappa_{\|}$. It is therefore the less interesting branch in this limit, because the \slowmodename\ branch  produces much larger growth rates.

\vspace{-0.5cm}
\subsection{Short(er) Wavelength Instability: The ``\Slowmodename'' Mode}
\label{sec:slow}

We now consider the \slowmodename\ mode branch of the high-$k$ limit of $\omega$, with leading-order $\omdimless = \kappa_{\|}$ (the free-drift dispersion relation). Assuming $\omdimless = \omdimless_{\slowmodesubscript} = \kappa_{\|} + \omegaZ + \mathcal{O}(\kdimless^{-1})$, and expanding to leading order in $\kdimless$, we obtain the leading-order cubic relation 
\begin{align}
0 = &\omegaZ\,(\omegaZ+\iimag)\,(\omegaZ + \iimag\,\tildeCoeffTSv)\,(1-\driftvelmag^{2}\,\cos^{2}{\theta}) - \mu\,(\iimag\,(\tildeCoeffTSv-\coeffTSrho)\,\driftvelmag^{2}\,\cos^{2}{\theta} \nonumber\\
&+ \omegaZ\,(1-\tildeCoeffTSv + (\tildeCoeffTSv\,(1+\driftvelmag^{2})-\driftvelmag^{2}\,\coeffTSrho - 1)\,\cos^{2}{\theta})).\label{eqn:slowmode.dispersion.relation}
\end{align}

Equation~\eqref{eqn:slowmode.dispersion.relation} is solvable in closed form but the expressions are tedious and unintuitive.\footnote{Eq.~\eqref{eqn:slowmode.dispersion.relation} does provide a simple closed-form solution if $\cos\theta=\pm 1$ (parallel modes), or $\coeffTSv=0$; in these cases the growing mode solutions are:
\begin{align}
\nonumber \omdimless_{\slowmodesubscript}(|\cos{\theta}| =  1) &\approx \kappa_{\|} + \iimag\,\frac{\tildeCoeffTSv}{2}\,\left[-1 + \left({1 + \frac{4\,{\mu}\,(\tildeCoeffTSv-\coeffTSrho)}{\tildeCoeffTSv^{2}\,(1-\driftvelmag^{-2})}} \right)^{1/2}\right] \\ 
\nonumber \omdimless_{\slowmodesubscript}(\coeffTSv=0) &\approx \kappa_{\|} + \iimag\,\frac{1}{2}\,\left[-1 + \left({1 + \frac{4\,\mu\,(1-\coeffTSrho)}{1-(\driftvelmag\,\cos{\theta})^{-2}}} \right)^{1/2} \right] 
\end{align}
}
For clarity of presentation, if we consider $\mu \ll 1$, the expression factors into a damped solution with $\omegaZ = -\iimag$, and a quadratic that gives a damped and a growing solution which simplifies to: 
\begin{align}
\label{eqn:omega.slow} \omdimless_{\slowmodesubscript}(\mu\ll1) &\approx \kappa_{\|} + \iimag\,\frac{(\driftvelmag\,\cos{\theta})^{2}\mu}{(\driftvelmag\,\cos{\theta})^{2}-1}\,\left(1 - \frac{\coeffTSrho}{\tildeCoeffTSv} \right)
\end{align}

This illustrates the general form of the full expression. In particular, we see that the expressions become invalid ($\Im (\omega) \rightarrow \infty$)
at the resonant angle $\driftvelmag^{2}\cos^{2}\theta  =1$, which will be treated separately below (\S~\ref{sec:resonance}). 

The requirement for instability (from the general version of Eq.~\eqref{eqn:omega.slow}) is:
\begin{align}
\label{eqn:slowmode.stability} (\driftvelmag^{2}\,\cos^{2}{\theta}-1)\,(1-\coeffTSrho/\tildeCoeffTSv) \ge 0
\end{align}
We thus see that if $\coeffTSrho/\tildeCoeffTSv < 1$ (the more common physical case), this mode is unstable for $\driftvelmag\,|\cos{\theta}| > 1$; if $\coeffTSrho/\tildeCoeffTSv > 1$, however, the mode is stable for $\driftvelmag\,|\cos{\theta}| > 1$ but becomes unstable for $\driftvelmag\,|\cos{\theta}| < 1$. 

Away from resonance (i.e., with $|\driftvelmag\cos{\theta}| \ne 1$), we see that, like the \intermediatemodename\ mode, the \slowmodename\ mode is described by the long-wavelength solution from \S~\ref{sec:long.wavelength}, with a growth rate that increases with $k$ until it saturates at the constant value of Eq.~\eqref{eqn:omega.slow}: roughly $\sim \driftvelmag^{2}\,\mu$ for $\driftvelmag < 1$ or $\sim \mu$ for $\driftvelmag > 1$. Comparing the growth rates (Eq.~\eqref{eqn:omega.slow} and Eq.~\eqref{eqn:longwave.mode}) we see this occurs at $\kdimless\gtrsim \mu\,\driftvelmag^{2} / (1 + \driftvelmag^{3})$ (i.e. $\sim \driftvelmag^{2}\,\mu$ for $\driftvelmag < 1$, $\sim \mu/\driftvelmag$ for $\driftvelmag > 1$). 

In App.~\ref{sec:hydrostatic.generalized}, we note that in an arbitrarily stratified background, a constant correction to the growth rate of this mode appears at leading-order, with the form $\omega_{\slowmodesubscript} \rightarrow \omega_{\slowmodesubscript} - \iimag\,\nabla\cdot\driftvel$ (or $\omega_{\slowmodesubscript} \rightarrow \omega_{\slowmodesubscript} + \iimag\,\rho_{d,\,0}^{-1}\,\driftvel\cdot\nabla\rho_{d,\,0}$, since the dust density and velocity are related by continuity). Because this mode is moving with the mean dust motion ($\omdimless\approx\kappa_{\|}$ or $\omega \approx \driftvel\cdot{\bf k}$ to leading order), this is just the statement that, if there is a non-zero divergence of the background drift, the perturbation is correspondingly stretched or compressed along with the mean flow. The correction is important only if the timescale for the dust to ``drift through'' the global gradient scale-length (in $\rho_{d,\,0}$ or $\driftvel$) is short compared to the growth time. 

\vspace{-0.1cm}
%\subsubsection{Supersonic ($\driftvelmag \ge 1$) Case: Resonant Mode Growth}
\subsubsection{The \Slowmodename\ Mode at Resonance}
\label{sec:resonance}

When $\driftvelmag \ge 1$, then Eq.~\eqref{eqn:omega.slow} (and its generalization, valid at all $\mu$) diverge as $\cos{\theta} \rightarrow \pm 1/\driftvelmag$. In this case the ``saturation'' or maximum growth rate of the mode becomes infinite. What actually occurs is that the growth rate continues to increase {\em without limit} with increasing $k$. 

In this limit, our previous series expansion at high-$k$ is invalid: we must return to $B_{\omega}$ and insert $\driftvelmag\,\cos{\theta}=\pm1$; i.e.\ $\kdimless^{2}=\kappa_{\|}^{2}$ or ${\bf k}\cdot \driftvel = \omega_{\rm sound} \equiv \pm c_{s}\,k$, the resonance condition for the RDI. Note that when the resonant condition is met, the mode satisfies $\omega = \driftvel\cdot{\bf k} = \pm c_{s}\,k$ -- i.e.\ to leading order it simultaneously satisfies the dispersion relation of gas absent drag (a sound wave) {\em and} dust absent drag (free drift). This effectively eliminates the restoring forces in the system, so the resulting dispersion relation\footnote{If the resonant condition is satisfied and $\coeffTSrho=\coeffTSv=0$, the dispersion relation has the simple form $\omprime^{2}\,[\omprime+\iimag\,(1+\mu)]\,(\omega+\kdimless) = -\mu\,\kdimless^{2}$.} has growing solutions with ${\Im}(\omega_{\ast}) > 0$ for {\em all} $\kappa_{\|}$, and {\em the growth rate increases monotonically with $\kappa_{\|}$ without limit} (here and below we use $\omega_{\ast}$ to denote the resonant frequency).\footnote{Note that at long wavelengths, $\kdimless \ll \hat{\mu}$, the series expansion in Eq.~\eqref{eqn:longwave.mode}  is still accurate and we just obtain the solutions in \S~\ref{sec:long.wavelength}, even at resonance.}

There are two relevant regimes for this mode at resonance:

{\bf (1) The Intermediate-wavelength (``mid-$k$'' or ``low-$\mu$'') Resonant Mode:} 
If $\hat{\mu} \ll \kdimless \ll \hat{\mu}^{-1}$, the resonant solutions
 to $B_{\omega_{\ast}}=0$ give: %lead to  the ``mid-k'' resonant mode: 
\begin{align}
\nonumber \omdimless_{\ast}(\hat{\mu} &\ll \kappa_{\|} \ll \hat{\mu}^{-1}) \approx \kappa_{1}  + \frac{\iimag\pm1}{2}\,\left( {\Bigl |}1 - \frac{\coeffTSrho}{\tildeCoeffTSv} {\Bigr|}\,{\hat{\mu}\,\kappa_{\|}} \right)^{1/2} \\ 
 \kappa_{1} &\equiv \left[ 1 - \frac{\hat{\mu}}{4}\,\left(1 - \frac{\tildeCoeffTSv\,\coeffTSv + \coeffTSrho\,\driftvelmag^{2}}{\tildeCoeffTSv^{2}\,\driftvelmag^{2}} \right) \right]\,\kappa_{\|} - \iimag\frac{(\tildeCoeffTSv-\coeffTSrho)\,\hat{\mu}}{8\,\tildeCoeffTSv}.\label{eqn:longwave.mode.midk}
\end{align}
As expected,  to $\mathcal{O}({\mu}^{1/2})$, this matches the ``acoustic RDI'' expression derived in \paperone,  with the resonance  between the dust drift velocity and the natural phase velocity of an acoustic wave without dust (the 
exact correspondence is explained in detail in App.~\ref{app: matrix relationship}).

{\bf (2) The Short-wavelength (``high-$k$'') Resonant Mode:} 
At larger $\kappa_{\|}\gg \hat{\mu}^{-1}$, expanding $\omdimless \sim \mathcal{O}(\kdimless )$ to leading order in $\kdimless \gg 1$ shows that the leading-order term must obey $\omdimless = \pm \kappa_{\|} = \pm \kdimless$, as before. Now expand to the next two orders in $\kdimless$ as $\omdimless_{\ast}\approx \kdimless + \omdimless_{1/3}\,\kdimless^{1/3} + \omegaZ$, where again $\omegaZ$ denotes a $k$-independent part (it is easy to verify that with $\nu\ge0$, any  term $\omdimless = \kdimless + \omdimless_{\nu}\, \kdimless^{\nu}$, other than $\nu=0$ and $\nu=1/3$, must have $\omdimless_{\nu}=0$ to satisfy the dispersion relation to next-leading order in $\kdimless$). This gives $2\,\omdimless_{1/3}^{3} + (1+\coeffTSv/\driftvelmag-\coeffTSrho)\,\mu=0$, and a simple linear expression for $\omegaZ$. There is always  one purely real root, one decaying root, and one unstable ${\Im}(\omega) > 0$ root. Taking the unstable root, we obtain the ``high-$k$'' resonant mode:
\begin{align}
\label{eqn:omega.resonant}\omdimless_{\ast}(k \gg \hat{\mu}^{-1}) &\approx \kappa_{\|} + (\iimag\,\sqrt{3} \pm 1)\,\left( \frac{|\Theta|\,\mu\,\kappa_{\|}}{16} \right)^{1/3} - \iimag\,\omegaZ \\ 
\nonumber \Theta &\equiv 1+\frac{\coeffTSv}{\driftvelmag^{2}}-\coeffTSrho \\ 
\nonumber \omegaZ &\equiv \frac{(1+\Theta)\,\mu}{6} + \frac{1 + (\tildeCoeffTSv^{2}-1)/\driftvelmag^{2} - \tildeCoeffTSv\,\coeffTSrho}{3\,\Theta}, 
\end{align}
where the sign in the $\pm$ part of the real part of $\omdimless_{\ast}$ is ``$+$'' if $\Theta > 0$ and ``$-$'' if $\Theta < 0$. Again this is just the high-$k$ expression for the acoustic RDI derived in \paperone.

Note that, formally, the intermediate-wavelength (mid-$k$) and short-wavelenth (high-$k$) resonant modes do not necessarily represent the same branch of the dispersion relation (they are distinct modes even at resonance, one of which is the fastest-growing at intermediate $k$, the other at high $k$). However, for $\coeffTSrho \le 1$, they are degenerate, and the resonant mode behavior transitions smoothly between the two limits with increasing $k$.

Qualitatively, the resonant modes grow in a similar way to the long-wavelength instability Eq.~\eqref{eqn:longwave.mode}. We see that the slope  decreases with increasing $\kappa_{\|}$ from $\omdimless\sim \kappa_{\|}^{2/3}$ (for $\kappa_{\|} \ll \hat{\mu}$), to $\omdimless_{\ast}\sim \kappa_{\|}^{1/2}$ (for $\hat{\mu} \ll \kappa_{\|} \ll \hat{\mu}^{-1}$), to $\omdimless_{\ast}\sim \kappa_{\|}^{1/3}$  (for $\hat{\mu}^{-1}\ll \kappa_{\|}$). Comparison to the \intermediatemodename\ mode (Eq.~\eqref{eqn:omega.med}) or the \slowmodename\ 
mode away from resonance (Eq.~\eqref{eqn:omega.slow}) shows that the resonant mode (Eqs.~\eqref{eqn:longwave.mode.midk} and \eqref{eqn:omega.resonant}) always grows fastest.  
Because resonance requires $\driftvelmag\,\cos{\theta}=\pm1$, we have: $k_{\|} = k\,\cos{\theta} = \pm k/\driftvelmag$, $k_{\bot} = |{\bf k}_{\bot}| = k\,\sin{\theta} = k\,(1 - \driftvelmag^{-2})^{1/2}$, and $k_{\|}/k_{\bot} = \pm 1/\sqrt{\driftvelmag^{2}-1}$. For modest $\driftvelmag\gtrsim 1$, the resonant mode is primarily parallel ($\cos{\theta} \sim \pm1$), but for large $\driftvelmag \gg 1$, the resonant mode becomes increasingly perpendicular, with $\theta \rightarrow \pi/2$ and $k_{\bot} \gg |k_{\|}|$. 

We can estimate the width of the resonant angle in Fig.~\ref{fig:growth.rate.demo} -- i.e., the range of angles over which the growth rate is similar to maximum -- by combining the maximum growth rate  at resonance (Eqs.~\eqref{eqn:longwave.mode.midk}-\eqref{eqn:omega.resonant}) with the growth rate of the \slowmodename\ mode away from resonance (Eq.~\eqref{eqn:omega.slow}).
%which showed that (away from the resonant angle)  mode scales as $\sim \mu/(1-(\driftvelmag\,\cos{\theta})^{-2})$.
 This gives $\Delta\cos{\theta} \sim \mu/(\driftvelmag\,\omdimless_{\ast})$ where $\omdimless_{\ast} \sim (\mu\,\kdimless)^{1/2}$ (at mid $\kdimless$) or $\omdimless_{\ast} \sim (\mu\,\kdimless)^{1/3}$ (at high $\kdimless$). We see that the resonance is broader at larger $\mu$, lower $\driftvelmag$, and lower $\kdimless$.

Similar to the out-of-resonance \slowmodename\ modes, if we consider arbitrarily stratified, hydrostatic backgrounds (App.~\ref{sec:hydrostatic.generalized}) the dispersion relation differs (to leading order in $\sim 1/k$) only in a constant offset in the growth rate (i.e.\ in the $\kappa_{1}$ term in Eq.~\ref{eqn:longwave.mode.midk} or $\omegaZ$ term in Eq.~\ref{eqn:omega.resonant}) of order $\sim \nabla\cdot\driftvel$. This correction is un-important for the ``high-$k$'' resonant mode, and for the ``mid-$k$'' resonant mode over the upper range of $k$ in which that mode exists. But it can, in principle, be a significant correction at the lower-$k$ range of the ``mid-$k$'' mode ($\kdimless \sim \hat{\mu}$) especially if $\hat{\mu}$ is very small (see App.~\ref{sec:hydrostatic.generalized} for details). 

At high-$k$ and at resonance, anti-aligned solutions of the form $\omdimless = -\kdimless + \omegaZ + \mathcal{O}(\kdimless^{-1})$ are also admitted. These have the simple solution $\omegaZ \approx -\iimag\,(\coeffTSv + \driftvelmag\,\coeffTSrho)\,\mu/(2\,\driftvelmag)$, which is growing only if $\coeffTSv + \driftvelmag\,\coeffTSrho < 0$.

\vspace{-0.5cm}
\subsection{Subsonic ($\driftvelmag < 1$) Modes}
\label{sec:subsonic.long.wavelenghts}

In \S~\ref{sec:slow} above, we saw that when $\driftvelmag>1$ (and $\hat{\mu}\ll1$) the fastest growing modes 
will be the long-wavelength mode (at low $k$) and the acoustic RDI ``resonant'' modes (at high $k$). When 
the streaming is subsonic ($\driftvelmag<1$) this resonance is no longer possible and the \intermediatemodename\ mode (\S~\ref{sec:intermediate})  is also stabilized. It thus seems helpful to cover 
the subsonic mode structure in a self-contained manner, which is the purpose of this section.
We collect some of the results derived in \S~\ref{sec:long.wavelength}--\S~\ref{sec:slow} 
and derive a new limit of the subsonic \slowmodename\ mode.  

At sufficiently low $k$, the long-wavelength solutions from \S~\ref{sec:long.wavelength} continue to be unstable. Moreover, the ``\slowmodename'' mode in Eq.~\eqref{eqn:omega.slow} is still unstable if $\coeffTSrho > \tildeCoeffTSv$ (see Eq.~\eqref{eqn:slowmode.stability}; in this case all $k$ are unstable). The mode then grows as in Eq.~\eqref{eqn:longwave.mode} until saturating at a maximum growth rate given by Eq.~\eqref{eqn:omega.slow}: approximately $\Im{(\omdimless)} \sim \driftvelmag^{2}\,\mu$, for $\kdimless \gtrsim \driftvelmag^{2}\,\mu$. From the form of Eq.~\eqref{eqn:omega.slow} we can also see that for $\driftvelmag<1$ the most rapidly-growing mode has $\cos{\theta}=\pm1$, i.e.\ the modes are parallel. 

If  $\tildeCoeffTSv > \coeffTSrho$ (and $\driftvelmag < 1$), the \slowmodename\ mode is stabilized for  $\kdimless \gg1$. However it  persists for some intermediate range of $\kdimless$, which was not included in  Eq.~\eqref{eqn:omega.slow} due to our assumption $\kdimless \gg1$. Specifically,  the growth of $\Im(\omega)$ with $\kappa_{\|}$ saturates at a similar point, but then $\Im(\omega)$ turns over and vanishes at finite $\kdimless \gtrsim \driftvelmag$. Since we are interested in small $\driftvelmag$ and low-$\kdimless$, we assume $\omdimless\sim \omegaZ + \omdimless_{1}\,\driftvelmag + \omdimless_{2}\,\driftvelmag^{2}$ and $\kdimless \sim \mathcal{O}(\driftvelmag)$, and expand the dispersion relation to leading order in $\driftvelmag$. This gives two results: (i)  that $\omegaZ$ must vanish, and (ii) that $\omdimless_{1}$ must obey $\omdimless_{1} (\omdimless_{1}^{2}\,(1+\mu) - (\kdimless/\driftvelmag)^{2})=0$. This gives the leading-order solution $\omdimless = \pm \kdimless_{\|}/\sqrt{1+\mu}$. Plugging in either the $+$ or $-$ root (they give the same growth rate), we solve for the second-order term, to obtain the relation
\begin{align}
\nonumber \omdimless_{\rm subsonic} &\approx \kdimless_{\|}\,\left( \pm\frac{1}{(1+\mu)^{1/2}} + \frac{(\coeffTSrho+\coeffTSv\,\driftvelmag)\,\driftvelmag\,\mu}{2\,(1+\mu)\tildeCoeffTSv}\right) \\ 
\label{eqn:omega.lowk} &+ \iimag\,\frac{\mu}{2}\,\left( \driftvelmag^{2}\,(\tildeCoeffTSv-\coeffTSrho) - \frac{\kdimless_{\|}^{2}}{(1+\mu)^{2}} \right).
\end{align}
We see that this subsonic \slowmodename\ mode is unstable for $\kdimless_{\|}<\driftvelmag (1+\mu)(\tildeCoeffTSv-\coeffTSrho)^{1/2}$. We reiterate 
that Eq.~\eqref{eqn:omega.lowk} is valid only for  $\tildeCoeffTSv > \coeffTSrho$; otherwise Eq.~\eqref{eqn:omega.slow} is correct and all $k$ are unstable.

\vspace{-0.5cm}
\subsection{Mode Structure}
\label{sec:mode.structure}

In this section we discuss the structure of the eigenmodes in ($\delta \rho,\delta {\bf u},\delta \rho_{d},\delta {\bf v}$).  We focus on the most relevant (fastest-growing) modes in the three limits: (i) $\kappa_{\parallel}\ll \hat{\mu}$ (dispersion relation in Eq.~\eqref{eqn:longwave.mode}), (ii) $\hat{\mu}\ll \kappa_{\parallel}\ll \hat{\mu}^{-1}$ (Eq.~\eqref{eqn:longwave.mode.midk}), and (iii) $\kappa_{\parallel}\gg \hat{\mu}^{-1}$ (Eq.~\eqref{eqn:omega.resonant}). In the subsonic streaming limit $\driftvelmag<1$, the long-wavelength mode is the most relevant. Examples of each are shown in Fig.~\ref{fig:mode.structure}. 

\vspace{-0.2cm}
\begin{enumerate}

\item{\bf Long-Wavelength / Pressure-Free Mode} ($\kappa_{\|} \ll \hat{\mu}$; Eq.~\eqref{eqn:longwave.mode}): As $k\rightarrow 0$, the fastest-growing mode has ${\bf k} \propto \driftvel$ (i.e.\ $\cos{\theta}=\pm 1$), and the perturbed velocities are parallel: $\delta {\bf v} \propto  \delta {\bf u}  \propto {\bf k} \propto \driftvel$. Moreover $\delta {\bf v} \approx \delta{\bf u}$ and $\delta \rho_{d} \approx \mu\,\delta {\rho}$. In other words the mode simply features {\em coherent} oscillations of the dust and gas together, because these modes have wavelengths larger than the deceleration length of the dust. To leading order, the mode does not generate fluctuations in the dust-to-gas ratio. A second order phase offset does appear between the dust and gas perturbations, and this drives the growth. But this offset is weak and the growth rate is correspondingly small. 

However, as we noted above, the long-wavelength mode is not a perturbed sound wave (coupled dust-gas soundwaves exist at low-$k$, but these are damped). It is a unique, approximately one-dimensional, pressure-free, two-fluid mode. The phase and group velocities scale as $\sim \driftvel \,(k\,|\driftvel|\,\langle t_{s} \rangle/\mu)^{-1/3} \propto k^{-1/3}$, diverging as $k\rightarrow0$ because of the leading-order term in $\omega \propto k^{2/3}$. There is also a phase offset, whereby the velocity perturbations lead (follow) the density perturbations by a phase angle of $\sim \pi/6$ for $\driftvelmag > 1$ ($\driftvelmag < 1$).\footnote{The phase angle $\pi/6$ (the argument of $\iimag^{1/3}$) appears repeatedly because the dominant imaginary terms in the dispersion relation are cubic.} This implies that the gas density response to the velocity perturbations is distinct from a sound wave,  satisfing $\delta\rho/\rho_{0} \sim \driftvelmag^{-1}\,(\kappa_{|}/\mu)^{1/3}\,|\delta {\bf v}/c_{s}| \sim [\kdimless /(\mu\,\driftvelmag^{2})]^{1/3}\,|\delta {\bf v}/c_{s}|$. 

\item{\bf Resonant Mode}, {\bf Intermediate-Wavelengths} ($\hat{\mu} \ll \kappa_{\|} \ll \hat{\mu}^{-1}$; Eq.~\eqref{eqn:longwave.mode.midk}): For intermediate $k$ with $\driftvelmag \ge 1$, the fastest-growing mode has ${\bf k}$ oriented at the resonant angle $\cos{\theta} = \pm 1/\driftvelmag$ (i.e.\ $\kappa_{\|}=\kdimless$, with $k_{\|} = \pm k/\driftvelmag$), so for $\driftvelmag \gg1$ it is increasingly transverse ($k \approx k_{\bot}$). To leading order in $\kdimless$ and $\mu$, $\omega \approx c_{s}\,k$ so the wave phase/group velocity $=c_{s}\,\hat{\bf k}$. This is the key RDI resonance: the wavespeed (approximately) matches the natural wavespeed of the system without dust (in this case, the sound speed), with a wavevector angle $\cos{\theta}=\pm1/\driftvelmag$, such that the dust drift velocity ({in the direction of the wave propagation}) is also equal to that wavespeed: $\driftvel \cdot \hat{\bf k} = c_{s}$. In other words, the bulk dust is co-moving with the wave in the direction $\hat{\bf k}$. 

For $\mu\ll1$, the gas density response behaves like a sound wave, $\delta \rho/\rho_{0} \approx \hat{\bf k}\cdot \delta {\bf u}/c_{s}$, in-phase with the velocity in the $\hat{\bf k}$-direction. However, the dust density response now lags by a phase angle $\sim \pi/6$, and, more importantly, the resonance generates a strong dust density response: $|\delta \rho_{d}| \sim (2\,\mu\,\kappa_{\|})^{1/2}\, |\delta \rho|$.  We see the dust-density fluctuation is enhanced by a factor $\sim (2\,\kappa_{\|}/\mu)^{1/2} \gg 1$ relative to the mean ($\mu$), which is much stronger than for the long-wavelength mode (with $\delta \rho_{d} \sim \mu\,\delta\rho$). The resonant mode can  thus  generate very large dust-to-gas fluctuations even for otherwise weak modes, and the magnitude of the induced dust response increases at shorter wavelengths.

Effectively, as the dust moves into the gas density peak from the wave, it decelerates, producing a trailing ``pileup'' of dust density behind the gas density peak, which can be large. This dust-density peak then accelerates the gas, amplifying the wave. Because of the resonance with both drift and sound speeds, these effects add coherently as the wave propagates, leading to the exponential growth of the mode.

One further  interesting feature of this mode deserves mention: the velocities ($\delta {\bf v} \approx \delta {\bf u}$ here) are not fully-aligned with $\hat{\bf k}$ but have a component in the ${\bf k}_{\bot}$ direction,\footnote{Note that for $\driftvelmag \gg1$, the ${\bf k}_{\bot}$ direction is approximately the $\driftvelhat$ direction.} which leads the velocity in the $\hat{\bf k}$ direction by a phase angle $\sim \pi/4$. This is a response to the dust streaming in the ${\bf k}_{\bot}$ direction and  the amplitude of this term decreases with $k$.

\item{\bf Resonant Mode}, {\bf Short-Wavelengths} ($\kappa_{\|} \gg \hat{\mu}^{-1}$; Eq.~\eqref{eqn:omega.resonant}): At high-$k$ with $\driftvelmag \ge1$ the details of the resonant mode (and scaling of the growth rate) change. The resonant condition remains the same as at mid $k$, however, the mode propagates with wavespeed $c_{s}\,\hat{\bf k}$ along the resonant angle $\cos{\theta}=\pm 1/\driftvelmag$, and the gas behaves like a soundwave (the velocities are now aligned $\delta {\bf u} \propto \delta {\bf v} \propto {\bf k}$). This generates a strong dust response with the slightly-modified scaling $|\delta \rho_{d} |/|\delta \rho| \sim (4\,\mu\,\kappa_{\|})^{1/3} \gg 1$ (scaling like the growth rate), with $\delta \rho_{d}$ lagging the gas mode by a phase angle $\sim \pi/6$. Importantly, $|\delta \rho_{d}|/|\delta \rho|$ continues to increase indefinitely with $k$, and in this regime, the dust density perturbation  becomes {\em larger} than the gas density perturbation in absolute units (even though the mean dust density is smaller than gas by a factor $\mu$). The dust velocity $\delta {\bf v}$ is parallel to $\delta {\bf u}$, but with a smaller amplitude $|\delta {\bf v}| / |\delta {\bf u}| \sim (\mu\,\kappa_{\|}/2)^{-1/3} \ll 1$, and $\delta {\bf v}$ leads $\delta {\bf u}$ by a phase angle $\sim \pi/6$.

\end{enumerate}

\begin{figure*}
\plotsidesize{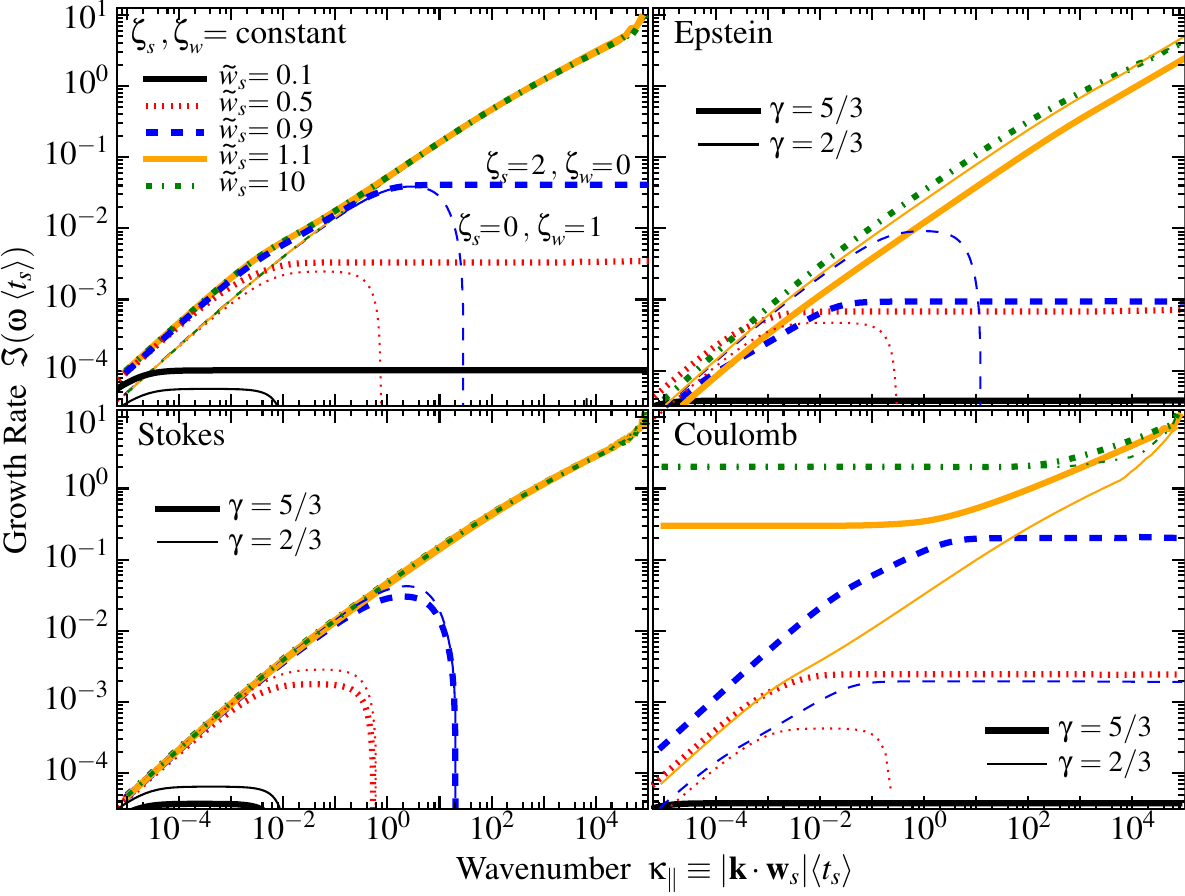}{0.85}
    \vspace{-0.25cm}
    \caption{Growth rates of the most-rapidly-growing unstable mode as a function of wavenumber and drift velocity, as Fig.~\ref{fig:growth.rate.demo}, for different drag laws (see \S~\ref{sec:draglaws}). Here we take $\mu=0.01$, and marginalize over angle (the most rapidly-growing cases $\cos{\theta}=1$ for $\driftvelmag<1$ or $\cos{\theta}=\pm1/\driftvelmag$ for $\driftvelmag \ge 1$). 
    {\em Top Left:} Arbitrary constant $\coeffTSrho$, $\coeffTSv$ parameterization of $t_{s}$ (Eq.~\eqref{eqn:ts.general}) with $\coeffTSrho=2$, $\coeffTSv=0$ (thick lines) or $\coeffTSrho=0$, $\coeffTSv=1$ (thin lines). As shown in \S~\ref{sec:general.modes} the dependence on these parameters is weak; the largest effect is to determine, when $\driftvelmag < 1$, whether all $k$ are unstable (if $\coeffTSrho > 1+\coeffTSv$) or only small-$k$ ($\coeffTSrho < 1+\coeffTSv$), but the maximum growth rates in these cases are very similar.
    {\em Top Right:} Epstein drag (\S~\ref{sec:epstein}), with gas equation-of-state parameters $\gamma=5/3$ (thick) or $\gamma=2/3$ (thin). The qualitative behavior is identical, with modest normalization differences, and the transition between regimes for $\driftvelmag<1$ ($\coeffTSrho = 1+\coeffTSv$) occurring at $\gamma^{-1} = 1 - 9\,\pi\,\driftvelmag^{2}/64$ ($\coeffTSrho$, $\coeffTSv$ depend on $\gamma$ and $\driftvelmag$). Note the low saturation value of the $\gamma=5/3$, $\driftvelmag=0.9$ case occurs because it is very close to this singular value ($(1+\coeffTSv)-\coeffTSrho \approx 0.02$). 
    {\em Bottom Left:} Stokes drag (\S~\ref{sec:stokes}). The dependence on $\gamma$ is weak and for all $\gamma < 3$, 
    high-$k$ modes with $\driftvelmag<1$ are stable.
    {\em Bottom Right:} Coulomb drag (\S~\ref{sec:coulomb}; here $\Gamma=1$). For long-wavelength modes with $\driftvelmag < 1$, and all high-wavelength modes, the qualitative behavior is similar to other laws although normalization differences are more obvious. The high growth-rate, low-$k$ modes with $\driftvelmag > 1$ are a different instability which manifests because when $\driftvelmag > 1$ in Coulomb drag, increasing the dust-gas velocity {\em decreases} the drag acceleration, so the dust speeds up and the system ``self-decouples.'' Physically Epstein or Stokes drag should be dominant over Coulomb drag in this limit.
        \vspace{-0.25cm}
        \label{fig:growth.rate.draglaw}}
\end{figure*}

\vspace{-0.5cm}
%\section{Different Drag Physics}
\section{Drag Physics}
\label{sec:draglaws}

In this section, we  consider different physical drag laws. This involves inserting specific forms of $\coeffTSrho$ and $\coeffTSv $ into the dispersion relations derived in \S~\ref{sec:general.modes}. Numerically calculated growth rates for representative cases are shown for comparison in Fig.~\ref{fig:growth.rate.draglaw}. 
We also show as illustrative cases two arbitrary but constant, order-unity choices: $(\coeffTSrho,\coeffTSv) = (0,1)$ and $(\coeffTSrho,\coeffTSv) =(2,0)$. The former case illustrates that with $\coeffTSv < \tildeCoeffTSv$, the qualitative behavior of the modes are largely similar to the constant-$t_{s}$ case in Fig.~\ref{fig:growth.rate.demo}. The latter shows that when $\coeffTSv > \tildeCoeffTSv$, the dominant effect is to extend the instability of sub-sonic ($\driftvelmag<1$) cases to high-$k$.
% (this also makes the transition between the long-wavelength and mid-$k$ resonant mode slightly more visually obvious, owing to their different dependence on $\coeffTSv$, $\coeffTSv$). 
For simplicity of notation, we again use the dimensionless variables of Eq.~\eqref{eqn:dimensionless vars} in this section.

\vspace{-0.5cm}
\subsection{Constant Drag Coefficient}
\label{sec:constantdrag}

The simplest case is $t_{s} = $\,constant, so $\delta {t_{s}} = 0$ -- i.e.\ $\coeffTSrho=\coeffTSv = 0$ (and $\tildeCoeffTSv=1$). The characteristic polynomial simplifies to $B_{\omega}=A_{\omega}\,B^{\prime}_{\omega}$ with $B^{\prime}_{\omega}\equiv\omprime\,(\omprime+\iimag)\,(\omdimless^{2}-\kdimless^{2}) + \iimag\,\mu\,(\omdimless^{2}\,\omprime - \kappa_{\|}^{2}\{ \omprime + \iimag\})$. Since $\tildeCoeffTSv=1$, all pure-perpendicular modes are damped or stable. 

The long-wavelength modes are unstable with growth rates,
\begin{align}
\omdimless(\kappa_{\|} \ll \hat{\mu}) & = \kappa_{\|} +  \frac{\pm \sqrt{3} + \iimag}{2}\,\hat{\mu}^{1/3}\,\kappa_{\|}^{2/3}. %\tag*{($\kappa_{\|} \ll \hat{\mu}$)}
\end{align}
For $\driftvelmag < 1$, these cut off at high-$k$ with $\omdimless \approx (\mu/2)\,(\driftvelmag^{2} - \kdimless_{z}^{2}/(1+\mu)^{2})$ (Eq.~\eqref{eqn:omega.lowk}). For $\driftvelmag \ge 1$, at large $k$ the \intermediatemodename\ mode (Eq.~\eqref{eqn:omega.med}) is present with growth rate $\Im{(\omdimless)} =\mu\,(\driftvelmag\,|\cos{\theta}| - 1)/2$ so the most rapidly-growing mode is parallel. The \slowmodename\ mode (Eq.~\eqref{eqn:omega.slow}) is present with growth rate $\Im{(\omdimless)} \sim \mu/[1 - (\driftvelmag\,\cos{\theta})^{-2}]$. At resonance ($\cos{\theta}\rightarrow \pm 1/\driftvelmag$),  the growth rate is,
\begin{align}
\omdimless_{\ast} &= 
\begin{cases}
	{\displaystyle \kappa_{\|}\,\left(1-\frac{\hat{\mu}}{4}\right) - \iimag\frac{\hat{\mu}}{8} + \frac{(1+\iimag)}{2}\,({\hat{\mu}\,\kappa_{\|}})^{1/2} \ \ \hfill{(\hat{\mu} \ll \kappa_{\|} \ll \hat{\mu}^{-1})} } \\ 
	\\
	{\displaystyle \kappa_{\|} - \iimag\,\frac{1+\mu}{3} + (1+\iimag\,\sqrt{3})\,\left(\frac{\mu\,\kappa_{\|}}{16}\right)^{1/3}  \ \ \ \ \hfill{(\kappa_{\|} \gg \hat{\mu}^{-1})}. }\\
\end{cases}
\end{align}

Examples of this case ($\coeffTSrho=\coeffTSv = 0$) are shown in Fig.~\ref{fig:growth.rate.demo}, but they are similar to the other cases with $\coeffTSv < \tildeCoeffTSv$ in Fig.~\ref{fig:growth.rate.draglaw}.

\vspace{-0.5cm}
\subsection{Epstein Drag}
\label{sec:epstein}

The general expression\footnote{Equation~\eqref{eqn:ts.epstein} is actually a convenient polynomial approximation, given in \citet{draine.salpeter:ism.dust.dynamics}, to the more complicated dependence on $|{\bf v}-{\bf u}|$. However using the more complicated expression yields  negligible ($\sim 1\%$) differences for any parameters considered here.} (including physical dimensions) for the drag coefficient in the Epstein limit is:
\begin{align}
\label{eqn:ts.epstein} t_{s} &= \sqrt{\frac{\pi\,\gamma}{8}}\,\frac{\bar{\rho}_{d}\,R_{d}}{\rho\,c_{s}}\,\left( 1 + \epsteincoeff\,\frac{|{\bf v}-{\bf u}|^{2}}{c_{s}^{2}} \right)^{-1/2}, \ \ \ \ \epsteincoeff \equiv \frac{9\,\pi\,\gamma}{128}.
\end{align}
Where $\bar{\rho}_{d}$ is the internal material density of the aerodynamic particle and $R_{d}$ is the particle (grain) radius. For astrophysical dust, $\bar{\rho}_{d}\sim 1-3\,{\rm g\,cm^{-3}}$,  and $R_{d} \sim 0.001 - 1\,\mu$m in the ISM,  or in denser environments $R_{d} \sim 0.1-1000\,\mu\,$m (e.g., protoplanetary disks, SNe ejecta, or cool star atmospheres; \citealt{draine:2003.dust.review}). Note that Epstein drag depends on the {\em isothermal} sound speed, $c_{\rm iso} \equiv \sqrt{k_{B}\,T/m_{\rm eff}}$ (where $m_{\rm eff}$ is the mean molecular weight). However, because we work in units of the sound speed $c_{s}\equiv \sqrt{\partial P/\partial \rho}$, we relate the two via the usual equation-of-state parameter $\gamma$,
\begin{align}
\gamma &\equiv \frac{c_{s}^{2}}{c_{\rm iso}^{2}} = \frac{\rho}{P}\,\partialAB{P}{\rho},
\end{align}
and will assume $\gamma$ is a constant under linear perturbations. We emphasize that the $\gamma$ here is the appropriate $\gamma$ describing how the temperature responds to compression or expansion on a wave-crossing time -- roughly the same $\gamma$ appropriate for a sound wave. This means that external heating or cooling processes are only important for $\gamma$ if the heating/cooling time is shorter than the sound-crossing time (otherwise we typically expect adiabatic $\gamma$).

Note that because $t_{s}$ now depends on $\langle |{\bf v}-{\bf u}| \rangle = |\driftvel|$, Eq.~\eqref{eqn:mean.v.offset} for the drift velocity, $\driftvel = {\bf a}\,\langle t_{s} \rangle/(1+\mu)$, is implicit. Define $\driftvelmagi{0} \equiv |{\bf a}|\,t_{0}/(c_{s}\,(1+\mu))$ where $t_{0} \equiv (\pi\,\gamma/8)^{1/2}\,\bar{\rho}_{d}\,R_{d}/(\rho_{0}\,c_{s})$ is the stopping time at zero relative velocity. Then the solution of Eq.~\eqref{eqn:mean.v.offset} is 
\begin{equation}
\driftvelmag^{2} = \frac{1}{2\,\epsteincoeff}\left[ (1+4\,\epsteincoeff\,\driftvelmagi{0}^{2})^{1/2}-1\right],\label{eqn:ws.in.epstein}
\end{equation}
which reduces to $\driftvelmag \approx \driftvelmagi{0}$ for  $|{\bf a}| \ll  c_{s}/t_{0}$, 
or $\driftvelmag \approx \epsteincoeff^{-1/4}\,\driftvelmagi{0}^{1/2}$
for  $|{\bf a}| \gg c_{s}/t_{0}$.

With Eq.~\eqref{eqn:ts.epstein} for $t_{s}$ and Eq.~\eqref{eqn:ws.in.epstein} for $\driftvelmag$, $\delta{t_{s}}$ follows Eq.~\eqref{eqn:ts.general} with
\begin{align}
%\delta \tilde{t_{s}} &= -\frac{\gamma+1+2\,\epsteincoeff\,\driftvelmag^{2}}{2\,(1+\epsteincoeff\,\driftvelmag^{2})}\,\delta\tilde{\rho}
%-\frac{\epsteincoeff\,\driftvelmag}{1+\epsteincoeff\,\driftvelmag^{2}}\,\driftvelhat\cdot\left(\delta{\bf \tilde{v}} - \delta{\bf \tilde{u}} \right) \\ 
\coeffTSrho &= \frac{\gamma+1+2\,\epsteincoeff\,\driftvelmag^{2}}{2\,(1+\epsteincoeff\,\driftvelmag^{2})}\ , \ \ \ \ \ 
 \coeffTSv = \frac{\epsteincoeff\,\driftvelmag^{2}}{1+\epsteincoeff\,\driftvelmag^{2}}.
\end{align}
From this we can derive the relevant instability behavior for different $\gamma$ and $\driftvelmag$. Note $\coeffTSrho>0$ and $\coeffTSv>0$, so the ``decoupling'' instability (which requires $\tildeCoeffTSv < 0$) is not present. 

In Fig.~\ref{fig:growth.rate.draglaw}, for this case (as well as Stokes and Coulomb drag), we show values of $\Im(\omega)$ for two values of $\gamma=2/3,\,5/3$ (and a range of $\driftvelmag$), which determine $\coeffTSrho$, $\coeffTSv$. The two values of $\gamma$ are chosen to bracket the range where the behavior changes ($\coeffTSrho < \tildeCoeffTSv$ and $\coeffTSrho > \tildeCoeffTSv$) and be qualitatively representative of cases where cooling (on the mode-crossing time) is either inefficient ($\gamma=5/3$, i.e.\ adiabatic) or efficient ($\gamma=2/3$, approximately valid in the dense/cold ISM of GMCs, see \citealt{glover:effective.ism.eos.cooling}, although not extremely dense cases such as proto-planetary disks, where cooling is again inefficient, \citealt{lin:cooling.protoplanetary.disks}).

\vspace{-0.5cm}
\subsubsection{Super-sonic streaming ($\driftvelmag \gg 1$)}
\label{sec:epstein:supersonic}

In the $\driftvelmag \gg 1$ limit, $\coeffTSrho\rightarrow1 + \mathcal{O}(\driftvelmag^{-2})$ (independent of $\gamma$) and $\coeffTSv\rightarrow 1$. This stabilizes the \intermediatemodename\ modes (Eq.~\eqref{eqn:omega.med}) because at high-$\driftvelmag$, the $\coeffTSv$ term dominates over ($1-\coeffTSrho$), viz., the stronger coupling from at high relative velocity stabilizes the modes. 
The long-wavelength modes (Eq.~\eqref{eqn:longwave.mode}) are present and saturate in the \slowmodename/resonant mode, with growth rate $\Im(\omdimless) \sim \mu\,[1-(\driftvelmag\,\cos{\theta})^{-2}]^{-1}\,(1-\coeffTSrho/\tildeCoeffTSv)$, which approaches $ \Im(\omdimless)\sim \mu/2$ for $\driftvelmag \gg 1$ out-of-resonance. 

At resonance, we insert  the full expressions for $\coeffTSrho$ and $\coeffTSv$ into Eq.~\eqref{eqn:longwave.mode.midk} and Eq.~\eqref{eqn:omega.resonant}. This gives
\begin{align}
\label{eqn:resonant.epstein.mid}\omdimless_{\ast} \approx&\,\kdimless\,\omegaZ_{\Re} - \frac{\iimag\hat{\mu}}{8}\left(\frac{\tildeCoeffTSv-\coeffTSrho}{\tildeCoeffTSv}\right) + \frac{\iimag\pm1}{2}\left( \left| \frac{\tildeCoeffTSv-\coeffTSrho}{\tildeCoeffTSv}\right| \hat{\mu} \,\kdimless \right)^{1/2}, \\
& \frac{\tildeCoeffTSv-\coeffTSrho}{\tildeCoeffTSv}  = \frac{1+2\epsteincoeff \driftvelmag^{2}-\gamma}{2+4\epsteincoeff \driftvelmag^{2}} 
= \frac{1}{2}+\mathcal{O}(\driftvelmag^{-2}),\nonumber \\
%= \frac{1}{2}-\frac{\gamma}{4\epsteincoeff \driftvelmag^{2}}+\mathcal{O}(\driftvelmag^{-4}),\nonumber \\
& \omegaZ_{\Re}= 1+ \frac{3\hat{\mu}}{16}\,\left( 1 + \mathcal{O}(\driftvelmag^{-2}) \right),\nonumber
\end{align}
in the ``mid-$k$'' regime (we show the lowest order terms in $\driftvelmag^{-1}$ for simplicity),  and 
\begin{align}
\label{eqn:resonant.epstein} \omdimless_{\ast} \approx&\, \kdimless - \iimag\,\omegaZ + (\iimag\,\sqrt{3}+1)\,\left( \frac{|\Theta|\,\mu\,\kdimless}{16} \right)^{1/3} \\ 
\nonumber &\Theta =\, \frac{1-\gamma+2\,\epsteincoeff}{2\,(1 + \epsteincoeff\,\driftvelmag^{2})} = \frac{1-\gamma+2\,\epsteincoeff}{2\,\epsteincoeff\,\driftvelmag^{2}} + \mathcal{O}(\driftvelmag^{-4}), \\ 
\nonumber & \omegaZ =  -\frac{2\,\epsteincoeff\,\driftvelmag^{2}}{3\,(1-\gamma+2\,\epsteincoeff)} + \mathcal{O}(\driftvelmag^{0}),
\end{align}
in the ``high-$k$'' regime.
We see that in the mid-$k$ regime, the growth rate is mostly independent of $\driftvelmag$ and $\gamma$, while in the high-$k$ regime the growth rate decreases, ${\Im}(\omega_{\ast}) \propto \driftvelmag^{-2/3}$, at large $\driftvelmag$.

The dependence on $\gamma$ is weak. At mid $k$, we see from Eq.~\ref{eqn:resonant.epstein.mid} that the growth rate declines as we approach the
point where $\tildeCoeffTSv-\coeffTSrho =0$, which occurs at $\driftvelmag^{2} = 64(\gamma-1)/(9\pi \gamma)$. This implies that unless the gas equation of state is very stiff -- specifically, $\gamma>64/(64-9\pi)\approx 1.8$ -- this ``stable point'' does not exist for $\driftvelmag>1$ (a necessary condition for  resonant modes). Even for $\gamma \gtrsim 1.8$, the point of stability 
occurs only at a specific $\driftvelmag$, and so is unlikely to be of physical significance.

At high-$k$, we see somewhat similar behavior, with  the growth rate declines as $\gamma$ approaches the point where $\Theta =0$ (and $\omegaZ$ diverges), at $\gamma = 64/(64-9\pi)\approx 1.8$. In fact, at this point exactly, our series expansion is incorrect (since $\omegaZ$ diverges), and a resonant mode still exists, but with a growth rate that increases more slowly with $k$: 
\begin{equation}
\label{eqn:resonant.epstein.high.reduced}\omdimless_{\ast} = \kdimless + \left(\sin\frac{\pi}{8}+ \iimag\,\cos\frac{\pi}{8} \right)\,\left(\frac{(\driftvelmag^{2}-1)\,\epsteincoeff\,\mu\, \kdimless}{2\,(1+\epsteincoeff\,\driftvelmag^{2})}\right)^{1/4}.\end{equation}
Again, it seems unlikely that this specific point, $\gamma\approx 1.8$ is of particular physical significance (and in any case, the system is still unstable, just with the reduced growth rate in Eq.~\eqref{eqn:resonant.epstein.high.reduced}).

\vspace{-0.5cm}
\subsubsection{Sub-Sonic ($\driftvelmag \ll 1$)}
\label{sec:epstein.subsonic}

Now consider $\driftvelmag\ll1$. In this limit $\coeffTSrho= (\gamma+1)/2+ \mathcal{O}(\driftvelmag^{2})$ and  $\coeffTSv= \epsteincoeff\,\driftvelmag^{2}+ \mathcal{O}(\driftvelmag^{4})$; i.e.,  the velocity-dependent terms in $t_{s}$  become second-order, as expected. For $\driftvelmag < 1$ the resonant and \intermediatemodename\ modes are stabilized. We also see that the type of unstable mode will depend on the value of $\gamma$: if $\gamma > 1$ then $\coeffTSrho/\tildeCoeffTSv \approx (\gamma+1)/2> 1$, which implies the ``subsonic'' mode at low-$k$ from Eq.~\eqref{eqn:omega.lowk} is stabilized, but ``\slowmodename'' mode from Eq.~\eqref{eqn:omega.slow} is unstable; if $\gamma< 1$, 
the ``\slowmodename'' mode at $\kdimless \gtrsim 1$ becomes damped at  high $k$, and the ``subsonic'' low-$k$ expression from Eq.~\eqref{eqn:omega.lowk} is unstable. 
 
 The ``\slowmodename'' modes, relevant for  $\gamma \gtrsim 1$, have growth rates that increase with $k$ for $\kdimless \ll \hat{\mu}$ (the long-wavelength mode; Eq.~\eqref{eqn:longwave.mode}), then saturate to a constant maximum  for $\kdimless \gtrsim 1$ (i.e.\ all modes shorter-wavelength than the length scale $\sim c_{s}\,\langle t_{s} \rangle$ have similar growth rate). For large $k$ and $\driftvelmag \ll 1$ the growth rate from Eq.~\eqref{eqn:omega.slow} is $\Im(\omdimless)\approx \driftvelmag^{2}\,\cos^{2}{\theta}\,\mu\,(\gamma-1)/2$. 
The ``subsonic'' mode (Eq.~\eqref{eqn:omega.lowk}), relevant for very soft equations of state with $\gamma\lesssim1$, has a maximum growth rate $\Im(\omdimless) \approx \driftvelmag^{2}\,\mu\,(\tildeCoeffTSv-\coeffTSrho)/2 \approx \driftvelmag^{2}\,\mu\,(1-\gamma)/4$, which again occurs for parallel modes. The mode is stabilized at short wavelengths, $\kdimless \gtrsim (1+\mu)\driftvelmag \sqrt{1-\gamma}$.

Overall, we see that for {\em all} $\gamma$, there is an unstable parallel mode at low $\driftvelmag \ll 1$, with maximum growth rate $\sim \driftvelmag^{2}\,\mu$. The difference is that for $\gamma>1$ the unstable modes are  \slowmodename\ modes, which are unstable at all $k$ and propagate with velocity ${\driftvel}$ when $\kdimless \gg 1$; for $\gamma<1$ the instability only exists for long wavelength modes $\kdimless \lesssim \driftvelmag$, which propagate with velocity $\pm c_{s}\,\driftvelhat/\sqrt{1+\mu}$.

Again there is one critical point when $\tildeCoeffTSv-\coeffTSrho=0$, or $\driftvelmag^{2} = 64\,(\gamma-1)/(9\pi\,\gamma)$, where the standard long-wavelength instability vanishes. This occurs only for some specific $\driftvelmag$ at a given $\gamma$, so is unlikely to be of physical significance for most $\gamma$. Again, at this point, there is in fact still an instability, albeit with a reduced growth rate (see footnote~\ref{foot:when.instab.vanishes}, near Eq.~\eqref{eqn:longwave.mode}; the instability only truly vanishes at $\coeffTSv=0$, $\coeffTSrho=1$ exactly). However, one does approach this vanishing-point for $\gamma=1$ as $\driftvelmag\rightarrow0$ becomes sufficiently small.

This leads to a cautionary note: it is common in some sub-sonic ($\driftvelmag \ll c_{s}$) applications to drop the term in $|{\bf v}-{\bf u}|^{2}/c_{s}^{2}$ in Eq.~\ref{eqn:ts.epstein} (i.e.\ simply taking $t_{s} \propto 1/\rho\,c_{s}$), for simplicity. If the gas is also isothermal ($\gamma=1$), this would give $\coeffTSv=0$, $\coeffTSrho=1$ exactly and the instabilities would vanish for $\driftvelmag \ll 1$. However, this can be mis-leading: although the term in $|{\bf v}-{\bf u}|^{2}/c_{s}^{2}$ is small, it does give rise to a non-zero (albeit small) growth rate. Moreover if the equation of state is even slightly non-isothermal (e.g.\ $\gamma=0.9,\,1.1$), the instability is not suppressed strongly. Also, we caution that the appropriate equation-of-state here is that relevant under local, small-scale compression by dust and sound waves (not necessarily the same as the effective equation-of-state of e.g.\ a vertical atmosphere).

\vspace{-0.5cm}
\subsection{Stokes Drag}
\label{sec:stokes}

The expression for drag in the Stokes limit -- which is valid for an intermediate range of grain sizes, when $R_{d} \gtrsim (9/4)\,\lambda_{\rm mfp}$ but  $\mathrm{Re}_{\mathrm{grain}}\equiv R_{d}|\driftvel|/(\lambda_{\mathrm{mfp}} c_{s})\lesssim 1$
-- is given by multiplying the Epstein expression (Eq.~\eqref{eqn:ts.epstein}) by $(4\,R_{d}) / (9\,\lambda_{\rm mfp})$. Here  $\lambda_{\rm mfp} \propto 1/(\rho\,\sigma_{\mathrm{gas}})$ is the gas mean-free-path, $\sigma_{\mathrm{gas}}$ is the gas collision cross section, and $\mathrm{Re}_{\mathrm{grain}}$ is the
Reynolds number of the streaming grain.

We can solve implicitly for the dust streaming velocity  $\driftvel$, which is the same as in the Epstein case (since $t_{s}$ depends on $|{\bf v}-{\bf u}|$ in the same manner). However, the absolute value of $t_{s}$ only determines our units, and the behavior of interest  depends only on the coefficients $\coeffTSrho$ and $\coeffTSv$. Since $R_{d}$ is a material property of the dust and $\sigma_{\rm gas}$ an intrinsic property of the gas, the important aspect of the Stokes drag law is that it multiplies the Epstein law by one power of $\rho$. Although it is certainly possible $\sigma_{\rm gas}$ might depend on density and/or temperature, lacking a specific physical model for this we will take it to be a constant for now. This simply gives $\coeffTSrho\rightarrow \coeffTSrho - 1$, relative to the scalings for Epstein drag.

When $\driftvelmag \ll 1$ (c.f., \S~\ref{sec:epstein.subsonic} for Epstein drag), $\coeffTSrho= (\gamma-1)/2+ \mathcal{O}(\driftvelmag^{2})$ and $\coeffTSv= \epsteincoeff\,\driftvelmag^{2}+ \mathcal{O}(\driftvelmag^{4})$, and \intermediatemodename\ and resonant modes are stabilized (because $\driftvelmag<1$). The \slowmodename\ (high-$k$) mode is stabilized for $1-\coeffTSrho/\tildeCoeffTSv \approx (3-\gamma)/2 > 0$, viz., so as long as $\gamma < 3$ (which is expected in almost all physical situations) the \slowmodename\ mode is damped.
% (if $\gamma > 3$, it is unstable with growth rate $\driftvelmag^{2}\,\mu\,(\gamma-3)/2$). 
However for all $\gamma < 3$, the subsonic low-$k$ mode (Eq.~\eqref{eqn:omega.lowk}) is unstable for $\kdimless \lesssim \driftvelmag$, with maximum growth rate $\Im(\omdimless)\approx \driftvelmag^{2}\,\mu\,(3-\gamma)/4$. This is larger (smaller) than the Epstein drag growth rate for $\gamma <5/3$ ($\gamma>5/3$).

In the limit $\driftvelmag \gg 1$, the Stokes drag expression cannot formally apply because   $R_{d}>\lambda_{\mathrm{mfp}}$  then implies $\mathrm{Re}_{\mathrm{grain}}=R_{d}|\driftvel|/(\lambda_{\mathrm{mfp}} c_{s})\gtrsim 1$.
When this is the case, either because $\driftvelmag$ is large  or (more commonly) $R_{d}$ is large, there is no longer a simple drag law because the grain develops a turbulent wake. This 
will tend to increase the drag above the Stokes estimate (the turbulence increases the drag)
with a stronger and stronger effect as $\mathrm{Re}_{\mathrm{grain}}$ increases. Given some empirically determined scaling of $t_{s}$ with $R_{d}$, $\rho$, $\driftvelmag$ etc. (see, e.g., 
\citealt{clair.hamielec:sphere.drag} for subsonic drag), one could still qualitatively consider such a turbulent drag 
within the framework above, with the properties of the turbulence determining $\coeffTSrho$ and $\coeffTSv$. We do not do this here, but note that because $\mathrm{Re}_{\mathrm{grain}}$
increases with $\driftvelmag$ and $\rho$ (through $\lambda_{\mathrm{mfp}}$), we  
expect $t_{s}$ to decrease with  $\driftvelmag$ and $\rho$, viz., $\coeffTSrho>0$ and $\coeffTSv>0$. 
The general scalings are thus likely similar to the Epstein case, but 
with a larger $\coeffTSv$ for $\driftvelmag\ll 1$, because the velocity dependence of the drag will 
be significant, even for subsonic streaming. 

Of course we can still simply calculate what the mode growth rates would be, if the usual Stokes expression applied even for $\driftvelmag \gtrsim 1$. This is shown in Fig.~\ref{fig:growth.rate.draglaw}, for the sake of completeness.

\vspace{-0.5cm}
\subsection{Coulomb Drag}
\label{sec:coulomb}

The standard expression\footnote{Again, Eq.~\eqref{eqn:drag.coulomb} is a polynomial approximation for more complex dependence on $|{\bf v}-{\bf u}|$, given in \citet{draine.salpeter:ism.dust.dynamics}. However using this approximation versus the full expression makes no important difference to our results.} (in physical units) for $t_{s}$ in the Coulomb drag limit is 
\begin{align}
\label{eqn:drag.coulomb} t_{s} &= \sqrt{\frac{\pi\,\gamma}{2}}\,\frac{\bar{\rho}_{d}\,R_{d}}{\rho\,c_{s}\,\ln{\Lambda}}\,\left(\frac{k_{B}\,T}{z_{i}\,e\,U} \right)^{2}\,\left[ 1 + \coulombcoeff\,\frac{|{\bf v}-{\bf u}|^{3}}{c_{s}^{3}} \right] \\ 
\nonumber \Lambda &\equiv \frac{3\,k_{B}\,T}{2\,R_{d}\,z_{i}\,e^{2}\,U}\,\sqrt{\frac{m_{i}\,k_{B}\,T}{\pi\,\rho}} \ \ \ \ \ , \ \ \ \ \ 
\coulombcoeff \equiv \sqrt{\frac{2\,\gamma^{3}}{9\,\pi}}
\end{align}
where  $\ln{\Lambda}$  is the Coulomb logarithm, $e$ is the electron charge, $z_{i}$ is the mean gas ion charge, $m_{i}$ is the mean molecular weight, $T\propto \rho^{\gamma-1}$ is the gas temperature, and $U$ is the electrostatic potential of the grains, $U\sim Z_{\rm grain}\,e/R_{d}$ (where $Z_{\rm grain}$ is the grain charge). The behavior of $U$ is complicated and depends on a wide variety of environmental factors: in the different regimes considered in \citet{draine.salpeter:ism.dust.dynamics} they find regimes where $U\sim$\,constant and others where $U \propto Z_{\rm grain} \propto T$, we therefore parameterize the dependence by $U\propto T^{\Gamma}$. 

With this ansatz, we obtain 
\begin{align}
\nonumber \coeffTSrho &=  1 +2\,(\gamma-1)\,\Gamma
- \frac{3\,(\gamma-1)}{2\,(1+\coulombcoeff\,\driftvelmag^{3})}
- \frac{1-(3-2\,\Gamma)\,(\gamma-1)}{2\,\ln{\Lambda}}, \\ 
 \coeffTSv &= -\frac{3\,\coulombcoeff\,\driftvelmag^{3}}{1+\coulombcoeff\,\driftvelmag^{3}} < 0.
\end{align}
For relevant astrophysical conditions, $\ln{\Lambda} \sim 15-20$, so the $\ln{\Lambda}$ term in $\coeffTSrho$ is unimportant.

In general, Coulomb drag is sub-dominant to Epstein or Stokes drag under astrophysical conditions when the direct effects of magnetic fields on grains (i.e.,  Lorentz forces) are not important. Nonetheless, the qualitative structure of the scaling  produces similar features to the Epstein and Stokes drag laws, and we consider it here for completeness. In fact, grains influenced by Coulomb drag are significantly ``more unstable'' than those influenced by Epstein or Stokes drag. 
For $\driftvelmag \ll 1$, $\coeffTSrho\rightarrow [(3\,\gamma-4) + (5-3\,\gamma)\,\log{\Lambda}]/(2\,\log{\Lambda}) \approx (5-3\,\gamma)/2$ if $\Gamma=0$, and $\coeffTSrho\rightarrow [(\gamma-2) + (1+\gamma)\,\log{\Lambda}]/(2\,\log{\Lambda}) \approx (1+\gamma)/2$ if $\Gamma=1$. 
Since $\tildeCoeffTSv\rightarrow1$, the ``\slowmodename'' mode is unstable if $\coeffTSrho > 1$ (for $\Gamma=0$ this requires $\gamma < (-4 + 3\,\log{\Lambda})/(3\,(-1 + \log{\Lambda})) \approx 0.98$; for $\Gamma=1$ this requires $\gamma > (2+\log{\Lambda})/(1+\log{\Lambda}) \approx 1.05$). As noted above for the Epstein case (\S~\ref{sec:epstein.subsonic}), because $\coeffTSv\rightarrow0$ at small $\driftvelmag$, the scaling of the ``subsonic'' low-$k$ mode is essentially reversed from the ``\slowmodename'' high-$k$ mode: when the ``\slowmodename'' mode is stable at high-$k$ ($\coeffTSrho < 1$) the ``subsonic'' mode is unstable at low-$k$, and when the ``\slowmodename'' mode is unstable ($\coeffTSrho > 1$) the ``subsonic'' mode is stable. In either case, whichever of the two is unstable has growth rate $\Im(\omdimless)\sim \driftvelmag^{2}\,\mu\,|\coeffTSrho|/2$. 

For $\driftvelmag \gg 1$, the drag force {\em decreases} rapidly for $|{\bf v}-{\bf u}| \gg c_{s}$ (i.e.\ $\coeffTSv \lesssim -1$ when $\driftvelmag \gg 1$). In this regime, one never  expects  Coulomb drag to dominate over Epstein drag (which becomes more tightly-coupled at high $\driftvelmag$), and in fact Coulomb drag alone does not allow self-consistent solutions for the equilibrium $\driftvel$ in Eq.~\eqref{eqn:mean.v.offset} without an additional Epstein or Stokes term when $\driftvelmag \gg 1$, but we consider the case briefly for completeness. 
We see that $\coeffTSrho \approx 1$ for $\Gamma=0$, and $\coeffTSrho\approx 2\,\gamma-1$ for $\Gamma=1$. 
More importantly, $\coeffTSv \rightarrow -3$. This produces the fast-growing ``decoupling instability'' (\S~\ref{sec:decoupling}), which affects {\em all} wavenumbers and has a growth rate ${\Im}(\omdimless) \approx -\tildeCoeffTSv\,(1+\mu) \approx 2\,(1+\mu)$. These modes arise from decoupling of the gas and dust: if the dust starts to move faster relative to the gas, $t_{s}$ increases (the coupling becomes weaker), so the terminal/relative velocity increases further, and so on. If we ignore 
the decoupling mode, we see that each of the other modes we have discussed are still present: the high-$k$ resonant mode (Eq.~\eqref{eqn:omega.resonant}) has $\Theta=(4-3\,\gamma)/(2\,\log{\Lambda})$ for $\Gamma=0$ and $\Theta\approx2\,(1-\gamma)$ for $\Gamma=1$.

\begin{figure*}
\plotsidesize{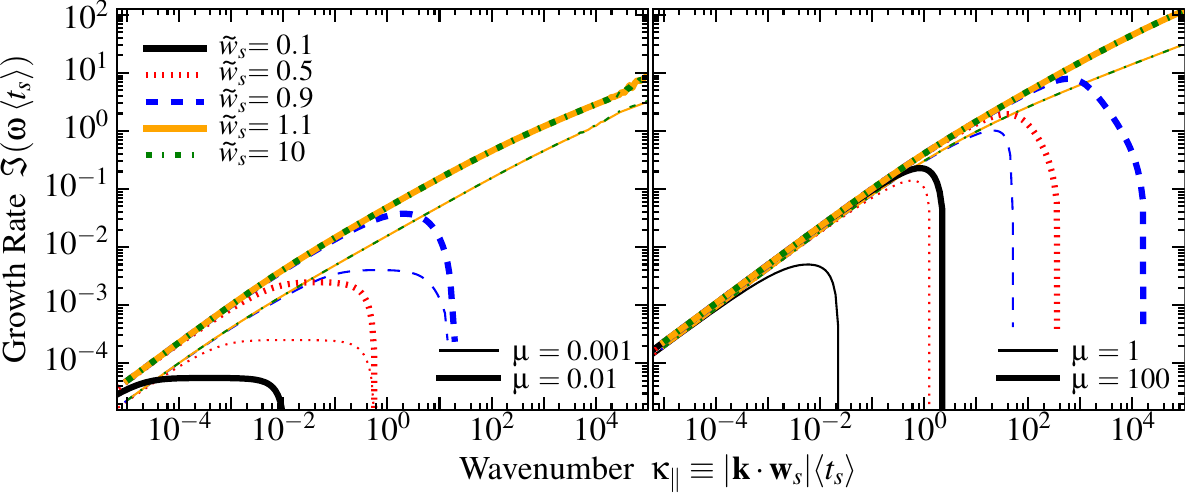}{0.95}
    \vspace{-0.25cm}
    \caption{Growth rates of the most-rapidly-growing unstable mode as a function of wavenumber and drift velocity, as Fig.~\ref{fig:growth.rate.demo}, for different dust-to-gas ratios $\mu=0.001,\ 0.01,\ 1,\ 100$ (the $\mu=0.1$ case is in Fig.~\ref{fig:growth.rate.demo}). For simplicity we take a constant drag coefficient ($\coeffTSrho=\coeffTSv=0$, as Fig.~\ref{fig:growth.rate.demo}), and marginalize over angle at each $\kappa_{\|}$. As shown in \S~\ref{sec:general.modes}, the dependence on $\mu$ at a given $\kappa_{\|}$ is quite weak. At low $\mu \ll 1$, the low and high-$k$ growth rates scale $\propto \mu^{1/3}$, with the slightly stronger $\propto \mu^{1/2}$ dependence around $\kappa_{\|}\sim 1$. At large $\mu \gtrsim 1$, the low and intermediate-$k$ growth rates become independent of $\mu$ (because they scale with $\hat{\mu}\equiv\mu/(1+\mu) \rightarrow 1$ for large $\mu$); the high-$k$ growth rate continues to increase weakly with $\mu^{1/3}$. In the sub-sonic ($\driftvelmag<1$) case, however, the maximum wavenumber where the growth rate either saturates or the mode becomes stable increases with $\mu$ so that the maximum growth rate (marginalizing over $k$) increases roughly $\propto \mu^{2/3}$. For the super-sonic ($\driftvelmag > 1$) case all wavelengths are unstable independent of $\mu$, so there is no such dependence.
        \vspace{-0.25cm}
        \label{fig:growth.rate.mu}}
\end{figure*}

\vspace{-0.5cm}
\section{Non-Linear Behavior \&\ Turbulence}
\label{sec:nonlinear}

The non-linear behavior of the coupled dust-gas system is complex and chaotic, and  will be studied in future work with numerical simulations (Moseley et al., in prep.).  Here, we  briefly speculate on some possible saturation mechanisms of the acoustic RDI and subsonic instabilities. 

For $\driftvelmag \ge 1$, the resonant mode at the shortest wavelengths will grow fastest, with the dust density aligning locally into crests at the phase peaks with orientation $\cos{\theta} = \pm 1/\driftvelmag$. These will launch small-scale perturbations in the tranverse directions in the gas. Because it is short-wavelength, we do not expect the modes to be coherent on large scales, so this will drive small-scale turbulence in the gas in the transverse directions, while in the $\driftvelhat$ direction, the modes will be stretched by the drift. For $\driftvelmag < 1$, the modes grow more slowly, and, depending on $\coeffTSrho$ and $\coeffTSv$ (see \S~\ref{sec:subsonic.long.wavelenghts}), either saturate to a constant growth rate or turn over above a critical $\kdimless \gtrsim \driftvelmag$. Thus, most of the power on large scales will be in modes of order this wavelength ($k^{-1}\sim c_{s}^{2}/(\mu\,|{\bf a}|)$). If $\mu\ll1$, dust will go strongly non-linear before the gas does, but eventually the non-linear terms will likely lead to turbulence in the gas and dust, at least for $\mu$ not too small. Gas turbulence can then enhance dust-to-gas fluctuations (see e.g.\ numerical experiments with dust in super-sonic turbulence in \citealt{hopkins.2016:dust.gas.molecular.cloud.dynamics.sims,lee:dynamics.charged.dust.gmcs}). Eventually sharp dust-filaments will form, and as the modes grow beyond this point, dust trajectories will cross and the fluid approximation for the dust will break down. Rayleigh-Taylor type secondary instabilities will likely appear, as regions with higher gas density are accelerated more rapidly, while those without dust are not dragged efficiently.
It also seems possible that for $\mu\ll1$ and/or $\driftvelmag$ not very large, the modes saturate in a laminar way (e.g., by changing shape, or if the dust fluid approximation breaks down).

We can crudely guess the saturation amplitude of the non-linear turbulence by comparing the energy input (per unit mass) from the imposed acceleration (without including the bulk acceleration of the system), \begin{equation}
\frac{dE_{\mathrm{accel}}}{dm\,dt} \sim \frac{d(m_{\rm dust}\,v_{\rm dust-gas}^{2})/dt}{m_{\rm dust} + m_{\rm gas}} \sim \frac{m_{\rm dust}\,\langle {\bf v}_{\rm dust-gas} \rangle\cdot {\bf a}}{m_{\rm dust} + m_{\rm gas}} \sim \frac{\mu\,|\driftvel|^{2}}{(1+\mu)\,\langle t_{s}\rangle},\label{eqn:forcing.energy.input}\end{equation}
to the specific energy decay rate of turbulence 
\begin{equation}
\frac{dE_{\mathrm{turb}}}{dm\,dt}  \sim -\frac{v_{\rm eddy}^{2}}{t_{\rm eddy}} \sim -\frac{\delta v_{\rm sat}^{3}}{\lambda},\label{eqn:turb.energy.diss}\end{equation}
where $\lambda$ is the driving scale of the turbulence. Equating Eq.~\eqref{eqn:forcing.energy.input} and Eq.~\eqref{eqn:turb.energy.diss} gives $\delta v_{\rm sat} \sim \,(\hat{\mu}\,|\driftvel|^{2}\,\lambda/\langle t_{s}\rangle)^{1/3}$. For each range of the RDI, we can then equate the turbulent dissipation rate $t_{\mathrm{diss}}^{-1}\sim t_{\mathrm{eddy}}^{-1}\sim  v_{\mathrm{eddy}}/\lambda\sim (\mu |\driftvel|^{2}/\langle t_{s}\rangle)^{1/3}\lambda^{-2/3}$ to the growth rate $\Im(\omega)$, which should (in principle) allow for the estimation of a characteristic scale and saturation amplitude in the resulting turbulence.  However, one finds that: (i) in the low-$k$ regime, with $\Im(\omega)\sim (\hat{\mu}/\langle t_{s}\rangle )^{1/3}(|\driftvel| k)^{2/3}$, the two are identical and there is no obvious characteristic $\lambda$; (ii) in the mid-$k$ regime, with  $\Im(\omega)\sim (\hat{\mu}\, c_{s} k/\langle t_{s}\rangle)^{1/2}$, the characteristic scale is $\lambda/(c_{s}\langle t_{s}\rangle )\sim \driftvelmag^{4}\hat{\mu}^{-1}$, which is outside of the range of validity of the mid-$k$ regime; and (iii) in the high-$k$ regime, with $\Im(\omega)\sim (\hat{\mu}\,c_{s} k/\langle t_{s}\rangle^{2})^{1/3}$, the characteristic scale is $\lambda/(c_{s}\langle t_{s}\rangle )\sim \driftvelmag^{2}$, which is outside of the range of validity of the high-$k$ regime (if $\hat{\mu}<1$). Thus, we see that there is no obvious way for the system to choose a scale for resonant modes in \emph{any} wavelength regime.
What we instead expect is that turbulence will begin on small scales and grow to larger and larger $\lambda$, up to the scale of the system (if the given sufficiently long time periods).   One might also expect that this the characteristic scale would increase in time, in some way proportional to the growth rate at a given $\lambda$. This suggests that  $\lambda\propto t^{3}$ ($\delta v \propto t$) at early times (with the instability growing in the high-$k$ regime), $\lambda\propto t^{2}$ ($\delta v \propto t^{2/3}$) at intermediate times (in the mid-$k$ regime), then slowing to $\lambda \propto t^{3/2}$ ($\delta v \propto t^{1/2}$) at longer times (in the long-wavelength regime).\footnote{Of course, actually resolving this shift in simulations would generally require an unfeasibly large dynamic range.} 
This qualitative behavior -- viz., turbulence that moves to larger and larger scales as a function of time -- is observed in simulations of  cosmic-ray-driven instabilities, which have some similar characteristics to the dust-gas instabilities studied here (see, e.g., \citealt{2009ApJ...694..626R,2017MNRAS.469.1849M}).

\vspace{-0.5cm}
\section{Scales where our analysis Breaks Down}
\label{sec:breakdown}

We now briefly review the scales where our analysis breaks down. 

\begin{enumerate}

\item{\bf Non-Linearity \&\ Orbit-Crossing:} If there is sufficiently sharp structure in the velocity or density fields, the dust trajectories become self-intersecting and the fluid approximation is invalid (for dust). In this limit numerical simulations must be used to integrate particle trajectories directly. This should not occur in the linear regime (see App.~A of \citealt{Jacquet:2011cy} for more discussion).

\item{\bf Smallest Spatial Scales:} At sufficiently short wavelengths (high $k$) approaching the gas mean-free-path, dissipative effects will be important.\footnote{More precisely, the fluid viscosity is important when $\omega\, u \sim \nu_{\mathrm{vis}} k^{2}u$, where $u$ is the perturbed gas velocity, and $ \nu_{\mathrm{vis}}\sim c_{s}\lambda_{\rm mfp}^{\rm gas}$ is the kinematic viscosity. For $\omega\sim c_{s} k$, as is the case for the acoustic RDI here, we find that viscosity is important when $k\sim1/ \lambda_{\rm mfp}^{\rm gas}$. } For ionized gas, this scale is $\lambda_{\rm mfp}^{\rm gas} \sim 10^{12}\,{\rm cm}\,(T/10^{4}\,{\rm K})^{2}\,(n_{\rm gas}/{\rm cm^{-3}})^{-1}$. If we assume Epstein drag with modest $\driftvelmag\sim 1$, this gives a dimensionless $\kappa_{\rm max} \sim (2\pi\,c_{s}\,\langle t_{s} \rangle/\lambda_{\rm mfp}) \sim 10^{9}\,(R_{d}/\mu\,{\rm m})\,(T/10^{4}\,{\rm K})^{-2} \gg 1$. 

In the dust, the fluid approximation breaks down on scales comparable to the dust-particle separation $\lambda_{\mathrm{sep}}^{\mathrm{dust}}\sim 10^{5}\,{\rm cm}\,(R_{d}/{\rm \mu\,m})\,(n_{\rm gas}/1\,{\rm cm^{-3}})^{-1/3}\,(\mu/0.01)^{-1/3}$, which is much smaller than  $\lambda_{\rm mfp}^{\rm gas}$ under most astrophysical conditions. 
%Because we assumed the dust to be pressure-free (and collisionless), there is no notion of a dust mean free path.
Because each of these minimum scales (for the gas and the dust) are small, very small wavelengths  (e.g., up 
to $\kappa_{\|} \sim k_{\rm max} c_{s} \langle t_{s}\rangle \sim 10^{9}$ in Figs.~\ref{fig:growth.rate.demo}, \ref{fig:growth.rate.draglaw}, and \ref{fig:growth.rate.mu}) are astrophysically relevant. 

\item{\bf Largest Spatial Scales:} At low $k$, we eventually hit new scale lengths (e.g.\ the gas pressure-scale-length). The physical scale where $\kappa_{\|}\sim 1$, i.e.,\ where $k\sim c_{s}\,\langle t_{s} \rangle$, can be large. For example, with Epstein drag at $\driftvelmag\sim 1$ this is $k^{-1} \sim 10^{20}\,{\rm cm}\,(R_{d}/\mu\,{\rm m})\,(n_{\rm gas}/{\rm cm^{-3}})^{-1}$. For  relatively low-density starburst regions or GMCs affected by massive stars, this is only $\sim 100$ times smaller than the system scale, so the long-wavelength instability ($k c_{s}\,\langle t_{s} \rangle \ll \mu$) will likely require a global analysis. However, in e.g.\ cool stars the densities are much higher and  the scales correspondingly smaller; e.g., for $\rho \sim \rho_{-12}\,10^{-12}\,{\rm g\,cm^{-3}}$ we obtain $k_{\rm min} c_{s}\,\langle t_{s} \rangle \sim 10^{-5}\,(R_{\rm min}/100\,R_{\rm sun})^{-1}\,(R_{d}/\mu\,{\rm m})\,\rho_{-12}^{-1}$ (see \S~\ref{sec:applications} for more details).

\item{\bf Maximum Timescales}: Dust with speed $|\driftvel|$ will drift through a system of size $L_{0}$ on a timescale $t_{\rm drift} \sim L_{0} / |\driftvel|$. An instability must grow faster than this to be astrophysically relevant. In App.~\ref{sec:hydrostatic.generalized} we show that this is equivalent to the condition for background dust stratification terms to be sub-dominant. In units of the stopping time, the relevant timescale is $L_{0}/(|\driftvel|\,\langle t_{s} \rangle) = (\driftvelmag/c_{s}) \, L_{0}/(c_{s}\,\langle t_{s} \rangle)$ -- i.e.\ the timescale criterion is closely related to the requirement that we consider modes smaller than the largest spatial scales. Another maximum timescale is set by the time for the equilibrium solution (dust+gas) to be accelerated out of the system of size $\sim L_{0}$, i.e.\ $t_{\rm acc} \sim (2\,L_{0}/|\hat{\mu}\,{\bf a}|)^{1/2}$ (or similarly, for e.g.\ a free-accelerating wind to expand and change density). Noting $|\driftvel|\sim |{\bf a}|\,t_{s}/(1+\mu)$, we have $t_{\rm acc}/t_{s} \sim \hat{\mu}^{-1/2}\,(t_{\rm drift}/t_{s})^{1/2}$, so (since $\hat{\mu}\ll 1$) this is generally a less-stringent criterion.

\end{enumerate}

\vspace{-0.5cm}
\section{Relation to Previous Work}
\label{sec:previous.work}

%\vspace{-0.5cm}
\subsection{Winds from Cool Stars}
\label{sec:previous.work:coolstar.winds}

In the context of dust-driven winds from red giants and other cool stars, there has been extensive work on other dust-related instabilities (involving thermal instability, dust formation, Rayleigh-Taylor instabilities, magnetic cycles, etc; see \citealt{macgregor:grains.cool.star.eventually.decouple,hartquist:bfield.dust.coupling.cool.star.winds,
sandin:agb.wind.sims,soker:agb.star.magnetic.cycle.instabilities,soker:2002.arcs.around.agb.stars.from.instability,simis:2001.shells.around.dust.agb.winds,woitke:2d.agb.wind.simulations,woitke:2d.rad.pressure.dust.agb.wind.models}), but these are physically distinct from the instabilities studied here. Of course, simulations with the appropriate physics -- namely, (1) explicit integration of a drag law with gas back-reaction (and compressible gas), (2) trans-sonic $\driftvelmag$, (3) multi-dimensional (2D/3D) domains, and (4) sufficient resolution (for the high-$k$ resonant modes) -- should see the instabilities studied here. Most studies to date to not meet these conditions. Moreover they often include other complicated physics (e.g.\ opacity and self-shielding, dust formation) which are certainly important, but make it difficult to identify the specific instability channel we describe here. 

However, some authors have previously  identified aspects of the instabilities described in this paper. \citet{morris:1993.cool.wind.dust.drag.instability.slow.saturated.mode} performed a much simpler linear stability analysis on a two-fluid mixture subject to drag (see also \citealt{mastrodemos:dust.gas.coupling.cool.winds.spherical.symmetry}), and noted two unstable solutions whose growth rates saturated at high-$k$: these are the ``\slowmodename'' and ``\intermediatemodename'' modes identified here. However, they assumed: (1) zero gas pressure (effectively $\driftvelmag\rightarrow \infty$), preventing identification of stability criteria; (2) a constant coupling coefficient; and (3) spherical symmetry (of the perturbations) which eliminates the resonant modes. \citet{deguchi:1997.dust.envelope.pne.spherical.drag.instability.quasi.resonant} followed this up allowing for non-zero gas pressure, but retaining spherical symmetry and imposing the assumption that the dust always exactly follows the local equilibrium drift velocity. This suppresses all instabilities except the resonant mode at $\driftvelmag=c_{s}$ exactly. To our knowledge, the scaling of these instabilities and the existence of the resonant instability for all $k$ and all $\driftvelmag > 1$ has not been discussed previously in the literature.

\vspace{-0.5cm}
\subsection{Starburst and AGN Winds}
\label{sec:previous.work:agn.winds}

In models of starbursts and AGN, there is a long literature discussing radiation pressure on grains as an acceleration mechanism for outflows or driver of turbulence \citep[see e.g.][]{heckman:1990.sb.superwinds,scoville:2001.dust.pressure.in.sb.regions,thompson:rad.pressure,krumholz:2009.rad.dom.region.dynamics,hopkins:twostage.feedback,hopkins:rad.pressure.sf.fb,murray:molcloud.disrupt.by.rad.pressure,kuiper:2012.rad.pressure.outflow.vs.rt.method,wise:2012.rad.pressure.effects}. But almost all calculations to date treat dust and gas as perfectly-coupled (so the instabilities here cannot appear). The instabilities in this paper are not related to the ``radiative Rayleigh-Taylor'' instability of a radiation pressure-supported gas+dust fluid \citep{krumholz:2012.rad.pressure.rt.instab,davis:2014.rad.pressure.outflows}, nor to non-linear hydrodynamic instabilities generated by e.g.\ pressure gradients or entropy inversions ultimately sourced by dust ``lifting'' material \citep[e.g.][]{berruyer:dust.wind.unstable.pressure.gradient}, nor the dust sedimentation effects in ambipolar diffusion in molecular clouds discussed in \citet{cochran:dust.bounded.hII.regions.outflow,sandford:radiatively.driven.dust.bounded.gmc.globules}. Each of these other classes of instability do not involve local dust-to-gas ratio fluctuations.

There recently has been more work exploring dust-gas de-coupling in molecular cloud turbulence and shocks (integrating the explicit dust dynamics; see \citealt{hopkins.2016:dust.gas.molecular.cloud.dynamics.sims,lee:dynamics.charged.dust.gmcs,monceau:shock.cloud.dust.disperson}) which has shown this can have important effects on cooling, dust growth, and star formation. However, these studies did not identify instabilities, or include the necessary physics  to capture the instabilities here, because they treated dust as a ``passive'' species (did not include its back-reaction on the momentum of gas).

\vspace{-0.5cm}
\subsection{Proto-Planetary Disks}
\label{sec:previous.work:disks}

There has been extensive study of  dust-gas instabilities and dynamics in proto-planetary disks \citep{youdin.goodman:2005.streaming.instability.derivation,johansen:2007.streaming.instab.sims,carballido:2008.grain.streaming.instab.sims,bai:2010.grain.streaming.sims.test,bai:2010.grain.streaming.vs.diskparams,pan:2011.grain.clustering.midstokes.sims,dittrich:2013.grain.clustering.mri.disk.sims,jalali:2013.streaming.instability.largescales,hopkins:2014.pebble.pile.formation,2017arXiv170802945L}. As mentioned  in \paperone,  the well-studied ``streaming instability'' \citep{youdin.goodman:2005.streaming.instability.derivation} is in fact an example of an RDI (although this has not been noted before in this context), a connection that is explored in detail in \citet{squire:rdi.ppd}. 
However, in the streaming instability, the wave with which the dust drift ``resonates'' is not a sound wave, but epicyclic oscillations of the gas. Similarly, as shown in \paperone\ (see also App.~\ref{sec:hydrostatic.generalized}), \BV\ oscillations create an RDI, which may be of importance in proto-planetary disks \citep[this is likely the cause for the instability seen in][]{lambrechts:bv.rdi}. 
The acoustic RDI has not been explored in this literature. In fact, it is common in these studies to simplify by assuming incompressible gas (enforcing $\delta\rho=0$), in which case all of the acoustic instabilities studied here vanish. Finally, it is worth noting that   dust-induced instabilities that occur  due to 
the mass loading of the gas caused by  dust \citep[see, e.g.,][]{Garaud:2004,2012ApJ...744..101T} or from changes to its thermodynamic properties (e.g., \citealt{2015MNRAS.453L..78L}, and some of the instabilities discussed in \citealt{2017arXiv170802945L}), are not in the RDI class, because they do not rely on the finite drift velocity between the dust and gas phases.

\vspace{-0.5cm}
\subsection{Plasma Instabilities}
\label{sec:previous.work:plasma}

As noted in \paperone, the most general RDI is closely related to instabilities of two-fluid plasmas (see, e.g.,\ \citealt{tytarenko:two.fluid.drift.intabilities} for an in-depth analysis of a closely related coupled neutral gas-MHD instability). These  include the \citet{wardle:instability.mhd.shocks.with.slip} instability and cosmic ray streaming instabilities \citep{kulsrud.1969:streaming.instability,Bell.cosmic.rays}. However, these are quite distinct physical systems and the instabilities have different linear behaviors.

%To our knowledge, the closest analogues which have been studied in the literature are the known instabilities of two-fluid plasmas, with e.g.\ a neutral and ion fluid and strong magnetic field\citep[see][for an overview]{tytarenko:two.fluid.drift.intabilities} including e.g.\ the \citet{wardle:instability.mhd.shocks.with.slip} instability (see also \citealt{stone:wardle.instability.simulations,maclow:wardle.instability.simulations}) and cosmic ray streaming instabilities \citep{kulsrud.1969:streaming.instability}; potentially some instabilities studied in terrestrial micro-fluid suspensions as well \citep{childress:instability.pattern.formation.suspensions}. Under the right conditions, these represent special sub-cases of the general \paperone\ class of resonant instabilities, but they are obviously distinct physical systems from those studied here.

\vspace{-0.5cm}
\section{Astrophysical Applications}
\label{sec:applications}

There are a number of astrophysical contexts where this specific example of the \paperone\ instability may be important, which we review here. In the discussions below, 
we estimate the radiative acceleration of the dust from ${\bf a}\sim{\bf F}_{\lambda}\,Q_{\lambda}\,\bar{\rho}_{d}/ (c\,R_{d})$, where $|{\bf F}|_{\lambda}\sim L/ r^{2}$ is the incident flux of radiation from a source of luminosity $L$ at distance $r$, $c$ is the speed of light, and $Q_{\lambda}$ is the absorption efficiency ($Q_{\lambda}\sim 1$ for very large grains, $Q_{\lambda}\propto R_{d}$ for smaller grains; see \S~\ref{sec:dust.species})

\begin{enumerate}

\item{\bf AGN-Driven Outflows and the AGN ``Torus'':} Around a luminous AGN, gas and dust are strongly differentially accelerated by radiation pressure. There is some dust sublimation radius close to the AGN, interior to which dust is destroyed.  The instabilities must occur outside this region in the dusty ``torus,'' or further out still, in the galactic narrow-line region.

We assume the AGN has luminosity $L \sim L_{46}\,10^{46}\,{\rm erg\,s^{-1}}$, and normalize the radius $r$ of the dusty torus to the dust sublimation radius, i.e.,  $r\sim \tilde{r}\,r_{\mathrm{sub}}\sim 0.3\,{\rm pc}\,\tilde{r}\,L_{46}^{1/2}$. For a  midplane column density  $ n_{\rm gas}\,r \sim N_{26}\,10^{26}\,{\rm cm^{-2}}$, and  gas temperature  $T\sim 1000\,$K, we find that we are in the highly super-sonic regime  with $\driftvelmag \sim 100\,L_{46}^{1/4}\,(\tilde{r}\,N_{26})^{-1/2}$ (dust is in the Epstein regime; see Eq.~\ref{eqn:ws.in.epstein}). For grains with size $R_{d} \sim R_{d,\mu}\,\mu{\rm m}$, the  stopping time is $\langle t_{s} \rangle\sim 0.01\,{\rm yr}\,R_{d,\mu}\,L_{46}^{1/4}\,\tilde{r}^{3/2}\,N_{26}^{-1/2}$ and the characteristic length scale is $c_{s}\,\langle t_{s} \rangle\sim 6\times10^{10}\,{\rm cm}\,R_{d,\mu}\,L_{46}^{1/4}\,\tilde{r}^{3/2}\,(T_{1000}/N_{26})^{1/2}$ (this is  $\sim 10^{-7}\,r$, and $\sim 1000$ times the viscous scale). Thus the large-scale dynamics are in the long-wavelength regime ($\kdimless \ll \hat{\mu}$), with growth timescales (see Eq.~\ref{eqn:longwave.mode})  $\Im(\omega)^{-1}\sim 30\,{\rm yr}\,R_{d,\mu}^{1/3}\,L_{46}^{-1/12}\,N_{26}^{1/6}\,\tilde{r}^{5/6}\,(Z/Z_{\sun})^{-1/3}\,(\lambda / 0.1\,{\rm pc})^{2/3}$ (where $\lambda$ is the mode wavelength and we assume the dust-to-gas mass ratio scales with $Z/Z_{\sun}$). This is faster than the dynamical time, and the turbulent eddy turnover time, on essentially every scale inside the torus. Much smaller-scale modes ($\lambda \ll {\rm au}$) fall into the mid-$k$ resonant regime, with the fastest growth timescales of $\Im(\omega)^{-1}\sim 10-100\,{\rm hours}$ for modes approaching the viscous scale ($\lambda\sim 10^{7-8}\,{\rm cm}$).

%$\driftvelmag \sim 80\,L_{46}^{1/4}\,(N_{26}\,T_{1000}\,x/Q)^{-1/2}$
%$\langle t_{s} \rangle \sim 3\times10^{5}\,{\rm s}\,a_{\mu}\,L_{46}^{1/4}\,x^{3/2}\,(Q\,N_{26})^{-1/2}$
%$c_{s}\,\langle t_{s} \rangle \sim 6\times10^{10}\,{\rm cm}\,a_{\mu}\,L_{46}^{1/4}\,x^{3/2}\,(T_{1000}/N_{26}\,Q)^{1/2}$
%$t_{\rm grow} \sim 1/\Im(\omega) \sim 80\,{\rm yr}\,L_{46}^{1/4}\,N_{26}^{1/6}\,(x/Q)^{3/2}\,a_{\mu}^{1/3}\,Z_{\sun}^{-1/3}\,(\lambda/r)^{2/3}$
%$(t_{\rm grow}\,\Omega) \ll 1$
%all low-k on observable scales. $c_{s}\,\langle t_{s} \rangle$ is something like $\sim 1000$ times the viscous scale, so mid-k as approach viscous and dust-collisional scales (maybe bit of high-k). As approach viscous scale, ($\sim 10^{7}-10^{8}\,{\rm cm}$) growth times in high-$k$ mode get to $\sim 10-100\,{\rm hr}\,a_{0.1}^{2/3}\,(L_{46}/T_{1000})^{1/2}\,(x/N_{26})\,Z_{\sun}^{-1/3}$.

Thus, essentially all luminous AGN ($L \gtrsim 10^{42}\,{\rm erg\,s^{-1}}$) should exhibit regions in the ``clumpy torus'' surrounding the AGN, as well as  radiation-pressure-driven AGN outflows, which are subject to the super-sonic instabilities described above. This may provide a natural explanation for clumpiness, velocity sub-structure, and turbulence in the torus \citep[see e.g.][]{krolik:clumpy.torii,mason:ngc1068.torus.obs,sanchez:circinus.torus.mass,nenkova:clumpy.torus.model.1,thompson:dust.em.from.unobscured.agn,mor:2009.torus.structure.from.fitting.obs,hoenig:clumpy.torus.modeling,hopkins:m31.disk,hopkins:torus,hopkins:qso.stellar.fb.together,deo:2011.z2.clumpy.torii}, as well as observed time-variability in AGN obscuration \citep{mckernan:1998.agn.occultation.by.clumpy.outflow,risaliti:nh.column.variability}. It of course is critical to understand whether this directly alters the AGN-driven winds in the torus region, a subject that will be addressed in future numerical simulations \citep[see e.g.][]{ciottiostriker:recycling,murray:momentum.winds,elitzur:torus.wind,miller:2008.clumpy.agn.disk.wind,roth:2012.rad.transfer.agn,wada:torus.mol.gas.hydro.sims}.

As noted above, the instability  requires only a dust-gas drift velocity, and this can instead be sourced by AGN line-driving of the \emph{gas} in the narrow/broad line regions. In this case, the scaling of $\driftvelmag$ depends on the opacity of the gas, but for plausible values in the narrow-line region, and similar luminosities and densities to those used above, we find $\driftvelmag \gtrsim 10^{2}-10^{3}$. 

\item{\bf Starburst Regions, Radiation-Pressure Driven Winds, and Dust in the ISM around Massive Stars:} Similarly, consider dusty gas in molecular clouds and HII regions surrounding regions with massive stars. It has been widely postulated that radiation pressure on dust (either single-scattering from optical/UV light or multiple-scattering of IR photons) can drive local outflows from these regions, unbinding dense clumps and GMCs, and stirring GMC or ISM-scale turbulence. 

Assuming geometric absorption of radiation by the dust ($Q_{\lambda}\sim 1$), a random patch of gas in a GMC (with temperature $T\sim T_{100}\,100\,$K, density $n\sim  n_{10}10\,{\rm cm^{-3}}$) at a distance $r\sim r_{\rm pc}\,{\rm pc}$ from a source with luminosity $L\sim L_{1000}\,1000\,L_{\sun}$ has $\driftvelmag \sim 10\,L_{1000}^{1/2}\,n_{10}^{-1/2}\,r_{\rm pc}^{-1}$. Similarly, consider a GMC of some arbitrary total mass $M_{\rm cl}$ and total size $r\sim r_{10}\,10\,{\rm pc}$, which has converted a fraction $\sim 0.1\,\epsilon_{0.1}$ of its mass into stars. If we assume a typical mass-to-light ratio for young stellar populations ($\sim 1100\,L_{\sun}/M_{\sun}$), we find  $\driftvelmag \sim 10\,r_{10}^{1/2}\,\epsilon_{0.1}^{1/2}$. For smaller (typical ISM) $R_{d}\sim 0.1\,R_{d,0.1}\,\mu{\rm m}$, 
the corresponding (Epstein) stopping time is $\langle t_{s} \rangle \sim 10^{4}\,{\rm yr}\,R_{d,0.1}\,\Sigma_{100}^{-1}\,(r_{10}/\epsilon_{0.1})^{1/2}$ (where $\Sigma_{100} = \Sigma/100\,M_{\sun}\,{\rm pc}^{-2}$ is the cloud surface density), with scale $c_{s}\langle t_{s} \rangle \sim 0.006\,{\rm pc}\,T_{100}^{1/2}\,(\langle t_{s} \rangle/10^{4}\,{\rm yr})$. So depending on grain size and gas temperature/density, directly observable ($\gtrsim 0.1\,{\rm pc}$) scales fall in the resonant mid-$k$ regime (larger dust) or long-wavelength regime (smaller dust), with growth timescales $t_{\rm grow}/t_{\rm dyn} \sim 0.03\,R_{d,0.1}^{1/2}\,(\lambda/0.1\,{\rm pc})^{1/2}\,(Z/Z_{\sun})^{-1/2}\,(r_{10}\,T_{100}\,\epsilon_{0.1})^{-1/4}$ (where $t_{\rm dyn}=1/\sqrt{G\,\rho}\sim 10\,{\rm Myr}\,(r_{10}/\Sigma_{100})^{1/2}$). 

Therefore, we again expect these instabilities to be important. They may fundamentally alter the ability of radiation pressure from massive stars to drive outflows and source local turbulence \citep[a subject of considerable interest and controversy; see][]{murray:momentum.winds,thompson:rad.pressure,krumholz:2007.rhd.protostar.modes,schartmann:2009.stellar.fb.effects.on.torus,hopkins:rad.pressure.sf.fb,hopkins:dense.gas.tracers,hopkins:2013.fire,guszejnov.2015:feedback.imf.invariance,grudic:sfe.cluster.form.surface.density}. They will also directly source dust-to-gas fluctuations, which can in turn drive abundance anomalies in next-generation stars \citep{hopkins:totally.metal.stars,hopkins.conroy.2015:metal.poor.star.abundances.dust}, as well as altering the dust growth, chemistry, and cooling physics of the clouds \citep{goldsmith:molecular.dust.cooling.gmcs,dopke.2013:fragmentation.all.dust.levels.but.enhanced.with.crit.dust,ji:2014.si.dust.cooling.threshold.for.early.stars,chiaki:2014.critical.dust.abundance.for.cooling}.

\item{\bf Cool Star (AGB and Red Giant) Winds and PNe:} In the photospheres and envelopes of cool stars, dust forms and is accelerated by continuum radiation pressure. This contributes to the launching and acceleration of  winds, and potentially defines key wind properties, such as their ``clumpiness'' and variability in time and space. There has been extensive study of accelerating dust-gas mixtures in this context (see references in \S~\ref{sec:previous.work:coolstar.winds}). 

Consider an expanding photosphere/wind ($\rho=\dot{M}/(4\pi\,r^{2}\,v_{\rm wind})$) with $v_{\rm wind}\sim v_{10}\,10\,{\rm km\,s^{-1}}$, $\dot{M}\sim \dot{M}_{-3}\,10^{-3}\,\msun\,{\rm yr^{-1}}$, and gas temperature $T\sim T_{1000}\,1000\,$K (in the outflow) around a giant with  luminosity $L\sim L_{5}\,10^{5}\,L_{\sun}$. Assuming geometric absorption, we obtain $\driftvelmag \sim 2\,(L_{5}\,v_{10}/\dot{M}_{-3}\,T_{1000})^{1/2}$. We therefore expect $\driftvelmag\sim 1$ (but with a broad range, $\driftvelmag\sim 0.1\rightarrow 10$, or larger) for plausible parameters of different cool stars, and different locations of the grains within the photosphere and wind. The corresponding (Epstein) stopping time is $\langle t_{s} \rangle \sim 1\,{\rm sec}\,R_{d,0.1}\,r_{100}^{2}\,(v_{10}/L_{5}\,\dot{M}_{-3})^{1/2}$ (where $r_{100} \equiv r/100\,R_{\sun}$) and the relevant scales are $c_{s}\,\langle t_{s} \rangle \sim 3\times10^{5}\,{\rm cm}\,T_{1000}^{1/2}\,(\langle t_{s} \rangle/{\rm sec})$. So large-scale modes ($\lambda\gtrsim 10^{8}\,{\rm cm}$) are in the long-wavelength (low-$k$) limit. However,  the mean free path is very small $\lambda_{\rm MFP} \sim 10\,{\rm cm}\,r_{100}^{2}\,v_{10}/\dot{M}_{-3}$, implying that the full dynamic range of the mid-$k$ and high-$k$ resonant modes is also present when $\driftvelmag \ge 1$. 
The growth timescale for the largest (low-$k$) modes scales as $t_{\rm grow}/t_{\rm wind} \sim 0.02\,v_{10}^{4/3}\,(R_{d,0.1}\,r_{100}\,Z/\dot{M}_{-3}\,Z_{\sun})^{1/3}\,T_{1000}^{-1/2}\,(\lambda / r)^{2/3}$, where $t_{\rm wind}=r/v_{\rm wind} \sim 0.2\,{\rm yr}\,r_{100}/v_{10}$, suggesting all modes can grow in a wind dynamical time. 
Approaching the viscous scale (in the high-$k$ regime), $t_{\rm grow}$ reaches $\sim 0.1\,{\rm sec}\,R_{d,0.1}^{2/3}\,T_{1000}^{-1/2}\,(Z/Z_{\sun})^{-1/3}\,(\lambda_{\rm mfp}/10\,{\rm cm})$.

%$\driftvelmag \sim 1.4\,(L_{5}\,Q\,v_{10}/\dot{M}_{-3}\,T_{1000}) \sim 1.6\,(L_{5}\,Q\,v_{10}/\dot{M}_{-3}\,T_{1000})^{1/2}$
%$\langle t_{s} \rangle \sim 6\,{\rm s}\,a_{\mu}\,(R_{100}^{2}\,v_{10}/\dot{M}_{-3}\,T_{1000}^{1/2}) \sim 7\,{\rm s}\,a_{\mu}\,R_{100}^{2}\,(v_{10}/L_{5}\,\dot{M}_{-3}\,Q)^{1/2}$
%$c_{s}\,\langle t_{s} \rangle \sim 2\times10^{6}\,a_{\mu}\,R_{100}^{2}\,v_{10}\,\psi$ where $\psi=1,\,(T_{1000}/L_{5}\,\dot{M}_{-3}\,Q\,v_{10})^{1/2}$
%so large scales are low-$k$ modes, but $\lambda_{\rm MFP} \sim 10\,{\rm cm}\,R_{100}^{2}\,v_{10}/\dot{M}_{-3}$ so also have all mid-k and high-k represented as well. 
%large-scale (low-$k$) mode growth timescales:
%$t_{\rm grow} \sim 3\times10^{5}\,{\rm s}\,R_{100}^{4/3}\,(a_{\mu}\,v_{10}/\dot{M}_{-3}\,Z_{\sun})^{1/3}\,T_{1000}^{-1/2}\,(\lambda/R)^{2/3}$
%or $t_{\rm grow}\,v_{\rm wind}/R \sim 0.02\,v_{10}^{4/3}\,(a_{0.1}\,R_{100}/\dot{M}_{-3}\,Z_{\sun})^{1/3}\,T_{1000}^{-1/2}\,(\lambda/R)^{2/3}$. so grows much faster than wind expansion time.

%$\sim 0.1\,{\rm s}\,R_{0.1}^{2/3}\,r_{100}^{2}\,v_{10}^{3/2}\,L_{5}^{1/2}\,T_{1000}^{-1}\,Q^{1/2}\,\dot{M}_{-3}^{-3/2}\,Z_{\sun}^{-1/3}$. 

This places the instability  in perhaps the most interesting range, where certain regimes of the outflows (with $\driftvelmag \lesssim 1$, but not vanishingly small) would be subject to the long-wavelength instability, and other regimes  (with $\driftvelmag \gtrsim 1$) would be subject to the short-wavelength acoustic RDI.  The long-wavelength instability, which  grows fastest in the direction parallel to $\driftvel$,  could perhaps explain large-scale features such as dust ``shells'' or ``arcs'' \citep[similar to ideas proposed by][]{morris:1993.cool.wind.dust.drag.instability.slow.saturated.mode,1994A&A...288..255W,deguchi:1997.dust.envelope.pne.spherical.drag.instability.quasi.resonant}. In contrast,  regimes with $\driftvelmag\gtrsim 1$, where the  fastest-growing modes are short-wavelength and oblique,  would likely develop non-linearly into turbulence, seeding clumpy sub-structure in the winds and in emission \citep[a subject of considerable interest; see e.g.][]{1998A&A...333L..51W,2003ApJ...582L..39F,young:2003.clumpy.wind.models,2007Natur.447.1094Z,2010ApJ...724L.133A,2012A&A...537A..35C}. The latter would almost certainly trigger secondary non-linear instabilities by driving large dust-gas clumping; for example via radiative Rayleigh-Taylor instabilities, dust opacity/self-shielding effects, and dust collisions/growth in the wind.

\item{\bf Proto-planetary Disks:} As discussed in \S~\ref{sec:previous.work}, instabilities of the coupled dust-gas system in proto-planetary disks are particularly interesting, given their implications for planet formation and observable disk properties. In proto-planetary disks we expect drift velocities to be highly subsonic. For a disk with parameters following \citet{chiang:2010.planetesimal.formation.review} at radius $ r\sim r_{10}\,10\,{\rm au}$ and surface density $\Sigma \sim \Sigma_{\rm MMSN}\,1000\,{\rm g\,cm^{-3}}\,(r/{\rm au})^{-1.5}$, pebbles with size $R_{d}\sim R_{d,{\rm cm}}\,{\rm cm}$ will have $\driftvelmag\sim 0.005\,r_{10}^{25/14}\,R_{d,{\rm cm}}\,\Sigma_{\rm MMSN}^{-1}$ \citep{nakagawa:1986.grain.drift.solution}. Since $\driftvelmag \ll 1$ we expect the growth rate of the instabilities here to have a maximum value $\Im(\omega)\sim \driftvelmag^{2}\,\mu\,t_{s}^{-1}$. For plausible disk parameters this rate is much slower than the radial drift rate $\sim v_{\rm drift}/r$ for the grains to drift through the disk.

Given this relatively low growth rate, we do not expect this particular sound-wave resonance (the acoustic RDI) to be dominant. However, we {\em do} expect other examples from the broad class of RDI resonances to be interesting. For example, as noted in \paperone\ and above, the well-studied disk ``streaming instability'' is an RDI associated with the disk epicyclic frequency. Other wave families such as \BV\ oscillations, slow magnetosonic, and Hall magnetosonic-cyclotron waves are also present with slow phase velocities, which can give rise to much larger growth rates (as compared to the acoustic RDI studied here) when $\driftvelmag\ll 1$. These are explored in \citet{squire:rdi.ppd}.

\end{enumerate}

\vspace{-0.5cm}
\section{Conclusions}
\label{sec:summary}

\subsection{Summary}

We study the acoustic family of the class of \citet{squire.hopkins:RDI} ``resonant drag instabilities'' (RDI) discovered in \paperone, as well as a spectrum of related ``non-resonant'' instabilities first identified here. Such instabilities can occur when a relative drift velocity arises between the dust and gas in  a coupled dust-gas mixture  (due, for example, to  different radiative forces on the dust and the gas, or pressure support of the gas). \paperone\ studied a general gas system and showed that if the gas (absent dust) supports some undamped waves, a streaming velocity that ``resonates with''  the wave phase velocity usually creates an instability (the RDI). In this work, we focus on the case where the gas is governed by neutral hydrodynamics and supports sound waves, studying the ``acoustic RDI'' (resonance with sound waves) and a collection of other non-resonant unstable modes (these are important in certain regimes, e.g., at long-wavelengths or high dust-to-gas ratios). Although  neutral hydrodynamics  is perhaps the simplest gas system possible, these instabilities have not (to our knowledge) been studied or identified in previous literature, despite their likely relevance for  a wide variety of astrophysical systems.

We identify a spectrum of exponentially-growing linear instabilities which {\em directly} source fluctuations in the dust-to-gas ratio.  Under certain conditions {\em all} wavelengths feature unstable modes, some of which have growth rates that increase without limit with increasing wavenumber. We show that the basic qualitative behaviors (dimensional scalings and nature of the fastest-growing modes) are not sensitive to the gas equation-of-state, the form of the drag law (constant drag coefficient, Epstein, Stokes, or Coulomb drag), the dust-to-gas ratio, or other details, although these do quantitatively alter the predictions. We derive stability conditions and simple closed analytic expressions for the growth rates of the instability (\S~\ref{sec:general.modes}).

There is one critical dimensionless parameter that determines the system's qualitative behavior, viz.,  ratio of the mean dust drift velocity ($|{\bf v}_{\rm dust} - {\bf u}_{\rm gas}|^{\rm drift}$) to the gas sound speed $c_{s}$:
\begin{align}
\driftvelmag &\equiv \frac{|\driftvel|}{c_{s}} =  \frac{|{\bf v}_{\rm dust} - {\bf u}_{\rm gas}|^{\rm drift}}{c_{s}}
= \frac{|\Delta {\bf a}_{\rm dust-gas}|\,\langle t_{s}({\bf a},\,\rho,\,...) \rangle}{c_{s}\,(1+\mu)}.
\end{align}
Here, the drift velocity $\driftvel$ is  the ``terminal'' velocity when the dust and gas experience accelerations which differ by some amount $\Delta {\bf a}_{\rm dust-gas}$,  $t_{s}$ is the drag coefficient or ``stopping time'' (determined by the drag law), and $\mu$ is the dust-to-gas mass ratio.

%Physically, drift can occur because the gas is pressure supported against gravity (or some other force) while the dust is not, or because one or the other experiences some additional acceleration (for example, continuum radiation pressure accelerating the dust, or resonant line radiation pressure accelerating the gas). 

When $\driftvelmag \ge 1$, i.e.\ when the dust is moving supersonically relative to the gas, the system is strongly unstable at {\em all} wavelengths. There are multiple unstable modes but the acoustic RDI from \paperone\ (\S~\ref{sec:resonance}) is the most rapidly growing. The 
%There are two unstable modes: first (1) the ``intermediate'' mode (\S~\ref{sec:intermediate}) which are longitudinal modes (propagating with the gas sound speed in the direction of the drift velocity) whose growth rate increases with wavenumber until saturating at a maximum growth rate at large $k\gtrsim 1/(c_{s}\,t_{s})$ with growth timescale $1/{\Im}(\omega) \sim t_{s} / (\mu\,\driftvelmag) \sim c_{s}/(\mu\,|\Delta {\bf a}|)$. Second, (2) the ``resonant'' mode from \paperone\ (\S~\ref{sec:resonance}), whose 
growth rate $\Im(\omega)$ increases {\em without limit} with increasing wavenumber $k$ as $\Im(\omega)\sim (\mu\,k\,c_{s} / t_{s})^{1/2}$ (in a mid range of $k$) or $\Im(\omega)\sim (\mu\,k\,c_{s} / t_{s}^{2})^{1/3}$ (at high $k$), independent of $\driftvelmag$. These  modes propagate at a critical angle $\cos{\theta} = \pm 1/\driftvelmag$ with respect to the drift direction; the wavespeed is the normal sound speed, and the drift velocity along the wavevector $\hat{\bf k}$ exactly matches this, allowing the dust to coherently push gas, and generate density perturbations. The denser gas then decelerates the dust further, causing a pileup, which runs away. For modes at angles that do not match the resonance condition ($\cos{\theta} \ne \pm 1/\driftvelmag$), the growth rates saturate at finite values (i.e., $\Im(\omega)$ does not increase indefinitely with $k$).

%with growth timescale $\sim t_{s}\,(\mu\,k\,c_{s}\,t_{s})^{-1/3}$. These are modes propagating at a critical angle with respect to the drift velocity $\cos{\theta} \equiv \hat{\bf k}\cdot \hat{\bf v}_{\rm drift} = \pm 1/\driftvelmag$; the wavespeed is the gas sound speed, but their speed in the drift velocity direction exactly matches the mean drift velocity. The resonance thus allows the dust to coherently push gas, and generate density perturbations; the denser regions in the gas then decelerate the gas, causing a pileup which further amplifies the drag force. Waves at angles outside of resonance are also unstable, but with growth rates that saturate at a finite value at high-$k$. 

When $\driftvelmag < 1$, i.e.\ when the dust is moving subsonically relative to the gas, the resonance above does not exist but there are still unstable, long-wavelength modes whose growth rate peaks or saturates above some wavenumber $k \propto \driftvelmag/(c_{s}t_{s})$,
%$\lambda \sim k^{-1} \sim c_{s}\,t_{s}/\driftvelmag  \sim c_{s}^{2}/ |\Delta {\bf a}|$, 
with maximum growth rate $\Im(\omega)\sim \driftvelmag^{2}\,\mu/t_{s}$.
%timescale $\sim t_{s}/(\driftvelmag^{2}\,\mu) \sim \mu^{-1}\,(\lambda / v_{\rm drift}) \sim c_{s}^{2} / (\mu\,|\Delta {\bf a}|^{2}\,t_{s})$. 

\vspace{-0.5cm}
\subsection{Implications, Caveats, \&\ Future Work}

In all cases, the instabilities drive dust-gas segregation and local fluctuations in the dust-to-gas ratio, compressible fluctuations in the gas density and velocity, and clumping within the dust (\S~\ref{sec:mode.structure}). Non-linearly, we expect them to saturate by breaking up into turbulent motions (in both dust and gas) which can be subsonic or supersonic, and in both cases can give rise to large separations between dense gas-dominated and dust-dominated regions. We provide simple estimates for the saturated turbulent amplitude (\S~\ref{sec:nonlinear}).

We discuss some astrophysical implications of these instabilities (\S~\ref{sec:applications}) and argue that the ``resonant'' instability is  likely to be important in the dusty gas around AGN (in the torus or narrow-line regions), starbursts, giant molecular clouds, and other massive-star forming regions, where $\driftvelmag \gg 1$ almost everywhere. In the winds and photospheres of cool stars, simple estimates suggest $\driftvelmag \sim 1$, with a broad range depending on the local conditions and location in the atmosphere. Thus, we again  expect these instabilities to be important. In each of these regimes, the instability may fundamentally alter the ability of the system to drive winds via radiation pressure (on the dust or the gas), and could source turbulence, velocity sub-structure, clumping, and potentially observable inhomogeneities in the winds. 

More detailed conclusions will require detailed numerical simulations to study the non-linear evolution of these systems. Our analytic results here make it clear what physics must be included to study such instabilities -- in particular, physical drag laws (with realistic density and velocity dependence) and backreaction from the dust to the gas -- and the range of  scales that must be resolved. Most previous studies of such systems either did not include the appropriate drag physics or lacked the resolution to treat these modes properly. This is especially challenging for the resonant mode: because the growth rate increases without limit at high $k$, it could (in principle) become more important and grow ever-faster as the simulation resolution increases.

We have focused on a relatively simple case here, namely gas with a pure acoustic wave in the absence of dust. This ignores, for example, magnetic fields, which alter the mode structure and could  influence the grain ``drag'' directly (if the grains are charged); this case is explored in more detail in a companion paper, \citet{hopkins:2018.mhd.rdi}. As shown in \paperone, the RDI generically exists for  systems that support undamped linear waves, so we expect a similar rich phenomenology of instabilities (both resonant and non-resonant) in other systems. However it is outside the scope of this work to explore these in detail.

Another topic which we will explore in more detail in future work is the influence of a broad size spectrum of dust grains. This is discussed in \S~\ref{sec:dust.species}, where we argue that under most conditions, we can think of the results of this work as being relevant for the large grains (specifically, the largest grains which contain a large fraction of the grain mass), because these dominate the mass and back-reaction on the gas. However as shown there, under some circumstances there is a complicated mix of terms dominated by small grains and others dominated by large grains, which could couple indirectly. Moreover, because the RDI can resonate with any wave family, it is possible that (for example) small, tightly-coupled grains (which may be more stable if considered in isolation) generate wave families to which larger grains can couple via the RDI (or vice versa).

\vspace{-0.7cm}
\acknowledgments 
We would like to thank our referee, Andrew Youdin, as well as E.~S.~Phinney and E.~Quataert for helpful discussions. Support for PFH \&\ JS was provided by an Alfred P. Sloan Research Fellowship, NASA ATP Grant NNX14AH35G, and NSF Collaborative Research Grant \#1411920 and CAREER grant \#1455342. JS was funded in part by the Gordon and Betty Moore Foundation
through Grant GBMF5076 to Lars Bildsten, Eliot Quataert and E. Sterl
Phinney.\\
%Numerical calculations were run on the Caltech compute cluster ``Zwicky'' (NSF MRI award \#PHY-0960291), allocation TG-AST130039 granted by the Extreme Science and Engineering Discovery Environment (XSEDE) supported by the NSF, and NASA HECC allocation SMD-16-7592. \\
\vspace{-0.2cm}
\bibliography{/Users/phopkins/Dropbox/Public/ms}

\appendix

\section{Relation to the matrix formalism of Squire \& Hopkins (2017)}\label{app: matrix relationship}
% Better name

Throughout the main text, our analysis was carried out through asymptotic expansions of the dispersion 
relation, so as to allow investigation into non-resonant modes (e.g., for $|\driftvel | < c_{s}$, and the ``long-wavelength'' modes).
To clarify the link to the RDI derivation in  \paperone, in this appendix, we  calculate the acoustic RDI
growth rates using the Jordan-form perturbation theory formalism of \paperone. 
We use the dimensionless variables of \S~\ref{sec:general.modes} (Eq.~\eqref{eqn:dimensionless vars}),
and, for the sake of concreteness, set $\driftvelhat=\hat{\bf z}$ and $\hat{\bf k}_{\perp}=\hat{\bf x}$ 
(it was not necessary to choose a specific direction in derivation of the dispersion relation, Eq.~\eqref{eqn:dispersion.full}).
We also ignore $u_{y}$ and $v_{y}$ because these  are decoupled from the sound-wave 
eigenmodes (these propagate in the $\hat{\bf k}$ direction).

From Eq.~\eqref{eqn:linearized}, the  coupled dust-gas equations are
\begin{equation}
\omdimless\boldsymbol{\xi}  = \left(\begin{array}{ccc}
\kappa_{\parallel}& \tilde{\bf k}^{T} & 0 \\ 
0 & \kappa_{\parallel} I + D_{\mathrm{drag}} & C_{\bf v}\\
\mu T^{(1)}_{\rho_{d}} & \mu T^{(1)}_{\bf v} & \mathcal{F} + \mu T^{(1)}_{g} 
\end{array}\right)\boldsymbol{\xi},\label{eqn:matrix form}
\end{equation}
where ${\boldsymbol{\xi}} =(\delta \rho_{d}/\rho_{0}, \delta v_{x}/c_{s}, \delta v_{z}/c_{s}, \delta \rho/\rho_{0}, \delta u_{x}/c_{s}, \delta u_{z}/c_{s} )^{T}$, 
$\tilde{\bf k}^{T}=(\kdimless_{x},\kdimless_{z})$,  $ T^{(1)}_{\rho_{d}} = (0,0,\iimag\,\driftvelmag)^{T}$, $ T^{(1)}_{\bf v} $ and $T^{(1)}_{g} $ are not
needed, 
\begin{equation}
D_{\mathrm{drag}} = \left(\begin{array}{cc}
-\iimag & 0 \\ 0 & -\iimag\, \tildeCoeffTSv\end{array}\right),\quad C_{\bf v} = \left(\begin{array}{ccc}
0 & \iimag & 0 \\ -\iimag\, \driftvelmag \coeffTSrho & 0 & -\iimag\,\tildeCoeffTSv
\end{array}\right),
\end{equation}
and 
\begin{equation}
\mathcal{F} = \left(\begin{array}{ccc}
0 & \kdimless_{x} & \kdimless_{z} \\ \kdimless_{x} & 0 & 0 \\ \kdimless_{z} & 0 &0
\end{array}\right).
\end{equation}
When at resonance, i.e.\ $\kappa_{\|} = \kdimless$ (where $\omdimless=\kdimless$ is forward-propagating sound-wave eigenvalue of $\mathcal{F}$), the matrix in Eq.~\eqref{eqn:matrix form}  is defective. This means
that although $\omdimless=\kappa_{\parallel}$ has multiplicity $2$, it has only one associated eigenvector. This is associated with an RDI, the growth rate of which scales as $\sim \mu^{1/2}$ because the
matrix is singular (rather than $\sim\mu$ as for standard perturbation theory). From \paperone~ (their Eq.~10), the perturbed eigenvalues in the ``mid-$k$'' regime (before $\tilde{\bf k}^{T}$ dominates over $D_{\mathrm{drag}}$ in Eq.~\eqref{eqn:matrix form}) are
\begin{equation}
\omdimless = \kappa_{\parallel} \pm \iimag\, \mu^{1/2 }\left[ (\boldsymbol{\xi}_{\mathcal{F}}^{L} T^{(1)}_{\rho_{d}})\,(\tilde{\bf k}^{T}D_{\mathrm{drag}}^{-1}C_{{\bf v}}\boldsymbol{\xi}_{\mathcal{F}}^{R})\right]^{1/2}+ \mathcal{O}(\mu) \label{eqn: dust eval pert}
\end{equation}
Here 
\begin{equation}
\boldsymbol{\xi}^{L}_{\mathcal{F}} =  \frac{1}{\sqrt{2}\,k}\left(\begin{array}{ccc}
k & k_{x} & k_{z}
\end{array}\right),\quad \boldsymbol{\xi}^{R}_{\mathcal{F}} =  \frac{1}{\sqrt{2}\,k}\left(\begin{array}{c}
k \\ k_{x}\\ k_{z}
\end{array}\right)
\end{equation}
are the left and right eigenvectors of the (forward-propagating) sound wave.
Equation~\eqref{eqn: dust eval pert} is easily verified to be the same as Eq.~\eqref{eqn:longwave.mode.midk} 
from the main text, up to $\mathcal{O}(\mu^{1/2})$.

In the ``high-$k$'' regime, the eigenvalue $\omdimless = \kappa_{\parallel}$ is nearly {triply defective} (meaning 
it has multiplicity $3$ with one associated eigenvector), because $\tilde{\bf k}^{T}\gg D_{\mathrm{drag}}$.
The perturbed eigenvalue is then 
\begin{equation}
\omdimless = \kappa_{\parallel} + \mu^{1/3 }\left[ (\boldsymbol{\xi}_{\mathcal{F}}^{L} T^{(1)}_{\rho_{d}})\,(\tilde{\bf k}^{T}C_{{\bf v}}\boldsymbol{\xi}_{\mathcal{F}}^{R})\right]^{1/3}+ \mathcal{O}(\mu^{2/3}),\label{eq: dust eval pert 3}
\end{equation}
which matches Eq.~\eqref{eqn:omega.resonant} from the main text.

We cannot treat the ``long-wavelength'' instability (Sec.~\ref{sec:long.wavelength}) using this method, because 
$\mu \gtrsim \kappa_{\parallel}$ in this regime. In other words, $\mu T^{(1)}_{\rho_{d}}$, $\mu T^{(1)}_{\bf v}$, and $\mu T^{(1)}_{g}$ are no longer a small perturbation to the fluid, and there is no well-defined undamped sound wave with which the
dust can resonate (see \S~\ref{sec:mode.structure} and Fig.~\ref{fig:mode.structure} for further discussion).
The long-wavelength growth rate Eq.~\eqref{eqn:longwave.mode} can be derived from the matrix (Eq.~\eqref{eqn:matrix form}) by treating $\kappa_{\parallel}$ and $\mathcal{F}$ as a small perturbation to $D_{\mathrm{drag}}$, $C_{\bf v}$ and $T^{(1)}$ (i.e., assuming small $k$). However, the procedure is not particularly illuminating (or, for that matter, easier algebraically than using the dispersion relation), so we do not reproduce it here.

\vspace{-0.5cm}
\section{Relation Between Free-Falling and Stationary Frames}\label{sec:accel.frame}

In \S~\ref{sec:free.streaming}, we transformed to a free-falling frame to analyze the instability. Here we derive this transformation in greater detail, and relate the mode properties in the free-falling and stationary frames. 

In the stationary frame, the fluid equations (Eq.~\eqref{eqn:general}) have homogeneous steady-state solutions given in Eq.~\eqref{eqn:mean.v.offset}. Consider small perturbations in this frame: $\rho = \rho_{0} + \delta\rho$, $\rho_{d} = \mu\,\rho_{0} + \delta\rho_{d}$, ${\bf u} = u_{0} + \tilde{\bf a}\,t + \delta {\bf u}$, and ${\bf v} = {\bf u}_{0} + \tilde{\bf a}\,t + \driftvel + \delta{\bf v}$, where $\tilde{\bf a} \equiv {\bf g} + {\bf a}\,\mu/(1+\mu)$. Note that both ${\bf u}$ and ${\bf v}$ contain both an arbitrary constant velocity offset (${\bf u}_{0}$) and a linear acceleration $\tilde{\bf a}\,t$. 

Inserting these into Eq.~\eqref{eqn:general} and linearizing in the perturbative ($\delta$) terms, we obtain the perturbation equations in the stationary frame:
\begin{align}
\nonumber \left( \partialAB{}{t} + \tilde{\bf u}_{0}(t)\cdot\nabla \right)\,\delta\rho =& -\rho_{0}\,\nabla\cdot \delta{\bf u},\\
\nonumber \left( \partialAB{}{t} + \tilde{\bf u}_{0}(t)\cdot\nabla \right)\,\delta{\bf u} =& -c_{s}^{2}\,\frac{\nabla \delta \rho}{\rho_{0}}
+ \mu\,\frac{(\delta {\bf v}-\delta{\bf u})}{\langle t_{s} \rangle}, \\
\nonumber &- \mu\,\frac{\driftvel}{\langle t_{s} \rangle}\,\left( \frac{\delta t_{s}}{\langle t_{s} \rangle} + \frac{\delta \rho}{\rho_{0}} - \frac{\delta \rho_{d}}{\mu\,\rho_{0}} \right), \\ 
\nonumber \left( \partialAB{}{t} + \tilde{\bf u}_{0}(t)\cdot\nabla + \driftvel\cdot\nabla \right)\delta \rho_{d} =& -\mu\,\rho_{0}\,\nabla\cdot \delta{\bf v},\\
\nonumber \left( \partialAB{}{t} + \tilde{\bf u}_{0}(t)\cdot\nabla + \driftvel\cdot\nabla \right)\delta {\bf v} =& -\frac{(\delta {\bf v}-\delta{\bf u})}{\langle t_{s} \rangle} + \frac{\driftvel\,\delta t_{s}}{\langle t_{s} \rangle^{2}}, \\
\label{eq: linearized eqns stationary frame}\tilde{\bf u}_{0}(t) \equiv {\bf u}_{0} + \tilde{\bf a}\,t =&\, {\bf u}_{0} + \left[{\bf g} + {\bf a}\,\frac{\mu}{1+\mu} \right]\,t.
\end{align}
To see the relationship between these stationary-frame equations (where ${\bf u} = \tilde{\bf u}_{0} + \delta {\bf u}$) and those in the free-falling frame (Eq.~\eqref{eqn:linearized}, where ${\bf u} = \delta {\bf u}$), consider e.g.\ the gas continuity equation: $\partial \rho/\partial t + \nabla\cdot ({\bf u}\,\rho) = 0$. Compared to the free-falling equations (Eq.~\eqref{eqn:linearized}), we see that the time-derivative of $\rho$ is unchanged, but the term $\nabla\cdot ({\bf u}\,\rho) = {\bf u}\cdot (\nabla\rho) + \rho \,(\nabla\cdot{\bf u})$ gives rise to an additional term $({\bf u}_{0} + \tilde{\bf a}\,t)\cdot \nabla \delta \rho = (\tilde{\bf u}_{0} \cdot \nabla)\,\delta\rho$. Note that the time-derivatives of ${\bf u}_{0}$ which appear in ${\bf u}$ and ${\bf v}$ are part of the homogenous solution, so do not appear in the linearized equations (Eq.~\eqref{eq: linearized eqns stationary frame}).

In this stationary frame, if we make the usual Fourier ansatz, where the terms in $\delta \propto \exp{\left[\iimag\,({\bf k}\cdot {\bf x} - \omega\,t) \right]}$, the fact that $\tilde{\bf u}_{0}$ is time-dependent prohibits a time-independent solution for $\omega({\bf k})$. However, note that the time derivatives $\partial/\partial t$ in Eq.~\eqref{eq: linearized eqns stationary frame} appear exclusively in the combination $\partial/\partial t + \tilde{\bf u}_{0}\cdot \nabla$. Motivated by this, consider the modified Fourier ansatz of the form:
%However, if instead we make the ansatz that $\delta \propto \exp{\left[\iimag\,({\bf k}\cdot {\bf x} - \{ \omega + [{\bf u}_{0} + \tilde{\bf a}\,t/2]\cdot {\bf k}\right} \,t) \right]}$, then the partial 
\begin{align}
\label{eqn:fourier.ansatz.stationary.frame} \delta &\propto \exp{\left\{ \iimag\,{\bf k}\cdot {\bf x} - \iimag\,\left[\omega + \left({\bf u}_{0} + \frac{1}{2}\,\tilde{\bf a}\,t \right)\cdot {\bf k} \right]\,t   \right\}}  
\end{align}
Inserting this, one finds  that the time and spatial derivatives behave as:
\begin{align}
{\Bigl(} \partialAB{}{t}& + \tilde{\bf u}_{0}(t)\cdot\nabla {\Bigr)}\,\delta = -\iimag\,\omega\,\delta \\
\nabla\delta &= \iimag\,{\bf k}\,\delta
\end{align}
In terms of $\omega$ and ${\bf k}$, we therefore obtain identical expressions for the dispersion relations  as the those derived in the main text in the free-falling frame (Eq.~\ref{eqn:linearized}). 

In other words, transforming from the free-falling frame to the stationary frame is equivalent to simply taking $\omega \rightarrow \omega + {\bf u}_{0}\cdot {\bf k} + (\tilde{\bf a}\,t/2)\,\cdot {\bf k}$. Along the direction of motion, the position of a wave crest is simply given by $x = \omega/k + u_{0}\,t + (1/2)\,\tilde{a}\,t^{2}$. So we immediately see that the offset in $\omega$ simply corresponds to motion with the homogenous solution, which has position $u_{0}\,t + (1/2)\,\tilde{a}\,t^{2}$. Physically, transforming into any linearly accelerating  and/or uniformly boosted frame has no effect on the character of the solutions.

Another, simpler way of seeing this is to return to the original, fully-general nonlinear equations (Eq.~\eqref{eqn:general}), and boost to a free-falling (uniformly accelerating) frame with spatial and time coordinates $t^{\prime} = t$, ${\bf x}^{\prime} = {\bf x} + {\bf u}_{0}\,t + (1/2)\,\tilde{\bf a}\,t^{2}$. In a uniformly accelerating frame the local equations of motion are necessarily identical in these variables, up to the introduction of a fictitious force/acceleration (${\bf a}_{\rm fict} = -\tilde{\bf a}$) felt by both the gas and dust. This is equivalent, in Eq.~\eqref{eqn:general}, to taking ${\bf g} \rightarrow {\bf g} - \tilde{\bf a} = -{\bf a}\,\mu/(1+\mu)$. It is easy to verify that the steady-state, homogeneous solution in this frame is then $\rho^{\prime}=\rho_{0}$, $\rho_{d}^{\prime} = \mu\,\rho_{0}$, ${\bf u}^{\prime} = \mathbf{0}$, ${\bf v}^{\prime} = \driftvel = {\bf a}\,t_{s}/(1+\mu)$ (i.e.\ the same homogeneous solution as in the stationary frame, but co-moving with the gas). Perturbing in these variables, the fictitious force is exactly canceled by the other terms in the homogenous solution, and the perturbative equations are  identical to Eq.~\eqref{eqn:linearized} (up to the replacement ${\bf x} \rightarrow {\bf x}^{\prime}$, $t \rightarrow t^{\prime}$). In this frame, we Fourier decompose each variable $\delta \propto \exp[\iimag\,({\bf k}\cdot{\bf x}^{\prime} - \omega\,t^{\prime})]$, and obtain the  dispersion relation in Eq.~\eqref{eqn:dispersion.full}. But noting the definition of $t^{\prime}$ and ${\bf x}^{\prime}$ above, we immediately see that ${\bf k}\cdot{\bf x}^{\prime} - \omega\,t^{\prime} = {\bf k}\cdot[{\bf x} + {\bf u}_{0}\,t + (1/2)\tilde{\bf a}\,t^{2}] + \omega\,t = {\bf k}\cdot {\bf x} + [\omega + {\bf u}_{0}\cdot {\bf k} + (t/2)\,\tilde{\bf a}\cdot{\bf k}]\,t$. This is simply the same equivalence between frames as we obtained above.

Obviously, for the hydrostatic cases considered in the text (\S~\ref{sec:pressure.gradients} and below), the equilibrium gas motion is stationary ($\langle {\bf u} \rangle = 0$) so our derivation in the text is already in the stationary frame. In App.~\ref{sec:hydrostatic.generalized} below, we show that the resulting instabilities are similar to those derived in the free-falling frame.

\vspace{-0.5cm}
\section{Hydrostatic \&\ Stratified Systems: General Cases}
\label{sec:hydrostatic.generalized}

In \S~\ref{sec:pressure.gradients}, we briefly discussed cases where the gas is initially hydrostatic and/or had arbitrary background gradients in the equilibrium fluid. Here we explore these cases in more detail, demonstrating that such modifications do not fundamentally alter the instabilities described in the main text.

\vspace{-0.5cm}
\subsection{General \&\ Linearized Equations}

If the system is initially hydrostatic, we seek steady-state equilibrium solutions of Eq.~\eqref{eqn:general} with ${\bf u}=0$. This implies $\driftvel={\bf v}$ with 
\begin{align}
\nonumber \nabla P_{0} &= \rho_{0}\,{\bf g} + \rho_{d,\,0}\,\frac{\driftvel}{\langle t_{s} \rangle} ,\\ 
\nonumber \driftvel\cdot \nabla \rho_{d,\,0} &= -\rho_{d,\,0}\,\nabla \cdot \driftvel ,\\
\label{eqn:gradients} (\driftvel\cdot \nabla)\, \driftvel &= -\frac{\driftvel}{\langle t_{s} \rangle} + {\bf g} + {\bf a}.
\end{align}
For finite $\rho_{d}$ and ${\bf a}$, there are few (if any) simple solutions to these equations (e.g.\ fully specifying $P=P(z)$ for ${\bf g}$ or ${\bf a}$ in the $\hat{z}$ direction) that do not become unphysical at some point (e.g.\ producing negative temperature/pressure/density, or exponentially-diverging dust-to-gas-ratios). Such solutions also require a specific form of $t_{s}(\rho,\,P,\,{\bf v},\,...)$, and an equation-of-state for $P$. Of course, in reality, boundary conditions and global evolution of the system will become important eventually and must be specified for a given problem. Further, in many cases the system will only be locally in equilibrium over some spatial or time scale, with, for example, some slow net drift of the dust through gas.

We  therefore consider {\em local} solutions; i.e.\ expanding some quantity $U$ as $\langle U \rangle \approx U_{0} + \nabla U_{0} \cdot ({\bf x}-{\bf x}_{0})$. This is valid for $|{\bf x}-{\bf x}_{0}| \sim (2\pi/k) \ll k_{U}^{-1}$ where $k_{U}^{-1} \equiv |U_{0}|/|\nabla U_{0}|$ is the relevant gradient scale-length, so we must drop terms $\mathcal{O}(|k_{U}/k|)$. 

Including background pressure/entropy gradients, we must also explicitly include an entropy equation, which takes the form $Ds/Dt=0$ or $DP/Dt = c_{s}^{2}\,D\rho/Dt$ (where $D/Dt = \partial/\partial t + {\bf u}\cdot \nabla$). Note the entropy equation was implicit in the main text (Eq.~\ref{eqn:general}), because without background gradients it just trivially simplifies to $\delta P = c_{s}^{2}\,\delta \rho$ at linear order. Similarly, since pressure and density can vary independently, we de-compose the perturbations to $t_{s}$ into separate pressure and density terms, i.e.\ 
\begin{align}
\frac{\delta t_{s}}{\langle t_{s} \rangle} &= -
\coeffTSrhoonly\,\frac{\delta\rho}{\rho_{0}} - (\coeffTSrho-\coeffTSrhoonly)\,\frac{\delta P}{\rho_{0}\,c_{s}^{2}} -\coeffTSv\,\frac{\driftvel\cdot(\delta{\bf v}-\delta{\bf u})}{|\driftvel|^{2}}
,
\end{align}
where $\coeffTSrhoonly$ and $\coeffTSPonly\equiv \coeffTSrho-\coeffTSrhoonly$ represent perturbations to $t_{s}$ from density or pressure fluctuations, respectively (with the other fixed). 
Note that we explicitly write this in this manner so that $\coeffTSrho$ has the same meaning in the text: when $\delta P \approx c_{s}^{2}\,\delta\rho$ (as occurs without gradients in $P_{0}$ or $\rho_{0}$), one finds $\coeffTSrhoonly\,\delta\rho/\rho_{0} + (\coeffTSrho-\coeffTSrhoonly)\,\delta P/\rho_{0}\,c_{s}^{2} = \coeffTSrho\,\delta\rho/\rho_{0}$. We will show that to leading-order, only the ``total'' term $\coeffTSrho$ appears.

Combining this and Eq.~\ref{eqn:gradients} with Eq.~\ref{eqn:general}, and subtracting the steady-state solution, we obtain the linearized equations:
\begin{align}
\label{eqn:linearized.with.gradients} \partialAB{\delta\rho}{t} =& -\rho_{0}\,\nabla\cdot \delta{\bf u} - \delta{\bf u} \cdot \nabla\rho_{0}, \\
\nonumber \partialAB{\delta{\bf u}}{t} =& 
-\frac{\nabla \delta P }{\rho_{0}}
+\frac{\delta \rho\,\nabla P_{0}}{\rho_{0}^{2}}
%-c_{s}^{2}\,\frac{\nabla \delta \rho}{\rho_{0}}
+ \mu\,\frac{(\delta {\bf v}-\delta{\bf u})}{\langle t_{s} \rangle} \\
\nonumber &- \mu\,\frac{\driftvel}{\langle t_{s} \rangle}\,\left( \frac{\delta t_{s}}{\langle t_{s} \rangle} + \frac{\delta \rho}{\rho_{0}} - \frac{\delta \rho_{d}}{\mu\,\rho_{0}} \right),  \\ 
%\nonumber &-\frac{\delta\rho}{\rho_{0}^{2}}\,\left[(\gamma-1)\,\nabla P_{0} - c_{s}^{2}\,\nabla\rho_{0} \right] \\
\nonumber \partialAB{\delta P}{t} + \delta{\bf u}\cdot \nabla P_{0} &= c_{s}^{2}\,\left( \partialAB{\delta\rho}{t} + \delta{\bf u}\cdot\nabla\rho_{0} \right), \\
\nonumber \left( \partialAB{}{t} + \driftvel\cdot\nabla \right)\delta \rho_{d} =& 
-\mu\,\rho_{0}\,\nabla\cdot \delta{\bf v} - \left( \delta{\bf v} - \frac{\delta \rho_{d}}{\mu\,\rho_{0}}\,\driftvel \right)\cdot \nabla \rho_{d,\,0}, \\
%-\mu\,\rho_{0}\,\nabla\cdot \delta{\bf v} - \delta{\bf v}\cdot \nabla \rho_{d,\,0} - \delta \rho_{d}\,\nabla\cdot \driftvel \\
%\nonumber =& -\left[\mu\,\rho_{0}\,\nabla\cdot \delta{\bf v} + \left( \delta{\bf v} - \frac{\delta \rho_{d}}{\mu\,\rho_{0}}\,\driftvel \right)\cdot \nabla \rho_{d,\,0} \\ 
\nonumber \left( \partialAB{}{t} + \driftvel\cdot\nabla \right)\delta {\bf v} =& -\frac{(\delta {\bf v}-\delta{\bf u})}{\langle t_{s} \rangle} + \frac{\driftvel\,\delta t_{s}}{\langle t_{s} \rangle^{2}} - \left( \delta{\bf v} \cdot \nabla \right)\,\driftvel .
\end{align}

Note that if we take $\mu\rightarrow 0$, the gas equations immediately reduce to the familiar standard equations for acoustic perturbations in a stratified fluid \citep{bray.loughhead:stratified.atmosphere.scalings}.

\vspace{-0.5cm}
\subsection{Degrees of Freedom and Validity}
\label{sec:validity.gradients}

Locally, Eq.~\eqref{eqn:gradients} permits arbitrary 3D gradients in $P_{0}$, $\rho_{0}$, $\rho_{d,\,0}$, and each component of $\driftvel$, with only one constraint equation.\footnote{In Eq.~\ref{eqn:gradients}, the $\driftvel\cdot \nabla \rho_{d,\,0} = -\rho_{d,\,0}\,\nabla\cdot\driftvel$ equation removes one degree of freedom if it is to be a true equilibrium. The equations for $\nabla P_{0}$ and $(\driftvel\cdot\nabla)\driftvel$ only relate these quantities in equilibrium to the arbitrary input vectors ${\bf g}$ and ${\bf a}$, they do not reduce the number of degrees of freedom of the problem.} Moreover if the problem has arbitrary 3D asymmetry we must consider 3D wavevectors ${\bf k}$ (we cannot treat the ${\bf k}_{\bot}$ direction as symmetric in the plane perpendicular to $\driftvel$). Formally, therefore, this introduces 18 degrees of freedom into the dispersion relation.
Fortunately, as shown below, only a couple of these degrees-of-freedom have any influence on the modes, within the constraints required for our local derivation to be valid.

As noted above, lacking a global solution and/or boundary conditions, Eq.~\eqref{eqn:linearized.with.gradients} is  valid only up to leading-order in $\mathcal{O}(k_{U} / k)$ ($k\gg k_{U}$), where $k_{U}^{-1} \sim |U_{0}|/|\nabla U_{0}|$ is the gradient-scale length of some background quantity $U$. Moreover, if the velocities $v$ (e.g.\ $\driftvel$ or the mode phase/group velocities $v_{0}\sim c_{s}$) are non-zero, then our derivation is also only valid on a timescale $ \Delta t \ll 1/(v\,k_{U})$. Over timescales longer than this, the mode and/or incoming dust travels a distance greater than $ k_{U}^{-1}$, outside the domain where our local gradient expansion is valid. Thus, we also require $|\omega| \gg v \, k_{U}$ (although if $\omega\sim c_{s}\,k$ to leading order, this condition is identical to $k \gg k_{U}$).

Another obvious requirement is that the  dust stopping length $L_{\rm stop} \sim |\driftvel|\,\langle t_{s} \rangle$ (the distance the dust travels in one stopping time) is small compared to the gradient scale lengths of the system ($|\driftvel|\,\langle t_{s} \rangle \ll k_{U}^{-1}$).  Otherwise the dust simply drifts through a full scale-length without feeling significant coupling to the gas. In that case the system could never meaningfully reach local equilibrium and a global solution is clearly required. Considering the dust-density and drift-velocity scale-lengths, $k_{\rho_{d,\,0}} = |\nabla\rho_{d\,0}|/\rho_{d,\,0} \approx k_{w} = |\nabla\cdot\driftvel|/|\driftvel|$ (related by Eq.~\ref{eqn:gradients}), we see that  $L_{\rm stop} \ll k^{-1}_{\rho_{d,\,0}}$ or $L_{\rm stop} \ll k^{-1}_{w}$ is equivalent to $|\driftvel|\,\langle t_{s} \rangle \ll \rho_{d,\,0}/|\nabla\rho_{d,\,0}| \sim |\driftvel| / |\nabla\cdot\driftvel|$, i.e.\ $|\langle t_{s} \rangle\,\nabla \cdot \driftvel| \ll 1$. 

%For example, consider the term $\langle t_{s} \rangle \, \rho_{d,\,0}^{-1}\,\driftvel \cdot \nabla \rho_{d,\,0} = -\langle t_{s} \rangle \, \nabla\cdot \driftvel \sim \langle t_{s} \rangle \, L_{w}/|\driftvel| \sim \langle t_{s} \rangle /\Delta t$. If this is not small, the physical statement is that the timescale for the dust to drift through/to locations where the dust density and drift velocity have changed by a large factor ($\Delta t \sim |\nabla\cdot\driftvel|^{-1}$) is shorter than the stopping time $\langle t_{s} \rangle$. But in this limit, we can never meaningfully reach equilibrium in the first place.

\vspace{-0.5cm}
\subsection{Dispersion Relation \&\ Scalings (Simplified Case)}
\label{sec:dispersion.simplified.for.gradients}

Above we noted the full set of gradients introduces 18 degrees of freedom. Analyzing this is generally un-interesting, however, and many parameter combinations have no effect on the modes, or are formally allowed but unphysical.

The analysis is greatly simplified if we consider one of two cases:    (a) either gravity or the external acceleration dominates, i.e.\ $|{\bf g}|\gg |{\bf a}|$ (e.g.\ dust settling through a hydrostatic, self-gravitating atmosphere) or $|{\bf g}|\ll |{\bf a}|$ (e.g.\ radiative acceleration of a dust-driven wind); or (b) ${\bf g}$ and ${\bf a}$  are parallel. In either of these cases, the equilibrium solution should be symmetric about this preferred axis. Then the gradient terms can be expressed as: 
\begin{align}
\nonumber \nabla_{\|} \driftvel &= \frac{|\driftvel|}{c_{s}\,\langle t_{s} \rangle}	\,\gradscalew , \ \ \ \ 
\nabla_{\|} \rho_{d,\,0}  = -\frac{\rho_{d,\,0}}{c_{s}\,\langle t_{s} \rangle}	\,\gradscalew, \\
\nonumber \nabla_{\|} P_{0} &= \frac{c_{s}^{2}\,\rho_{0}}{c_{s}\,\langle t_{s} \rangle} \, \gradscaleP  , \ \ \ \ 
\nabla_{\|} \rho_{0} = \frac{\rho_{0}}{c_{s}\,\langle t_{s} \rangle}\,\gradscalerho, \\ 
\nonumber \gradscalew &= \frac{c_{s}\,\langle t_{s} \rangle\,\nabla\cdot \driftvel}{|\driftvel|} = \frac{c_{s}}{|\driftvel|}\, \left[\frac{({\bf g}+{\bf a})\cdot {\driftvelhat}\,\langle t_{s} \rangle}{|\driftvel|} - 1\right], \\ 
\label{eqn:lambda.definitions} \gradscaleP &= \frac{c_{s}\,\langle t_{s} \rangle\,\driftvelhat\cdot\nabla P_{0}}{c_{s}^{2}\,\rho_{0}} = \frac{({\bf g}\cdot \driftvelhat)\,\langle t_{s} \rangle + \mu\,|\driftvel|}{c_{s}},
\end{align}
where $\nabla_{\|} \equiv \driftvelhat\cdot \nabla$ is the gradient along the drift direction, and the latter two equations are constraints arising from the momentum equations. 
These equations define three dimensionless parameters, $\gradscalew$, $\gradscalerho$, $\gradscaleP$, which are proportional to the relevant gradient scale lengths in the parallel direction (e.g.\ $\gradscalew = k_{w}\,c_{s}\,\langle t_{s} \rangle = -k_{\rho_{d,\,0}}\,c_{s}\,\langle t_{s} \rangle$, 
$\gradscalerho = k_{\rho_{0}}\,c_{s}\,\langle t_{s} \rangle$, and $\gradscaleP = (P_{0}/c_{s}^{2}\rho_{0})\,k_{P_{0}}\,c_{s}\,\langle t_{s} \rangle  = (1/\gamma)\,k_{P_{0}}\,c_{s}\,\langle t_{s} \rangle$). Since we have allowed for arbitrary background entropy profiles, there is no equation to determine $\nabla \rho_{0}$ and $\gradscalerho$ is an arbitrary parameter. For an adiabatic (isentropic) background pressure gradient, $\gradscalerho=\gradscaleP$ (this is convenient below and the reason for our particular definition here), while for a pure entropy gradient (with constant background density), $\gradscalerho=0$.

Earlier we noted that $|\driftvel|\,\langle t_{s} \rangle\,|k_{w}| = \langle t_{s} \rangle\,|\nabla \cdot \driftvel| = \driftvelmag\,|\gradscalew|  \ll 1$ was required for our derivation to be valid. 
We typically expect $|\driftvel|\approx |{\bf g}+{\bf a}|\,\langle t_{s} \rangle$ (the normal terminal velocity if the gas is in hydrostatic equilibrium), so $|\gradscalew|\ll 1$, and this is satisfied (so long as $\driftvelmag$ is not extremely large, which is not usually expected in systems of interest). For streaming in a pressure-supported atmosphere that is only weakly perturbed by the dust (i.e.\ when $\nabla P_{0} \approx \rho_{0}\,{\bf g}$), we see that  $|\gradscaleP| \ll 1$ is usually satisfied if the dust stopping length ($L_{\mathrm{stop}}\sim|\driftvel|\,t_{s}$) is smaller than the pressure scale-length $c_{s}\,t_{s}\,|\gamma\,\gradscaleP|^{-1}$ (otherwise, a global solution is needed). If instead ${\bf g}$ is weak (e.g.\ for highly super-sonic streaming in a dust-driven wind), we see that $|\gradscaleP| \approx \mu\,\driftvelmag$. 

For convenience of notation below, we define the generic inverse scale length $\gradscale \equiv \max\{|\gradscalew|,\, |\gradscalerho|,\, |\gradscaleP|\}$, so the full set of conditions for a local derivation to be valid from \S~\ref{sec:validity.gradients} above become $|\driftvelmag\,\gradscale|\ll \min\{1,\,|\omdimless|\}$, and $\gradscale \ll \kdimless$.

As discussed in more rigorous mathematical detail in \citet{squire:rdi.ppd}, where we explore the \BV\ RDI, at this point it is in principle possible to consider a fully-general WKBJ analysis, assuming that the linear perturbations $\delta\rho$, etc., have the form $\exp[i\,\epsilon^{-1}\,\sum_{n=0}^{\infty}\,\epsilon^{n}\,Q_{n}({\bf x})]$, keeping all terms in the background and deriving an expression for the frequencies $\omega$ to lowest order in the expansion parameter $\epsilon \ll 1$ (with $\epsilon$ some appropriate function of $\gradscale / \kdimless$). However this is not enlightening: the expressions in full generality can only be expressed as complicated integro-differential functions of the background (which is unspecified), which can only be evaluated numerically (and then only if the background profiles are specified; see e.g.\ \citealt{bender:book.math.methods}). Moreover the ordering of the expansion is fundamentally ambiguous, since above we note multiple independent small parameters (e.g.\ $|\gradscale/\kdimless|$ and $|\driftvelmag\,\gradscale|$) as well as other parameters which may also be small under some circumstances (e.g.\ $\mu$ or $\driftvelmag$). And there is no unique or obvious ``preferred'' background as there is for common pure-hydrodynamic cases (e.g.\ an exponentially-stratified vertical atmosphere), since we have introduced stratification of the dust properties. So instead we will consider a simpler local approximation in which we assume $|\gradscale/\kdimless| \ll 1$, $|\driftvelmag\,\gradscale|$, and that each of the background gradient terms $\gradscalew,\, \gradscalerho,\, \gradscaleP$ is constant, so we can Fourier-decompose the perturbations keeping only the lowest-order WKBJ term in $\gradscale/\kdimless$ (i.e.\ our usual Fourier {\em ansatz} for the perturbations), and solve them ``locally'' in an infinitesimally small region about the ``origin'' where the background quantities $\rho_{0}$, etc., and their gradients are defined. 

Bear in mind, this means our solutions will only be valid to lowest order in this expansion, and should be regarded somewhat heuristically: but this still allows us to see if there are leading-order corrections which could be important when $|\gradscale/\kdimless| \ll 1$. 

Finally, then, the full dispersion relation in this simplified case is a 9th-order polynomial, with roots given by the eigenvalues of:
\begin{scriptsize}
\begin{align}
\nonumber	&\begin{bmatrix}
0 & 0 & \kdimless_{\bot} & 0 & \kdimless_{\|} - \iimag\,\gradscalerho & 0 & 0 & 0 & 0 \\
0 & b_{3} & 0 & 0 & 0 & \mu\,\kdimless_{\bot} & 0 & b_{4} & 0 \\
\kdimless_{\bot} & 0 & -\iimag\,\mu & 0 & 0 & \iimag\,\mu & 0 & 0 & \kdimless_{\bot} \\
0 & 0 & 0 & -\iimag\,\mu & 0 & 0 & \iimag\,\mu & 0 & 0 \\
b_{0} & \iimag\,\driftvelmag & 0 & 0 & -\iimag\,\mu\,\tildeCoeffTSv & 0 & 0 & \iimag\,\mu\,\tildeCoeffTSv & b_{5} \\
0 & 0 & \iimag & 0 & 0 & b_{1} & 0 & 0 & 0 \\
0 & 0 & 0 & \iimag & 0 & 0 & b_{1} & 0 & 0 \\
-\iimag\,\driftvelmag\,\coeffTSrho & 0 & 0 & 0 & \iimag\,\tildeCoeffTSv & 0 & 0 & b_{2} & -\iimag\,\driftvelmag\,\coeffTSrhoonly \\
0 & 0 & 0 & 0 & \iimag\,(\gradscalerho-\gradscaleP) & 0 & 0 & 0 & 0 \\
	\end{bmatrix} 
\end{align}
\end{scriptsize}
where
\begin{align}
\nonumber	b_{0} &= \kdimless_{\|} + \iimag\,\left[ \mu\,\driftvelmag\,(\coeffTSrho-1)+\gradscaleP \right] \, ,	\\ 
\nonumber	b_{1} &= -\iimag+\driftvelmag\,\kdimless_{\|} \ \ \ \ \ \ \ \ , \ \ \ 	b_{2} = -\iimag\,\tildeCoeffTSv + \driftvelmag\,(\kdimless_{\|}-\iimag\,\gradscalew) \, , \\
\nonumber   b_{3} &= \driftvelmag\,(\kdimless_{\|}-\iimag\,\gradscalew) \ \ \ , \ \ \  b_{4} = \mu\,(\kdimless_{\|} + \iimag\,\gradscalew) \, , \\ 
b_{5} &= \kdimless_{\|} + \iimag\,\mu\,\driftvelmag\,\coeffTSPonly \ \ \ \ , \ \ \  \kdimless_{\bot} = |\driftvelhat \times {\bf \kdimless}| = \kdimless\,\sin{\theta} \, , 
\end{align}
where  we use the same dimensionless units as in Eq.~\eqref{eqn:dispersion.full}.

\begin{figure*}
\plotsidesize{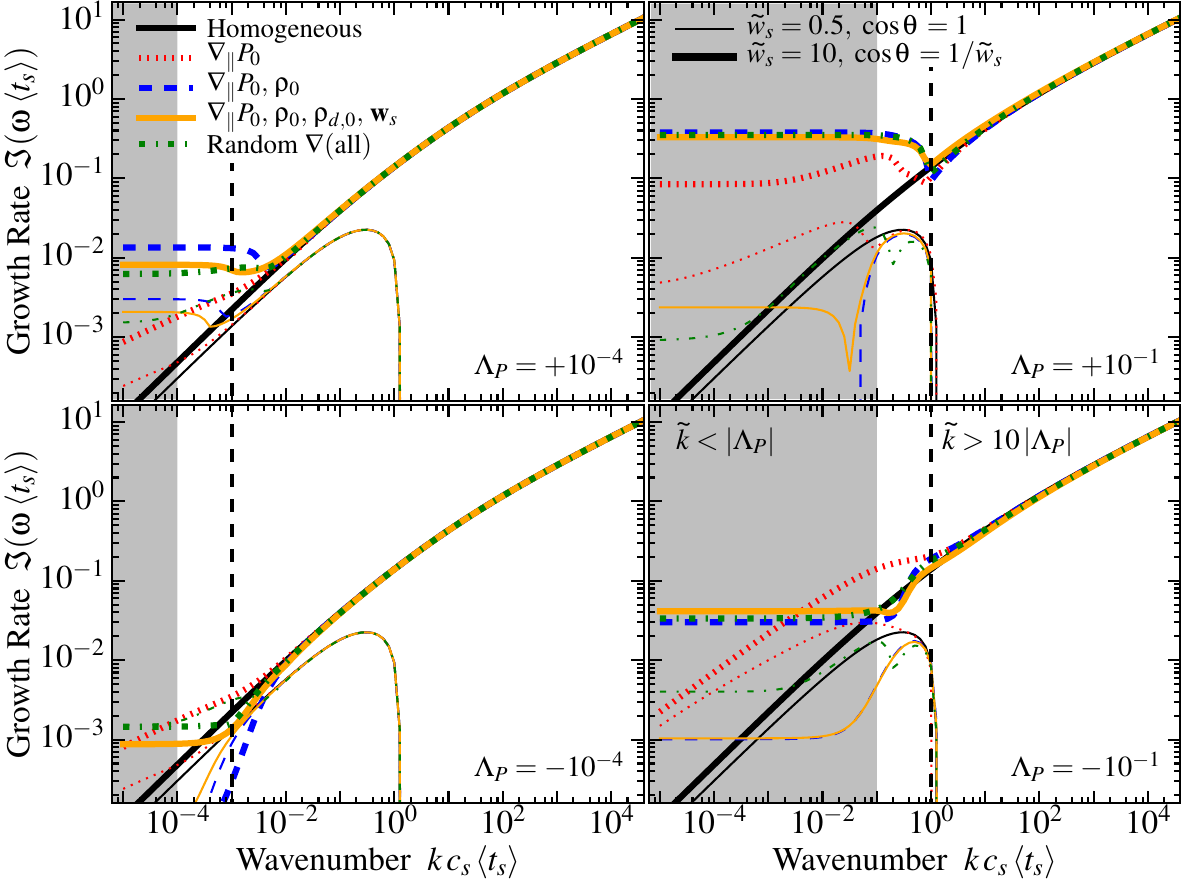}{0.8}
    \vspace{-0.25cm}
    \caption{Effects of stratification (background gradients) on the growth rates of the acoustic RDI. We show growth rates versus wavenumber  (as in Fig.~\ref{fig:growth.rate.demo}), calculated from the full solution to the 9th-order dispersion relation (Eq.~\ref{eqn:linearized.with.gradients}) allowing for arbitrary gradients in $P_{0}$, $\rho_{0}$, $\rho_{d,\,0}$, and each component of $\driftvel$. For simplicity we take $\coeffTSrho=\coeffTSv=0$, and show (with thick lines) a supersonic ($\driftvelmag=10$) case with $\hat{\bf k}$ oriented at the resonant angle ($\cos{\theta}=1/\driftvelmag$) and (with thin lines) a subsonic case ($\driftvelmag=0.5$) with parallel $\hat{\bf k}$ ($\cos{\theta}=1$). We consider five regimes as described in \S\ref{app:sub: numerical}: {\bf (i)} Homogeneous: the case from the main text (neglecting background gradients; $\gradscaleP=\gradscalerho=\gradscalew=0$). {\bf (ii)} $\nabla_{\|}P_{0}$: A hydrostatic system (external acceleration balanced by a pressure gradient obeying Eq.~\ref{eqn:gradients}), with negligible gradients in other quantities ($\gradscalerho = \gradscalew=0$, in Eq.~\ref{eqn:lambda.definitions}). We compare, as labeled at the bottom-right of each subfigure, two signs and two absolute values of $\gradscaleP \equiv c_{s}\,\langle t_{s} \rangle\,(\nabla_{\|} P_{0})/(\rho_{0}\,c_{s}^{2})$  (i.e.\ $(\gamma\,\gradscaleP)^{-1}$ is approximately the pressure-gradient scale length). $\nabla_{\|}$ refers to the gradient along the direction of $\driftvel$, so opposite signs correspond to pressure increasing ($+$) or decreasing ($-$) along the drift direction. {\bf (iii)} $\nabla_{\|}P_{0},\,\rho_{0}$: We include gradients in gas pressure and density with $\gradscalerho = 2\,\gradscaleP$, so the gas system (without dust) is stably stratified. {\bf (iv)} $\nabla_{\|}P_{0},\,\rho_{0},\,\rho_{d,\,0},\,\driftvel$: We include gradients along $\driftvelhat$ in all properties (gas pressure and density, dust density and streaming velocity), with $\gradscalerho=2\,\gradscaleP$, and $\gradscalew = -\gradscaleP$ (for $|\gradscaleP|=10^{-4}$ cases) or $\gradscalew = -0.1\,\gradscaleP$ (for $|\gradscaleP|=10^{-1}$ cases, because $|\driftvelmag\,\gradscalew| \ll 1$ is required for equilibria to exist).  {\bf (v)} Random $\nabla$(all): We impose gradients as in case {\bf (iv)}, but also impose a gradient in every non-parallel direction (18 total gradient components), each set to a random number with value between $-|\gradscale|$ and $+|\gradscale|$ (for the appropriate $\gradscale$ of each quantity). 
    The derivation in the main text requires $k \gg |\gradscale|$ --   i.e.\ without a global solution our dispersion relation is only valid on scales smaller than the gradient scale length -- so we indicate $\kdimless < |\gradscaleP|$ (i.e.\ $k\,c_{s}\langle t_{s} \rangle < |\gradscaleP|$, shaded) and $\kdimless = 10\,|\gradscaleP|$ (dashed vertical line) to show where the solutions are physical. 
    In all cases, we see the predictions rapidly converge to the homogeneous case for $\kdimless \gg |\gradscale|$, as expected.    
    \vspace{-0.25cm}
        \label{fig:growth.rate.stratification}}
\end{figure*}

\vspace{-0.5cm}
\subsection{Solutions Without Dust}
\label{sec:solution.mods.nodust.gradients}

Absent dust (i.e.\ for gas alone, $\mu=0$), the dispersion relation simplifies dramatically as one might expect. However the presence of background gradients still modifies the dispersion relation from $\omdimless_{0}^{2} = \kdimless^{2}$ (sound waves in a homogeneous background, in dimensionless units) to $\omdimless_{0}^{4} = \omdimless_{0}^{2}\,(\kdimless^{2} + \gradscaleP\,\gradscalerho) + \kdimless_{\bot}^{2}\,\gradscaleP\,(\gradscaleP-\gradscalerho)$ (where $\kdimless_{\bot}$ is the component of ${\bf \kdimless}$ perpendicular to $\nabla P_{0}$). This has the usual solution branches (e.g.\ \citealt{bray.loughhead:stratified.atmosphere.scalings}) given by $\omdimless_{0}^{2} = (1/2)\,[\kdimless^{2} + \gradscaleP\,\gradscalerho \pm \{(\kdimless^{2} + \gradscaleP\,\gradscalerho)^{2} + 4\,\kdimless_{\bot}^{2}\,\gradscaleP\,(\gradscaleP-\gradscalerho)\}^{1/2}]$, where at $\kdimless \gg |\gradscaleP|$ the ``$+$'' branch corresponds to a weakly-modified sound wave, with $\omdimless_{0}^{2} \approx \kdimless^{2} + \gradscaleP\,[\gradscalerho + (\kdimless_{\bot}/\kdimless)^{2}\,(\gradscaleP-\gradscalerho)]$, and the ``$-$'' branch corresponds to buoyancy oscillations with $\omdimless_{0}^{2} \approx (\kdimless_{\bot}/\kdimless)^{2}\,\gradscaleP\,(\gradscalerho-\gradscaleP)$. From this we see that, with our definitions, the usual \BV\ frequency is $N_{BV}^{2} = \gradscaleP\,(\gradscalerho-\gradscaleP)$.

Note that the leading-order terms in the dispersion relation (relevant for both the sound-wave and buoyancy oscillation regime) are correctly captured here by our local (leading-order) analysis. But the next-to-leading order term (in $|\gradscale/\kdimless|$) in the modified sound wave above does not match that usually derived from a more accurate WKBJ expansion for sound waves in an exponentially-stratified, plane-parallel atmosphere \citep[see e.g.][]{lighthill:book.waves.fluids,clarke:astro.fluid.dynamics.book}, except for special values of $\gradscalerho$. This owes to (1) different assumptions about what is held constant (e.g.\ we assume here the $\gradscale$ quantities are constant, whereas the usual pure-hydrodynamic analysis assumes $\nabla P_{0}/\rho_{0} = {\bf g}$ is constant), and (2) the local approximation described in \S~\ref{sec:dispersion.simplified.for.gradients} above, made for generality. We note this to remind the reader that sub-leading order terms here, while given for completeness, should be regarded as heuristic and more detailed conclusions require solutions that actually specify the background gradients.

\vspace{-0.5cm}
\subsection{Solutions With Dust: Numerical Examples}\label{app:sub: numerical}

In Fig.~\ref{fig:growth.rate.stratification}, we present numerical solutions for the full linearized equations including both dust and gas, comparing hydrostatic systems with arbitrary background gradients (Eq.~\ref{eqn:linearized.with.gradients}) to the homogeneous (free-falling) systems analyzed in the main text. For any given value of the gradients, we obtain from Eq.~\ref{eqn:linearized.with.gradients} a ninth-order dispersion relation for $\omega$, as a function of each of the gradients, as well as the independent variables studied in the hydrostatic case ($\driftvel$, $\mu$, $\coeffTSrho$, $\coeffTSrhoonly$, $\coeffTSv$, ${\bf k}$, etc.). We discuss analytic approximations to the solutions for each relevant mode below. 

We compare five different assumptions for the nature of the gradients in  Fig.~\ref{fig:growth.rate.stratification}, and for each assumption, compare four different actual values of the gradients. These different gradient assumptions are:
\begin{enumerate}
\item Homogeneous: This is the homogeneous (free-falling) case from the text (all gradients in the background quantities neglected). %Without a pressure gradient, the system is necessarily free-falling.

\item $\nabla_{\|}P_{0}$: Here we consider a hydrostatic system, which therefore must have a pressure gradient following Eq.~\eqref{eqn:gradients}, offsetting the net acceleration. But we neglect all other gradient terms, i.e.\ consider only a simple pressure gradient aligned along the drift/acceleration direction, of the form in Eq.~\eqref{eqn:lambda.definitions}, with value of the gradient (in our dimensionless units) of $\gradscaleP$. We note that since we include no density gradient, the \BV\ frequency in the gas is $N_{BV}^{2} = -\gradscaleP^{2}$, i.e.\ the system is hydrodynamically unstable. The effects of this gradient on the growth rates, relative to the homogeneous case, are small at $\kdimless \gg |\gradscaleP|$, but at smaller $k$ the sense is always to enhance instability (but a global solution is really required in this limit). 

\item $\nabla_{\|}P_{0},\,\rho_{0}$: We also include a gas density gradient along the same direction, of the form in Eq.~\ref{eqn:lambda.definitions} with $\gradscalerho = 2\,\gradscaleP$. Now, the \BV\ frequency is $N_{BV}^{2} = \gradscaleP^{2}$, so the hydrodynamic system (in the absence of dust) is unconditionally stable. We have experimented with a range of values of $|\gradscalerho/\gradscaleP|$, and find that our results at $\kdimless \gg \max\{|\gradscaleP|,\,|\gradscalerho|\}$ are very weakly sensitive to $|\gradscalerho/\gradscaleP|$, particularly at high $k$. At low-$k$, when $\gradscalerho \sim \gradscaleP < 0$, this actually produces closer agreement with the homogeneous case than (i) where we considered $\nabla_{\|}P_{0}$ alone (the density and pressure gradient effects partially cancel). For $\gradscalerho  > 0$, the growth rates are further enhanced at low-$k$, owing to the fact that $\nabla \mu$ along the drift direction is non-zero. 

\item $\nabla_{\|}P_{0},\,\rho_{0},\,\rho_{d,\,0},\,\driftvel$: Here we follow Eq.~\eqref{eqn:lambda.definitions} and impose gradients in the pressure, gas density, dust density, and drift velocity, all along the drift direction. We again take $\gradscalerho=2\,\gradscaleP$, and for $\gradscalew$ take $\gradscalew = -\gradscaleP$ for our ``low-$\gradscale$'' case ($|\gradscaleP|=10^{-4}$) or $\gradscalew = -0.1\,\gradscaleP$ for our ``high-$\gradscale$'' case ($|\gradscaleP|=10^{-1}$). These values of $|\gradscalew|$ ensure that the condition noted above for the solutions to exist, $|\driftvelmag\,\gradscalew| \ll 1$, is met (i.e.\ that the free-streaming scale is shorter than the gradient scale length). The sign of $\gradscalew$ is chosen such that gas and dust densities increase in the same direction. Adding dust-density and drift velocity gradients appears to make a small difference, relative to solutions that already include pressure and density gradients. We will show below that the dust-density gradients dominate the leading-order corrections to the growth rates of the modes at high-$k$; however, in the figure these corrections are small enough so as to be essentially invisible, even though they are technically the leading-order correction.

\item Random $\nabla$(all): In case (iv), we imposed gradients in the drift direction only, following Eq.~\ref{eqn:lambda.definitions}. For completeness here, we now set every component of every gradient  to a different non-zero value. There are 18 gradient components: we first set the four aligned components defined above by $\gradscaleP$, $\gradscalerho$, and $\gradscalew$ above, and then set all other components. These are drawn as uniform random numbers with values between $-|\gradscale|$ and $+|\gradscale|$, where $\gradscale=(\gradscaleP,\,\gradscalerho,\,\gradscalew,\, \gradscalew)$ for each component of ($\nabla P_{0}$, $\nabla \rho_{0}$, $\nabla \rho_{d,\,0}$, $\nabla_{i}  {\rm w}_{s,j}$), respectively. A couple of these components are re-drawn as necessary until a set is obtained which (1) ensures the hydrodynamic system (without dust) is stably stratified ($N^{2} > 0$), and (2) satisfies the constraint equation~\eqref{eqn:gradients}. We also randomly determine the orientation of ${\bf k}_{\bot}$ in the plane perpendicular to $\driftvelhat$. Despite adding a large number of degrees-of-freedom and complexity to the dispersion relation, we see that this has generally small effects on the solutions, compared to the much simpler cases above.

\end{enumerate}

For each of the gradient systems described above and shown in Fig.~\ref{fig:growth.rate.stratification}, we compare two absolute values of the gradients (labeled by $\gradscaleP$), one of which ($|\gradscaleP|=10^{-4}$) is sufficiently small that $\kdimless \sim |\gradscale|$ falls into the wavelength range where the ``long wavelength'' mode dominates, and one of which ($|\gradscaleP|=10^{-1}$) is much larger so that it falls around the ``mid-$k$'' resonant mode. We also compare two signs of the gradients along the $\driftvelhat$ direction: for $\gradscaleP > 0$, pressure, gas, and dust density increase along the drift direction, while for $\gradscaleP < 0$ they decrease.
For simplicity, we focus on the case with $\coeffTSrho=\coeffTSv=0$ (constant $t_{s}$), and consider a single value of $\mu=0.1$ and two representative values of $\driftvelmag$ (a supersonic case with $\driftvelmag=10$ and a subsonic case with $\driftvelmag = 0.5$). For the super-sonic case, we consider modes at the resonant angle $\cos{\theta} = 1/\driftvelmag$, while for the sub-sonic case we consider aligned modes $\cos{\theta}=1$ (which are the fastest-growing in the homogeneous case). 

Overall, the dispersion relations shown Fig.~\ref{fig:growth.rate.stratification} are sufficient to demonstrate the key qualitative behaviors that arise. At lower $\mu$, one does have to go to slightly higher $\kdimless /|\gradscale|$ before the growth rates converge to the homogeneous prediction, as we derive in more detail below. For Epstein or Stokes drag, with $\coeffTSrho$, $\coeffTSrhoonly$, and $\coeffTSv$ all nonzero and $\gamma$ in the range $\gamma \approx 0\rightarrow2$, the qualitative effects of gradients and magnitude of the deviations from the homogeneous case are very similar to the cases shown here. For Coulomb drag, the fact that at low-$k$ the ``decoupling mode'' already exists with high growth rates means that the effects of gradients at low-$k$ are even less important than the cases studied here.

As in the text, for a given $k$ and mode angle, Fig.~\ref{fig:growth.rate.stratification} only shows the most rapidly-growing mode. There are new, albeit slower-growing modes, which appear in the presence of stratification. At certain angles not studied here, the \BV\ RDI can also appear. This causes  sub-sonic streaming to be unstable at {\em all} $k$ with growth rates $\sim |\mu\,\driftvelmag\,\gradscale|^{1/2}$, at the \BV\ resonant angle. This is discussed in \S~\ref{app:subsub BV RDI} below, and in more detail in \citet{squire:rdi.ppd}.

\vspace{-0.5cm}
\subsection{Mode Structure}

The full 9th-order dispersion relation with 18 degrees of freedom is not helpful to write out in full. To understand the relevant behavior, here we consider  each of the key limiting regimes as analyzed in \S~\ref{sec:general.modes} of the main text, but including the leading-order corrections for arbitrary background gradients.

\vspace{-0.5cm}
\subsubsection{The Long-Wavelength / Pressure-Free (Low-$k$) Mode}

First consider behavior at low-$k$, following \S~\ref{sec:long.wavelength} from the text. Expand the dispersion relation to leading order in $\kappa_{\|} \ll \hat{\mu} \lesssim 1$, bearing in mind that we require $|\gradscale| \ll \kdimless$ for the validity of the derivation. 
The dispersion relation can then be written
\begin{align}
\label{eqn:modified.longwavelength.mode.gradients} \left(\frac{\omdimless}{\varomega}\right)^{3} &= \iimag\,\left( 1-\frac{\coeffTSrho}{\tildeCoeffTSv} \right) + \left(\frac{\omdimless}{\varomega}\right)\,\hat{\mu}^{1/3}\,\kappa_{\|}^{2/3}\,\frac{\gradscalemu}{\kdimless}  + \mathcal{O}(\kappa_{\|}^{2+n}\,\gradscale^{1+m}) , 
\end{align}
with $n>0$, $m>0$, and 
\begin{align}
\varomega &\equiv \hat{\mu}^{1/3}\,\kappa_{\|}^{2/3} \ \ \ \ , \ \ \ \ \gradscalemu \equiv - \frac{\driftvelhat \cdot\nabla\mu}{\mu} 
%= \frac{\nabla\cdot\driftvel}{\driftvelmag} + \frac{\driftvelhat\cdot\nabla\rho_{0}}{\rho_{0}} 
\approx \gradscalerho+\gradscalew,
\end{align}
where the latter equality ($\gradscalemu\approx \gradscalerho+\gradscalew$) arises from the general statement ($\gradscalemu = -\mu^{-1}\,\driftvelhat\cdot\nabla\mu = \driftvelhat\cdot [\rho_{0}^{-1}\nabla\rho_{0}-\rho_{d,\,0}^{-1}\nabla\rho_{d,\,0}]$) using the approximations of  \S~\ref{sec:dispersion.simplified.for.gradients}.

The dimensionless term $\epsilon_{\gradscale} \equiv \hat{\mu}^{1/3}\,\kappa_{\|}^{2/3}\,\gradscalemu/\kdimless \sim \mathcal{O}(\gradscalemu/\kdimless)$  gives the (fractional) correction to the mode growth rate. If this is small, this gives exactly the dispersion relation from the text in the homogeneous case (Eq.~\ref{eqn:longwave.mode}), with a small normalization correction $\omdimless \approx [\iimag\,(1-\coeffTSrho/\tildeCoeffTSv)]^{1/3}\,\varomega\,(1 + i_{1}\,\epsilon_{\gradscale}/3\,(1-\coeffTSrho/\tildeCoeffTSv)^{2/3})$
	(where $\iimag_{1}$ is a complex argument with $|\iimag_{1}|=1$, which depends on the signs of $1-\coeffTSrho/\tildeCoeffTSv$ and $\gradscalemu$, and the solution branch chosen). The correction is therefore small so long as $ \hat{\mu}^{1/3}\,\kappa_{\|}^{2/3}\,|\gradscalemu/\kdimless |/3 \ll 1$. However, for the local approximation to be valid we require $|\gradscalemu/ \kdimless | \lesssim |\gradscale/ \kdimless | \ll 1$,  we are explicitly taking the limit $\kappa_{\|} \ll \hat{\mu} \lesssim 1$, and physically we have $\hat{\mu}\ll 1$. Thus {\em every} term in the leading-order correction is small. Moreover, it is worth noting that the nature of Eq.~\eqref{eqn:modified.longwavelength.mode.gradients} is such that the correction term in $\mathcal{O}(\gradscalemu)$ is not stabilizing; solution branches always exist where it (weakly) increases the growth rate.

To summarize, if, in the first place, we meet the conditions required for our local derivation to be valid ($ \kdimless \gg \gradscale$) and for the long-wavelength mode to exist ($\kdimless \ll \hat{\mu}$), then we are almost always guaranteed to also meet conditions for the background gradient terms to be irrelevant for the mode. %The relevant question, physically, is therefore not whether background gradients modify the instability. Rather, it is whether or not the values of $k$ at which the long-wavelength mode is relevant are indeed smaller than the global pressure-scale lengths of the system (discussed in \S~\ref{sec:breakdown}). %It is tedious but straightforward to verify that a nearly-identical result obtains for oblique ($|\cos{\theta}| \ne 1$) modes at low-$k$, as well, although these are not the fastest-growing modes. 

\vspace{-0.5cm}
\subsubsection{Non-Resonant, Short-Wavelength (High-$k$) \Intermediatemodename\ \&\ \Slowmodename\ Modes} 

Now consider the dispersion relation in the high-$k$ limit as in \S~\ref{sec:intermediate}-\ref{sec:slow}. Off-resonance (far from $\cos{\theta} \approx \pm 1/\driftvelmag$) we obtain an identical expression to that in the main text for the ``\intermediatemodename'' mode (Eq.~\eqref{eqn:omega.med}; with leading-order real part $\omega \approx \pm\,c_{s}\,k$). More precisely, to third-from-leading order in $k$, no terms in $\gradscale$ appear. 

 For the off-resonant ``\slowmodename'' mode (Eq.~\eqref{eqn:omega.slow}; with leading-order real part $\omega \approx \driftvel\cdot{\bf k}$), we obtain a leading-order correction $\omega_{\slowmodesubscript} \rightarrow \omega_{\slowmodesubscript}(\gradscaleP=\gradscalew=0) 
 %- \iimag\,\nabla\cdot\driftvel 
 + \iimag\,\rho_{d,\,0}^{-1}\,\driftvel\cdot\nabla\rho_{d,\,0} 
 + \mathcal{O}(\gradscalew^{2},\,\mu\,\gradscalew)$ (where $\rho_{d,\,0}^{-1}\,\driftvel\cdot\nabla\rho_{d,\,0} \approx -\driftvelmag\,\gradscalew$). Because this mode (to leading order) is moving with the dust drift, the statement is simply that the mode (whose growth rate is proportional to the dust density $\rho_{d,\,0}$) grows (decays) in strength along with the mean dust density, as the dust drifts into regions of higher (lower) density. This amounts to a constant offset in the growth rate, important only if (a) the \slowmodename\ mode is present at high-$k$, and (b) the angle is sufficiently far from resonance (where the growth rates one would obtain with $\gradscalew=0$ become small, in our dimensionless units, compared to $\gradscalew$), since $\omega_{\slowmodesubscript}\rightarrow \infty$ as $\theta$ approaches the resonant angle. However, as noted above, we must have $|\driftvelmag\,\gradscalew| \sim |\langle t_{s} \rangle \rho_{d,\,0}^{-1}\,\driftvel\cdot\nabla\rho_{d,\,0}| \sim |\langle t_{s} \rangle\, \nabla\cdot \driftvel|\ll 1$ for the derivation to be valid, so the correction is necessarily small.

%Note that there is a formally unstable solution even when $|k\,\gradscale^{-1}|\ll 1$; however in this limit the instability may be spurious because the implied pressure $=P_{0} + \nabla P_{0}\,\cdot ({\bf x}-{\bf x}_{0})$ becomes negative over the mode wavelength (within $|{\bf x}-{\bf x}_{0}| \ll k^{-1}$). Some real instabilities do exist in this limit, in the few cases with exact background solutions (e.g.\ convective instabilities for a radiation-pressure supported atmosphere; see \citealt{x}), but properly analyzing them requires a global background solution (which in turn requires specifying boundary conditions, etc). 

\vspace{-0.5cm}
\subsubsection{The Intermediate-Wavelength (``Mid-$k$'') Resonant Mode} 

Following \S~\ref{sec:resonance}, now consider the mid-$k$ and high-$k$ modes at the ``resonant angle'' where $\driftvel \cdot {\bf k} = \omega_{0}$ and $\omega_{0}$ is the natural sound-wave frequency of the system without dust. As noted in \S~\ref{sec:solution.mods.nodust.gradients} this is modified, albeit weakly, from the pure sound-wave case by the background gradients to $\omdimless_{0}^{2} = (1/2)\,[\kdimless^{2} + \gradscaleP\,\gradscalerho \pm \{(\kdimless^{2} + \gradscaleP\,\gradscalerho)^{2} + 4\,\kdimless_{\bot}^{2}\,\gradscaleP\,(\gradscaleP-\gradscalerho)\}^{1/2}]$ or $\omega_{0} = \pm \kdimless\,[1 + (1/2)\,|\gradscaleP/\kdimless|^{2}\,[\gradscalerho/\gradscaleP + (\kdimless_{\bot}/\kdimless)^{2}\,(1-\gradscalerho/\gradscaleP)] + \mathcal{O}(|\gradscale/\kdimless|^{4})  ]$. This correspondingly shifts the resonant angle, $\cos{\theta} = \pm \driftvelmag^{-1}\,[ 1 +(1/2)\,|\gradscaleP/\kdimless|^{2}\,(1 + \{\gradscalerho/\gradscaleP-1 \}/\driftvelmag^{2}) + \mathcal{O}(|\gradscale/\kdimless|^{4})]$. 

With this $\omega_{0}$ and $\hat{\bf k}$, taking $\hat{\mu} \ll \kappa_{\|} \ll \hat{\mu}^{-1}$ where the mid-$k$ mode is relevant, we obtain the leading-order correction $\epsilon$ to the growth rate, 
\begin{align}
\nonumber \omdimless &= \kappa_{\|} + \frac{\iimag\pm1}{2}	\left(\left|1-\frac{\coeffTSrho}{\tildeCoeffTSv} \right|\,\hat{\mu}\,\kappa_{\|} \right)^{1/2}\,\left[ 
1 + \epsilon + \mathcal{O}\left( \hat{\mu}^{m}\,\left|\frac{\gradscale}{\kdimless} \right|^{1+n} \right) \right], \\ 
\label{eqn:midk.mode.corr.gradients} \epsilon &\equiv \pm 
\frac{(1+\iimag)}{2\,(1-\coeffTSrho/\tildeCoeffTSv)^{1/2}}
\,\frac{\driftvel\cdot\nabla\rho_{d,\,0}}{(\hat{\mu}\,\kdimless)^{1/2}\,\rho_{d,\,0}}
\approx \frac{\mp\,(1+\iimag)}{2\,(1-\coeffTSrho/\tildeCoeffTSv)^{1/2}}\,\frac{\driftvelmag\,\gradscalew}{(\hat{\mu}\,\kdimless)^{1/2}},
\end{align}
where $m\ge 0$, $n\ge 0$. 
Recall, $|\driftvelmag\,\gradscalew| \ll 1$ is required for our derivation, so the fractional correction $\epsilon$ should usually be small. However, unlike all the still-higher-order corrections from the $\gradscale$ terms, which are un-ambiguously small at all $k$ where our derivation is valid,\footnote{At third-to-leading order, the correction to $\omega$ in Eq.~\eqref{eqn:midk.mode.corr.gradients} becomes considerably more complicated, with $\epsilon \rightarrow \epsilon + \epsilon_{P}^{1} + \epsilon_{\rho}^{1} + \epsilon_{d}^{1} + \epsilon_{w}^{1}$, with 
\begin{align}
\nonumber 
\epsilon_{P}^{1} \equiv\, & \frac{\iimag}{2}\,\left( \frac{\tildeCoeffTSv-\coeffTSrhoonly}{\tildeCoeffTSv-\coeffTSrho}\,\hat{\bf k} - \frac{\driftvel}{c_{s}}\right)\cdot\frac{\nabla P_{0}}{k\,\rho_{0}\,c_{s}^{2}}
\approx \frac{\iimag}{2}\,\left[\frac{\tildeCoeffTSv-\coeffTSrhoonly}{\driftvelmag\,(\tildeCoeffTSv-\coeffTSrho)}-\driftvelmag \right]\frac{\gradscaleP}{\kdimless}, \\ 
\nonumber \epsilon_{\rho}^{1} \equiv\, &\, \frac{\iimag}{2}\,\frac{\coeffTSrhoonly\,\hat{\bf k}}{(\tildeCoeffTSv-\coeffTSrho)}\cdot\frac{\nabla \rho_{0}}{k\,\rho_{0}} 
\approx\, \frac{\iimag}{2}\,\left[\frac{\coeffTSrhoonly}{\driftvelmag\,(\tildeCoeffTSv-\coeffTSrho)}\right]\,\frac{\gradscalerho}{\kdimless}, \\ 
\nonumber \epsilon_{d}^{1} \equiv\, &\frac{\iimag}{2\,(\tildeCoeffTSv-\coeffTSrho)}\left[  \frac{(3\,\tildeCoeffTSv+5\,\coeffTSrho)\,\driftvel}{4\,c_{s}} - \tildeCoeffTSv\,\hat{\bf k}
\right]\cdot \frac{\nabla\rho_{d,\,0}}{k\,\rho_{d,\,0}} \\ 
\nonumber \,&-\, \frac{(\tildeCoeffTSv\,[\tildeCoeffTSv-\coeffTSv/\driftvelmag^{2}]-\coeffTSrho)}{4\,\tildeCoeffTSv\,(\tildeCoeffTSv-\coeffTSrho)} \, \frac{(\driftvel\,\langle t_{s} \rangle)\cdot \nabla\rho_{d,\,0}}{\rho_{d,\,0}} \\
\nonumber \approx\,& -\frac{\iimag\,[\tildeCoeffTSv\,(3\,\driftvelmag^{2}-4) + 5\,\driftvelmag^{2}\,\coeffTSrho]}{8\,\driftvelmag\,(\tildeCoeffTSv-\coeffTSrho)}\,\frac{\gradscalew}{\kdimless} + \frac{(\tildeCoeffTSv\,[\tildeCoeffTSv-\coeffTSv/\driftvelmag^{2}]-\coeffTSrho)}{4\,\tildeCoeffTSv\,(\tildeCoeffTSv-\coeffTSrho)}\,\driftvelmag\,\gradscalew, \\ 
\nonumber 
\epsilon_{w}^{1} \equiv \, & 
-\frac{
[\tildeCoeffTSv\,\hat{\bf k} - \coeffTSv\,\driftvelmag^{-1}\,\driftvelhat]
\cdot\left(\langle t_{s} \rangle\, \nabla \otimes \driftvel \right)\cdot
[\tildeCoeffTSv\,\hat{\bf k} - \driftvelmag\,\coeffTSrho\,\driftvelhat]
}{2\,(\tildeCoeffTSv-\coeffTSrho)} \\ 
 \approx\, & -\frac{(\tildeCoeffTSv/\driftvelmag^{2})-\coeffTSrho)}{2\,\tildeCoeffTSv\,(\tildeCoeffTSv-\coeffTSrho)}\,\driftvelmag\,\gradscalew
\end{align}
(note that $\nabla\otimes\driftvel$ is a tensor here). Although this is complicated, note that every term here is suppressed by a power of $|\gradscale/\kdimless|\ll 1$, or $|\driftvelmag\,\gradscale|\ll 1$, or both, with only order-unity pre-factors. For example for highly super-sonic Epstein drag we just have $\epsilon_{P}^{1} + \epsilon_{\rho}^{1} + \epsilon_{d}^{1} + \epsilon_{w}^{1} \rightarrow -(\iimag/2)\,\driftvelmag\,\gradscaleP/ \kdimless - (11\,\iimag/8)\,\driftvelmag\,\gradscalew/ \kdimless + (5/8)\,\driftvelmag\,\gradscalew$, so these terms (which appear as fractional corrections to the growth rate) are all small.}
the leading-order fractional correction here has a power of $\sim \hat{\mu}^{-1/2}$, so could be important at sufficiently small $\hat{\mu}$. 

Equivalently, we can take the imaginary part of Eq.~\eqref{eqn:midk.mode.corr.gradients} to write the growth rate as $2\Im{(\omdimless)} \approx (|1-\coeffTSrho/\tildeCoeffTSv|\,\hat{\mu}\,\kdimless)^{1/2} - \driftvelmag\,\gradscalew$. We see that the leading-order term in $\gradscale$ is the same (up to a constant pre-factor) constant offset in the growth rate that we saw in the off-resonant \slowmodename\ mode. Since the absolute correction to the growth rate is constant (or, equivalently, the fractional correction $\epsilon$ scales $\propto k^{-1/2}$), it must be negligible at high-$k$, specifically when $\kdimless \gg (\driftvelmag\,\gradscalew)^{2}/\hat{\mu}$. Now recall from \S~\ref{sec:intermediate.mode.at.resonance} that this mid-$k$ mode is present (and is the fastest-growing mode) for $k$ in the range $\hat{\mu} \ll \kdimless \ll \hat{\mu}^{-1}$. Since $|\driftvelmag\,\gradscalew| \ll 1$, this means there must always exist a range of $k$ where $(\driftvelmag\,\gradscale)^{2}/\hat{\mu} \ll \kdimless \ll 1/\hat{\mu}$ and thus the correction term $\epsilon$ is negligible.

However, if $\mu$ is very small, such that $\hat{\mu} \lesssim |\driftvelmag\,\gradscalew|^{2} \ll 1$, then at small $k$ where $\hat{\mu} \ll \kdimless \ll  (\driftvelmag\,\gradscale)^{2}/\hat{\mu}$, the growth rate of this mode can be modified significantly. The mode will then either grow faster or slower, depending on whether the dust is drifting into regions of higher or lower dust density on a timescale short compared to the mode-growth time.

\vspace{-0.5cm}
\subsubsection{The Short-Wavelength (High-$k$) Resonant Mode} 

Again taking the resonant condition and expanding the dispersion relation, now at high $k$ as in \S~\ref{sec:resonance}, we find it is identical to the homogeneous ($\gradscaleP=\gradscalew=0$) case at leading ($\mathcal{O}(\kappa_{\|})$) and next-to-leading ($\mathcal{O}(\kappa_{\|}^{1/3})$) orders. The first correction term from background gradients appears at third-to-leading order, in the constant ($\mathcal{O}(\kappa_{\|}^{0})$) correction to the growth rate $\omegaZ$ in Eq.~\eqref{eqn:omega.resonant}, where
\begin{align}
\nonumber \omegaZ &\rightarrow \omegaZ + \frac{\langle t_{s} \rangle}{3}\left[\frac{[\hat{\bf k} + (\Theta-1)\,\driftvelhat ] \cdot \left( \nabla \otimes \driftvel \right)\cdot \hat{\bf k}}{\Theta} -\frac{\driftvel\cdot\nabla\rho_{d,\,0}}{\rho_{d,\,0}}\right], \\ 
&\approx \omegaZ + \frac{1}{3}\,\driftvelmag\,\gradscalew\,(1 + \Theta^{-1}),
\end{align}
where  $\Theta \equiv 1-\coeffTSrho + \coeffTSv/\driftvelmag^{2}$ and $\otimes$ denotes the outer product.
This is not surprising, since the gradients in the gas properties only enter the resonant mode in the {\em gas} at $\mathcal{O}(|\gradscale/ \kdimless |^{2})$ at high-$k$, and (as noted for the ``\slowmodename'' mode above) a divergence in the dust velocity/density $\gradscalew$ enters as a constant offset in the growth rate for modes moving with the mean dust motion. 

Because $|\driftvelmag\gradscalew|\ll 1$, and  since this correction only appears in the constant term (while the dominant term in the growth rate is increasing with $k$), it becomes a vanishingly small correction to the mode at high-$k$.

\vspace{-0.5cm}
\subsubsection{New Instabilities: The \BV\ RDI} \label{app:subsub BV RDI}

In addition to the acoustic modes above, which we showed are not fundamentally altered by the background gradient terms, new unstable modes appear due to the stratification. As noted above, with these gradient terms, the dispersion relation for $\omega_{0}$ is modified to include two branches: both the usual sound wave modes ($\omega_{0} \sim \pm c_{s}\,k$) and buoyancy modes ($\omega_{0} \sim \pm N_{BV}$, the \BV\ frequency). As shown in \paperone, any  mode of the gas without dust introduces a corresponding RDI when $\driftvel\cdot{\bf k} = \omega_{0}$, and the \BV\ RDI is one of the examples discussed there (within the Boussinesq approximation, which eliminates the sound waves). These modes have $k_{\|} \approx \pm N_{BV}/(|\driftvel| \langle t_{s} \rangle) \sim \gradscale/ (|\driftvel| \langle t_{s} \rangle)$ (recall that $N_{BV}^{2} = \gradscaleP\,(\gradscalerho-\gradscaleP)$), and growth rates $\Im(\omdimless)\sim (\hat{\mu}\,\driftvelmag\,\gradscaleP)^{1/2}$ in our units. However the \BV\ RDI is fundamentally distinct from the acoustic RDI (the resonance is with buoyancy oscillations with $\omega_{0}\sim$\,constant, not sound waves), so we do not show or discuss them here, but instead explore them separately, in a more detailed analysis (which also allows for explicitly incompressible or compressible fluids) in \citet{squire:rdi.ppd}. We also note that they are also never the fastest-growing mode when the acoustic RDI resonance is possible ($\driftvelmag>1$) and $\kdimless\gg\gradscale$, although they could certainly be important and the fastest-growing mode if the acoustic RDI is not present.

\vspace{-0.5cm}
\subsection{Summary}

We have considered the dispersion relation allowing {\em every} component of the gradients of $P_{0}$, $\rho_{0}$, $\rho_{d,\,0}$ and $\driftvel$ to have arbitrary values, subject only to the constraints in \S~\ref{sec:validity.gradients} necessary for our local approximation to the equations of motion to be valid ($\kdimless \gg |\gradscale|$, $|\driftvelmag\,\gradscale|\ll 1$). It is worth noting that at leading order in $\gradscale/\kdimless$ (and up to third-from-leading order in the other relevant expansion parameters for each mode considered above) the  pressure gradient term, which allows the system to be hydrostatic and motivated this study,  does not appear. Likewise for any transverse gradient terms.

In fact, the leading-order corrections all follow from the derivative of the background {\em dust} properties (density or drift velocity) {\em along the direction of the drift}. These corrections, which  appear for those  modes that are (to leading order) ``moving with'' the drift, have a simple physical interpretation. Because the relevant mode growth rates depend on the dust-to-gas ratio (and drift velocity), the physical statement is simply that as a mode moves into regions of larger (smaller) dust-to-gas ratio, the mode growth rates correspondingly increase (decrease). However, these would represent significant corrections to the growth rates (relative to the spatially homogeneous case in the main text) only if the parameter $\driftvelmag\,\gradscalew \sim \langle t_{s} \rangle \nabla \cdot \driftvel \sim \langle t_{s} \rangle\,\rho_{d,\,0}^{-1}\, \driftvel\cdot \nabla \rho_{d,\,0}$ were large -- i.e.\ if the dust ``free-streaming'' length were larger than the gradient scale-length of the equilibrium dust distribution. Obviously in this regime our local expansion is invalid.

Finally, we note that these corrections do not fundamentally alter the character or dimensional scalings of the relevant acoustic RDI, provided $\kdimless \gg \gradscale$ (they only modify the growth rates by some numerical pre-factor). Most importantly, they do not stabilize the system in any systematic sense. 
In fact,  they can introduce more instabilities, for instance the \BV\ RDI (\paperone), which is explored in  detail in \citet{squire:rdi.ppd}.

\end{document}